\newcommand{\lnls}{log\,N($>$S)-log\,S}

\newcommand{\xmm}{XMM-Newton }
\newcommand{\nh}{N$_{\rm H}$}
\newcommand{\Halp}{H${\alpha}$}
\newcommand{\Hbet}{H${\beta}$}
\newcommand{\ergscm}{\,erg\,cm$^{-2}$\,s$^{-1}$}
\newcommand{\Fx}{$\rm F_{\rm X}$}
\newcommand{\ergs}{\,erg\,s$^{-1}$}

\newcommand{\Teff}{$T_{\rm eff}$\,}

\newcommand{\degree}{\degr}
\newcommand{\Lbol}{\rm L$_{\rm bol}$}
\newcommand{\Lx}{$\rm L_{\rm X}$}

\newcommand{\chisqr}{$\chi^2$}

\documentclass[traditabstract]{aa} 
\usepackage{psfig}
\usepackage{graphicx}
\usepackage{txfonts}
\usepackage[authoryear]{natbib}
\usepackage{lscape}
\usepackage{booktabs}
\usepackage{subfig}
\begin{document}
   \title{The X-ray source content of the XMM-Newton Galactic Plane Survey}

   \subtitle{}

   \author{C. Motch
          \inst{1}\fnmsep\thanks{Based on optical observations obtained at ESO under programmes 69.D-0143, 70.D-0227 and 71.D-0552}
          \and
          R. Warwick \inst{2}
	  \and 
	  M. S. Cropper \inst{3} 
	  \and
          F. Carrera \inst{4}
	  \and
	  P. Guillout \inst{1}
	  \and  
	  F.-X. Pineau \inst{1}
	  \and
	  M. W. Pakull \inst{1}
	  \and
	  S. Rosen \inst{2}
	  \and 
	  A. Schwope \inst{6}
	  \and
	  J. Tedds \inst{2}
	  \and	  
	  N. Webb  \inst{5}
	  \and
	  I. Negueruela \inst{7}
	  \and
	  M.G. Watson \inst{2}
          }
    \institute{
         CNRS, Universit\'e de Strasbourg, Observatoire Astronomique, 11 rue de l'Universit\'e, F-67000 Strasbourg, France
         \and
         Department of Physics and Astronomy, University of Leicester, LE1 7RH, UK
	 \and
	 Mullard Space Science Laboratory, University College London, Holmbury St. Mary, Dorking, RH5
         6NT Surrey, UK 
         \and
         Instituto de F\`isica de Cantabria (CSIC-UC), 39005 Santander, Spain
	 \and
	 Centre d'Etude Spatiale des Rayonnements, 9 Avenue du Colonel Roche, 
         31028 Toulouse Cedex 4, France
	 \and
	 Astrophysikalisches Institut Potsdam, An der Sternwarte 16, 14482 Potsdam, Germany
	 \and
        Departamento. de F\'{i}sica, Ingenier\'{i}a de Sistemas y Teor\'{i}a de la Se\~{n}al, Universidad de Alicante, Apdo. 99, E03080 Alicante, Spain}
	
   \date{Received ; accepted}

\abstract
{We report  the results  of an  optical campaign carried  out by  the XMM-Newton
Survey Science Centre with the  specific goal of identifying the brightest X-ray
sources in  the XMM-Newton  Galactic Plane  Survey of Hands  et al.  (2004).  In
addition to  photometric and spectroscopic observations obtained  at the ESO-VLT
and  ESO-3.6m, we  used cross-correlations  with the  2XMMi, USNO-B1.0,  2MASS and
GLIMPSE  catalogues  to progress  the  identification  process.  Active  coronae
account for 16 of the 30  positively or tentatively identified X-ray sources and
exhibit the  softest X-ray spectra.  Many  of the identified  hard X-ray sources
are  associated with  massive stars,  possibly in binary systems and emitting at  intermediate X-ray
luminosities  of 10$^{32-34}$\ergs.   Among  these are  a  very absorbed  likely
hyper-luminous star with X-ray/optical  spectra and luminosities comparable with
those of $\eta$  Carina, a new X-ray selected WN8 Wolf-Rayet  star in which most
of the  X-ray emission probably  arises from wind  collision in a binary,  a new
Be/X-ray  star belonging  to the  growing class  of $\gamma$-Cas  analogs  and a
possible supergiant  X-ray binary of  the kind discovered recently  by INTEGRAL.
One of  the sources,  XGPS-25, has a  counterpart of moderate  optical luminosity
which exhibits He{\sc II} $\lambda$4686 and Bowen C{\sc III}-N{\sc III} emission lines suggesting that this may be  a quiescent or  X-ray shielded Low Mass X-ray Binary, although its X-ray properties might also be consistent with a rare kind of cataclysmic variable (CV).   We also report the  discovery of three  new CVs, one  of which  is a  likely  magnetic system displaying strong  X-ray  variability.   The  soft  (0.4--2.0\,keV)  band  \lnls\  curve  is completely dominated by active stars in  the flux range of 1$\times$10$^{-13}$
to 1$\times$10$^{-14}$\ergscm . Several active coronae are also detected above
2\,keV  suggesting  that  the   population  of  RS  CVn  binaries  significantly
contributes  to the  hard X-ray  source  population. In  total, we  are able  to
identify a large fraction of the hard (2--10\,keV) X-ray sources in the flux range
of 1$\times$10$^{-12}$ to 1$\times$10$^{-13}$\ergscm\ with Galactic objects  at a rate consistent  with that expected for the Galactic contribution  only.}


   \maketitle
%

\section{Introduction}

The brightest X-ray sources discovered by the early collimator based experiments
in   operation  in  the   1970s  were   very  luminous   ($\sim$  10$^{38}$\ergs )
accretion-powered  X-ray binaries located  in our  Galaxy \citep[see e.g. the fourth Uhuru catalogue of X-ray sources;][] {forman1978}.  The dramatic
improvement  in  sensitivity  and   spatial  resolution  afforded  by  the  next
generation of instruments exploiting X-ray imaging revealed a Galactic landscape
with much larger numbers of lower  luminosity systems powered by a wide range of
physical processes.   For example, observations  first with Einstein,  and later
with ROSAT, showed  that low Galactic latitude regions are  crowded with a large
number  of low-luminosity sources  in  the X-ray  bands explored  by these  two
missions (0.5--4.5\,keV  and 0.2--2.4\,keV  for Einstein and  ROSAT respectively).
These  soft sources  are predominantly  identified with  active  stellar coronae
\citep[e.g.][]{hertz1984,motch1997}.  The  first imaging hard  (2--10\,keV) X-ray
survey  of  the  Galactic  Plane  was  performed  a  few  years  later  with ASCA
\citep{sugizaki2001}.   ASCA  revealed  the  presence of  a  genuinely  Galactic
population of  low-luminosity non-coronal  X-ray sources which emit  hard X-rays
and   are   therefore  detectable   through   large   columns  of   interstellar
absorption. However, the still relatively large positional errors affecting ASCA
sources combined with the effects of the strong interstellar
extinction in the Galactic Plane prevented their optical identification  in most cases. The prospects for
building a detailed  census of the low to  intermediate X-ray luminosity sources
has  now  greatly  improved  with  the  launch of  the  Chandra  and  XMM-Newton
observatories. Thanks to their unprecedented sensitivity and spatial resolution,
these observatories are  able to study the properties of  the faint X-ray source
populations  throughout the  Galaxy and  most  notably those  present in  large
numbers in the  central parts of the Galaxy \citep{wang2002,muno2003}.   The fields-of-view of the
Chandra and  XMM-Newton X-ray cameras  are such that many  serendipitous sources
are  seen in  low-latitude fields  in addition  to the  primary targets  and the
preliminary results of  such surveys have already emerged in  the context of the
Champlane and XMM-Newton SSC surveys
\citep{grindlay2003,motch2003,hands2004,grindlay2005,motch2006}. A number of
dedicated Galactic surveys involving either single deep observations or a mosaic
of shallower  exposures have
also been carried out or are in progress, \citep[e..g. the Galactic Centre region,][]{muno2003,wijnands2006,koenig2008,revnivtsev2009,muno2009}.

The nature of the intermediate- to low-luminosity Galactic hard X-ray sources discovered in the Galactic Plane surveys quoted above is only partially known. Cataclysmic variables (CVs) and quiescent X-ray binaries (both low- and 
high-mass types) account for a significant fraction, however, the exact nature of the majority of the sources detected in the central regions of the Galaxy remains doubtful in the absence of optical or infrared identifications,  which in most cases are hindered by the large interstellar absorption. 

The observation  of intermediate-  to low-luminosity Galactic X-ray sources can
address many issues related to binary-star evolution and to accretion physics at low rates. 
In addition surveys which extend the search boundaries have the 
potential to improve our knowledge of source population statistics 
and to unveil rare types of compact X-ray-emitting systems which
have a low space density in the local Galactic neighbourhood. For instance,  the determination of the space  density and  scale  height of CVs is relevant to estimates of the Galactic novae rate and connects to the origin of low-mass
X-ray  binaries and  type Ia  supernovae. Low  X-ray luminosity  but  long lived
evolutionary stages of classical low and  high mass X-ray binaries could also be
found  in  relatively large  numbers  if  the  predictions of  some  evolutionary
scenarios are  correct  \citep[see e.g.][]{pfahl2002,willems2003}.  
Massive X-ray  binaries are good proxies  of relatively
recent star-bursts and if detected in sufficient numbers might give insight into
the recent star-formation history of specific regions of our Galaxy. 

Moreover, no  X-ray telescope  planned in the  foreseeable future will  have the
necessary  spatial   resolution  and  sensitivity   to  study  in   detail  the
distribution  of  these low-  and  intermediate-luminosity  sources in  external
galaxies.  Consequently,  X-ray surveys of our  own Galaxy and its Magellanic
satellites will remain for  a long time the only means to  study how the various
X-ray emitting  populations relate to the  main Galactic structures  such as the
thin disc, the thick disc and the bulge.

However, such  studies presuppose the knowledge  of the nature of  the bulk of the X-ray sources  surveyed, a prerequisite
which can  only be fulfilled through multi-wavelength spectroscopic identifications. Nevertheless, the statistically
controlled cross-correlation with large catalogues such as  the 2MASS catalogue and/or with the recently made available  source lists from the GLIMPSE surveys, 
provide useful information  on the  intensity and  shape  of the  part of  the spectral  energy distribution in the observing
windows least affected by line-of-sight absorption in the interstellar medium.

The work  presented here was  carried out in  the framework of a  Galactic Plane
Survey programme instigated by the  XMM-Newton Survey Science Centre (SSC).  The
long term  objective of  this project  is to gather  a representative  sample of
identified  low-latitude X-ray  sources,  which  eventually could  be  used as  a
template  for identifying  and classifying  in a  statistical manner  the entire
catalogue  of  serendipitous  XMM-Newton  sources  detected  in  the  Milky  Way
region.  Companion programs  are being  carried out  at high  Galactic latitudes
using three separate samples of sources selected at faint
\citep{furusawa2008}, medium \citep{barcons2002} and bright
\citep{dellacecca2004} fluxes. 

In the first part of this paper, we recall the main properties of the  \xmm Galactic Plane Survey (XGPS) of \cite{hands2004} and describe the way a sub-sample of X-ray sources were selected for optical follow-up at the telescope. The following section explains the method applied to cross-correlate the XGPS source list with the USNO-B1.0 and 2MASS catalogues and how identification probabilities were assigned to each optical or near infrared candidate. After a description of the observing procedures, the next section presents in detail the results of our optical spectroscopic identification campaign. The last sections discuss the \lnls\ curves and the statistics of the various kinds of X-ray emitters identified in the soft and hard bands as well as their contributions to the genuine Galactic population of low to medium X-ray luminosity sources. 

\section{The XGPS source sample}

\begin{figure*}
\begin{tabular}{cc}
\psfig{figure=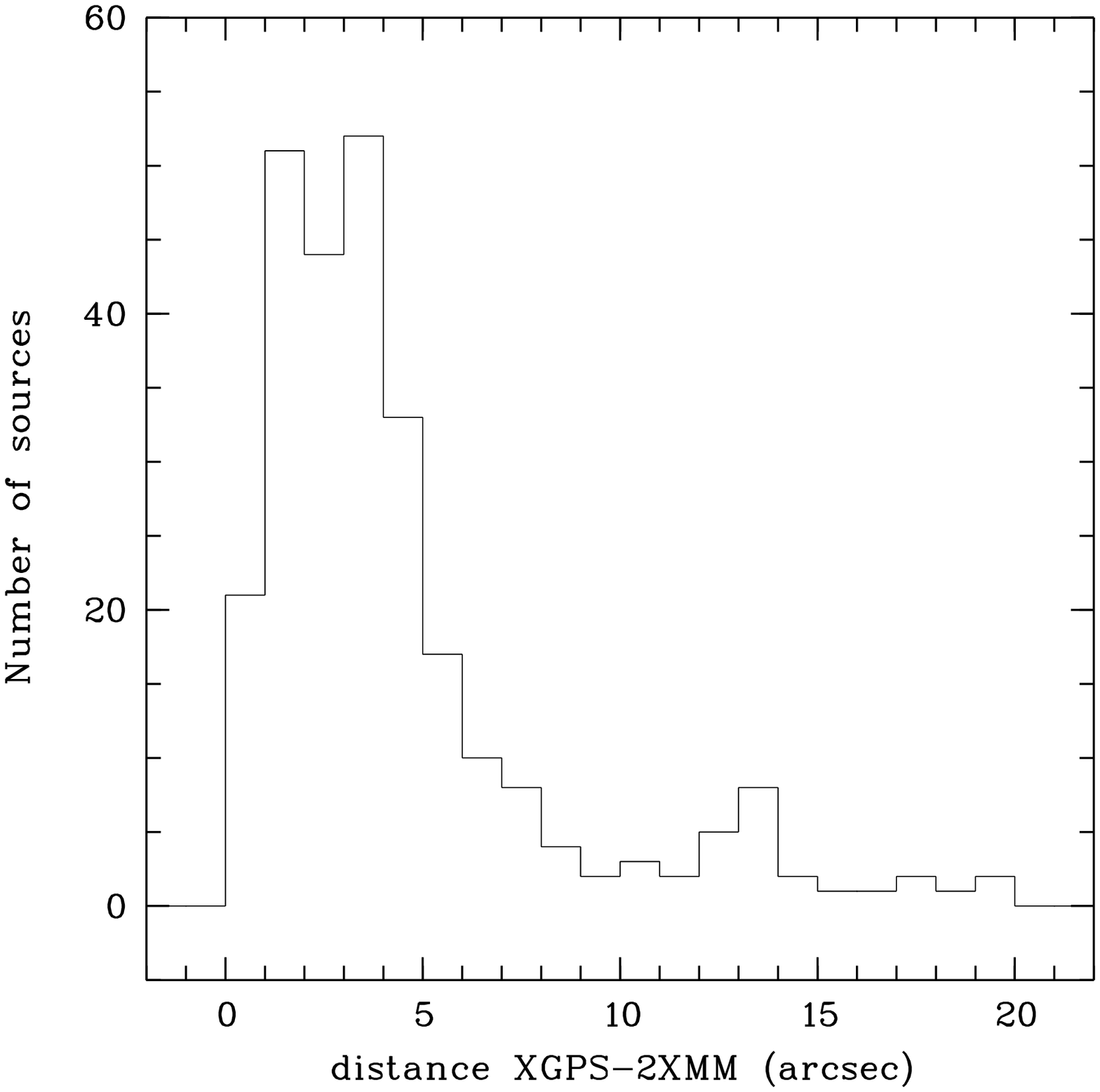,width=8.5cm,bbllx=1.0cm,bburx=21.5cm,bblly=1cm,bbury=21.5cm,angle=-90,clip=true} &
\psfig{figure=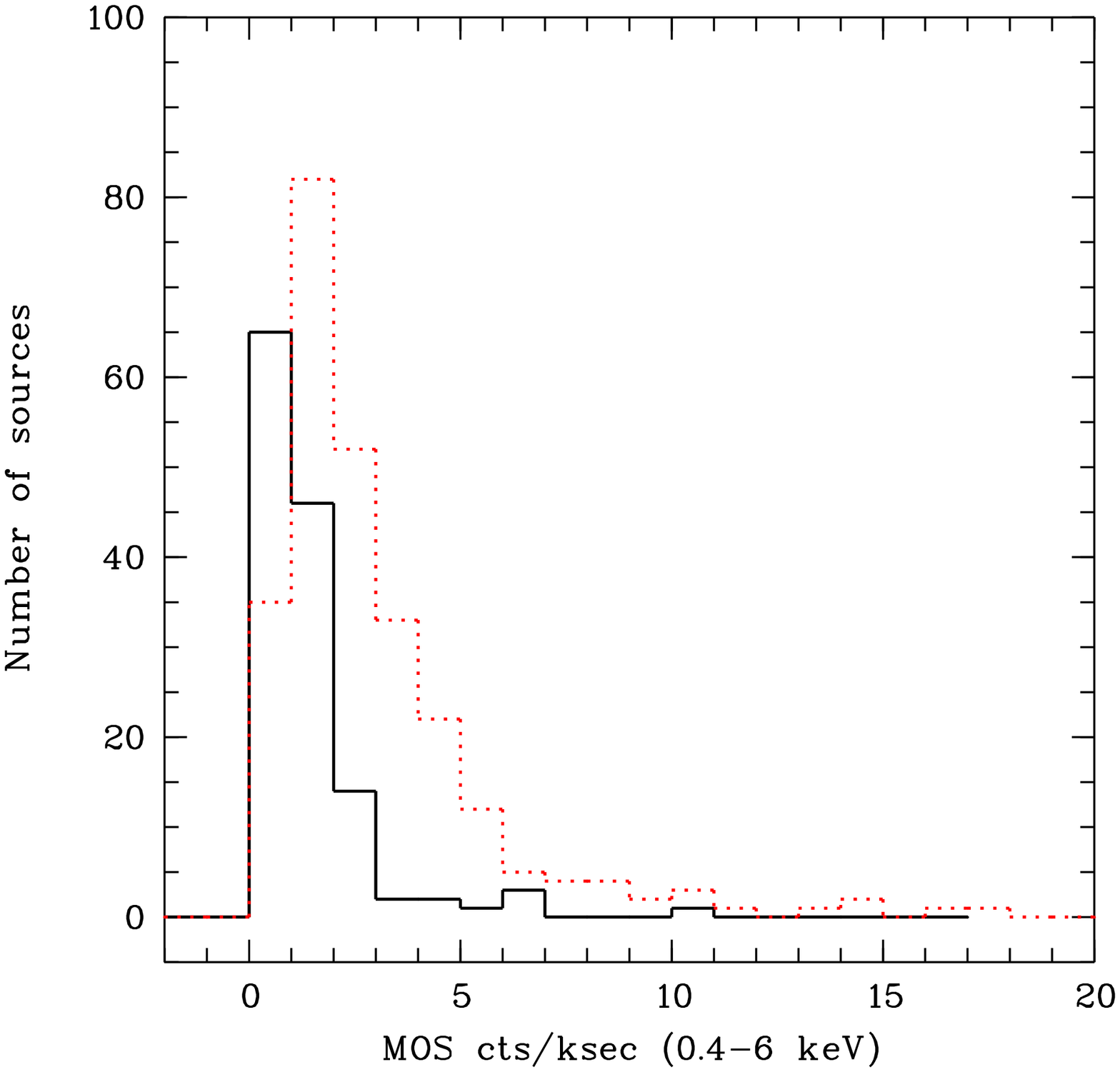,width=8.5cm,bbllx=1.0cm,bburx=21.5cm,bblly=1cm,bbury=21.5cm,angle=-90,clip=true}  \\
\end{tabular} 
\caption{Left: Histogram of distances between original XGPS positions and 2XMM 
positions. Right: Histogram of the XGPS MOS count rates for XGPS sources with  a matching 2XMM entry (dashed red line) and without (plain black). The XGPS-2 field which is not part of the 2XMM catalogue has been excluded from the sample.}
\label{plot_Histo_d_XGPS_2XMM}
\end{figure*}

The  XGPS was the  first survey  of an  extended segment  of the  Galactic Plane
carried  out  by XMM-Newton  \citep{hands2004}.  It  covers  a narrow  strip  of
1.2\degr\  in Galactic  latitude,  extending between  19\degr\  and 22\degr\  in
Galactic  longitude,   centered  on  the  Galactic  Plane   and  covering  about
3\,deg$^{2}$. With an  exposure time ranging between 6.6  and 13.7\,ksec for the
MOS cameras (but typically $\sim$ 2 ksec  less for the pn camera due to its longer
set-up  time),  the  limiting  sensitivity   is  of  the  order  of  4$\times$10$^{-15}$\ergscm\   in  the   0.4--2\,keV  band   and  about   2$\times$10$^{-14}$\ergscm\ in the 2--10\,keV  band. The soft band sensitivity is about
a  factor 30  deeper than that  of the  ROSAT all-sky  survey at low Galactic latitudes
\citep[e.g.][]{motch1997}.  In  hard X-rays, the XGPS reaches  limiting fluxes a
factor 10 or more fainter than the ASCA Galactic Plane Survey
\citep{sugizaki2001}.  Chandra is in principle able to reach significantly
fainter flux  limits. For instance, the  deep observation at  $l$ = 28.5\degr,
$b$ =  0\degr\ by \cite{ebisawa2005} detects  sources down to  $\sim$ 3$\times$10$^{-15}$\ergscm\ in the hard band and $\sim$ 2$\times$10$^{-16}$\ergscm\ in
the  soft band.  The  Champlane project  \citep{grindlay2005}  based on  archive
observations  with a  broad spread  of exposure  times spans a  range  of sensitivity extending down to this deep survey limit.  In summary, the XGPS can be described as  a medium  sensitivity  survey  offering unique  coverage  of several  square degrees of the Galactic Plane around  $l$ = 20\degr\ .  This is a direction in which the line-of-sight column density, \nh, integrated through the Galaxy, ranges 
from log(\nh) $= 22.5$ to $23.5$. It is also  a region unhindered by the presence of
one  or  more  very  bright  Galactic sources  \citep[see][for  more details]{hands2004}.

The catalogue of X-ray sources derived from the XGPS survey by \cite{hands2004}
was  based  on a  customized  background-fitting  and source-detection  software
package, that pre-dated the full  development of the XMM-Newton Science Analysis
Software (SAS). The very robust  XMM-Newton X-ray source catalogues that are now
available \citep[e.g. the 2XMM and 2XMMi catalogues,][]{watson2009}, which utilise the latest software
and calibration information, provide  an opportunity for consistency checking of
the original  XGPS data. In this  context we have  cross-correlated the original
XGPS  list of  sources  with the  2XMMi  catalogue in  order  to retrieve  fully
up-to-date X-ray source statistics and parameters. Among a total of 424 original
XGPS  sources,   269  have   a  match  within   20\arcsec\  with   a  2XMM/2XMMi
entry. This rather large search radius was chosen so as to cope with possibly large positional errors affecting some of the original XGPS source positions. Enlarging  the search radius beyond  20\arcsec\ only brings  a few extra matches. None  of the new fields entered in the  incremental version of the
2XMM catalogue (2XMMi)  overlaps with the XGPS area  and therefore, all matching
sources also belong to the original 2XMM catalogue{\footnote{The processing used to generate the incremental 2XMMi catalogue is the same as used for 2XMM. The 2XMMi catalogue has 626 more observations and about 17\% more detections than the 2XMM catalogue}. Among the 155 XGPS sources not present in
2XMM, 21 are located in the  XGPS-2 field (Obsid 0051940201) which due to an
ODF format problem could not be processed by the 2XMMi production pipeline.  The
distribution of the positional offsets between XGPS and 2XMM sources plotted in the left panel of  Fig. \ref{plot_Histo_d_XGPS_2XMM} shows  that most matches occur  at a distance less than  5\arcsec\ but that a small group of  XGPS sources tend to
be systematically offset by 12 to 14 arcsec.  The positions of only 22 of the 269 2XMM sources could be corrected for residual attitude errors by {\em eposcorr}. This SAS task cross-correlates the observation source list with the USNO-B1.0 catalogue and searches for systematic offsets and rotation of the EPIC field with respect to the optical astrometric references.  Typically, raw positions have 90\% confidence error radii of the order  of 4-5\arcsec\ while  attitude corrected  positions have  90\% confidence error  radii of the  order of  3\arcsec\ \citep{watson2009}. The relatively low EPIC exposure times and consequently small number of sources per observation combined with a high density of field stars at these low Galactic latitudes probably explain the failure of {\em eposcorr} to find significant correction patterns in most cases. This also implies that the XGPS-2XMM offsets are not due to the application/non-application of {\em eposcorr}, since no {\em eposcorr}-like task was used to correct the original XGPS-I coordinates,  but rather result from the different algorithms used for source centering and attitude reconstruction. XGPS sources without corresponding
2XMM entries tend to be significantly  fainter than those having a 2XMM match as
evidenced in  the right panel  of Fig. \ref{plot_Histo_d_XGPS_2XMM}. It  seems a
reasonable conjecture  that the 134 XGPS sources not  recovered in 2XMM have,
in effect, a  lower maximum likelihood than the threshold applied  in 2XMM (ML =
6). As a consequence the fraction of spurious sources in this fainter sample may
well  be  considerably higher  than  is the  case  for  the confirmed  XGPS/2XMM
matches.  

As a second cross-check, in March 2008, we searched for Chandra observations covering the field of view of the XMM-Newton pointings involved in the XGPS. Unfortunately, the overlap  is not large,  with only 4 of the 22 XMM-Newton  fields coincident with Chandra observations (obsids 5563, 6675 and 7479). We extracted from the Chandra  X-ray Center archive the source lists produced by the standard data processing and created by {\em celldetect} version DS7.6 for the total energy range.
The merged  ACIS source lists were then cross-correlated with the  XGPS catalogue
using  a search  radius of  20\arcsec. The  process resulted  in only  16 source
matches. We  list in  Table \ref{xcorxgpschandra}  the distance  to the  matching Chandra source of  the XGPS  source, of its  corresponding 2XMM  entry and of  the 2MASS
entry matching  the 2XMM source when  available (see below \S \ref{xcorsection}). The quoted errors on the offsets
between the 2XMM and Chandra source positions were derived by adding the 
individual positional errors in quadrature. Chandra positions have a typical one $\sigma$ error in the range of 0.1\arcsec\ to 0.6\arcsec. The 2XMM entries are in most cases
closer to  the Chandra position than  the original XGPS positions, reflecting the improved precision of 2XMM positions compared to those of the XGPS. Moreover, the 2XMM coordinates are in all cases but one consistent at the 90\% confidence with the Chandra values.  The group of XGPS  sources located  far (10  to 17\arcsec)  from their  2XMM  counterparts have Chandra positions consistent  with those of the 2XMM  source, clearly indicating that the  large offset is not  due to an  attitude error in the  2XMM catalogue. Interestingly, the  vast majority  of the 2MASS candidate counterparts are located less than 1\arcsec\  from the Chandra position, the two discrepant  cases being those for which  the probability  of identification, computed  by the  likelihood ratio algorithm, was less than 14\% (see \S \ref{xcorsection}). 

In planning the  optical follow-up observations (which were carried out in 2002
and 2003, see  \S \ref{s3}) it was necessary,  of course, to make use  of the
X-ray source flux and position information available at the time. In effect this
was  a combination of  preliminary XGPS catalogue  data and  information derived  from early versions of  SAS-pipeline processing. The error circles available at the time of the optical observations were systematically larger than those provided by the 2XMM catalogue. This mainly results from the improved astrometric accuracy provided by the latest SAS versions illustrated by the fact that the systematic positional error went down from 1.5\arcsec\ to 1.0\arcsec\ in the absence of {\em eposcorr} correction. In addition, the 2XMM catalogue now merges detections from distinct exposures to derive improved positions.
Consequently, we observed on occasions optical candidates that were found out later to be well outside the revised error circle. However, in describing the multiwaveband follow-up we use in the main, when available, source parameters derived from the most recent 2XMM
catalogues \citep{watson2009}, including most importantly  the positional information (e.g.  see \S 3).   That said, the  X-ray  bright,  sub-sample  of   XGPS  sources  studied spectroscopically  was  selected  on   the  basis  of  the  original  count-rate information from  \cite{hands2004} consisting  of MOS camera  count rates  in a soft (0.4--2\,keV),  hard (2--6\,keV) and broad (0.4--6\,keV) band. More specifically the  full preliminary XGPS  catalogue was  sorted into  a  list ordered  on the  basis of  a decreasing  MOS  count  rate  in  the  broad band  (0.4--6\,keV).   Objects  were subsequently  studied at  the telescope  in effect  by working  down  this list. Fortunately, the  post facto rationalisation  in terms of a  XGPS/2XMM selection does not have a  large impact on this bright source sub-sample.   In the event a total of  43 X-ray sources down  to a MOS  broadband count rate of 4.2 counts/ks
were investigated in detail (see \S 5 for further details).

\begin{table*}[ht]
\caption{XGPS sources with matches in archival Chandra catalogues. The columns 
give the following information: (1) - XGPS identifier. (2) - Name of the corresponding 2XMM source. (3) - 2MASS entry correlating with the 2XMM source. (4) - Probability of identification of the 2XMM source with the 2MASS entry. (5) - $d$ XGPS$-$Chandra. (6) - $d$ 2XMM$-$Chandra. (7) - 90\% confidence error on $d$ 2XMM$-$Chandra. (8) $d$ 2MASS$-$Chandra. (9) and (10) Chandra position. All position offsets are quoted in arcsec.}

\begin{tabular}{rccccccccc}
\hline \hline
 XGPS& 2XMM                & 2MASS          &Ident &$d$   & $d$  & error & $d$   & \multicolumn{2}{c}{Chandra} \\
   ID& Name                & Name           & Prob & XGPS & 2XMM &       & 2MASS & RA & DEC \\
  (1)& (2)                 & (3)            & (4)  & (5)  & (6)  & (7)   &  (8)  &    (9)  &  (10)    \\
\hline
   12&2XMM J182830.8-114514&18283106-1145159&0.44  &  3.8 &  4.0 & 2.4   &  0.2  &   18 28 31.05&-11 45 15.7\\
   23&2XMM J183017.0-095326&18301701-0953272&0.96  &  2.5 &  1.0 & 2.6   &  1.2  &   18 30 16.96&-09 53 26.8\\
   29&2XMM J182740.3-113953&18274040-1139532&0.94  &  2.3 &  0.4 & 1.7   &  0.6  &   18 27 40.44&-11 39 53.1\\
   73&2XMM J182732.2-113357&18273223-1133571&0.98  &  3.0 &  0.6 & 1.9   &  0.7  &   18 27 32.27&-11 33 57.4\\
   74&2XMM J182750.5-113549&		    &  -   &  4.8 &  1.0 & 2.2   &  -	 &   18 27 50.66&-11 35 48.6\\
  110&2XMM J182728.5-113741&18272856-1137400&0.92  &  3.2 &  1.7 & 2.3   &  0.7  &   18 27 28.61&-11 37 40.1\\
  124&2XMM J182811.1-114933&18281123-1149341&0.60  & 11.3 &  1.8 & 3.8   &  0.6  &   18 28 11.27&-11 49 34.3\\
  142&2XMM J182744.6-113957&18274465-1139576&0.98  &  3.1 &  0.2 & 3.0   &  0.2  &   18 27 44.66&-11 39 57.7\\
  161&2XMM J182812.3-114144&18281235-1141399&0.10  & 17.3 &  3.8 & 4.2   &  6.3  &   18 28 12.55&-11 41 45.5\\
  168&2XMM J182749.5-113725&18274944-1137264&0.80  &  2.5 &  1.4 & 2.7   &  0.2  &   18 27 49.44&-11 37 26.2\\
  190&2XMM J182821.1-114750&18282106-1147472&0.74  &  8.4 &  3.9 & 4.2   &  0.2  &   18 28 21.06&-11 47 47.0\\
  208&2XMM J182831.5-115145&18283165-1151464&0.80  & 14.6 &  2.2 & 3.5   &  0.5  &   18 28 31.68&-11 51 46.3\\
  237&2XMM J182804.6-113832&18280497-1138320&0.14  & 10.1 &  0.9 & 4.0   &  5.4  &   18 28 04.61&-11 38 33.1\\
  263&2XMM J182824.2-114153&		    &	-  & 14.3 &  2.3 & 3.7   &  -	 &   18 28 24.39&-11 41 55.0\\
  287&2XMM J182806.1-113734&18280597-1137354&0.82  &  0.6 &  2.5 & 4.9   &  0.3  &   18 28 05.97&-11 37 35.7\\
\hline
\end{tabular}
\label{xcorxgpschandra}
\end{table*}

The XGPS id number used throughout this paper refers to the rank of the source in the preliminary XGPS catalogue used at the time of the optical observations and sorted by decreasing MOS countrates in the broad band. This list differs from that published in \cite{hands2004} in the sense that it misses a few sources which were later appended to the final catalogue (see \S \ref{sms}) and also contains a small number of duplicated or spurious entries. As a consequence, the id number does not exactly follow the true broad band flux order of the final XGPS catalogue. Note that the naming of the XGPS XMM-Newton fields in \cite{hands2004} is distinct from the source numbering system used in this paper. 

All error circles plotted on finding charts correspond to a 90\% combined X-ray/optical confidence circle. For a number of interesting sources we retrieved X-ray data from the XMM-Newton archive at ESAC and processed them with  the  latest  version  of  the  SAS package  (version  8)  and  calibration files. Spectral analysis was performed using {\em XSPEC} version 12.1 and above. Unless specified otherwise, all fit parameters are given at the 90\% confidence level for one parameter of interest. Fit results are summarized in Table \ref{xrayspectralsummary}.

\section{Cross-correlation of the XGPS/2XMM list with optical/IR catalogues}

Optical crowding makes the identification  of XMM-Newton sources in the Galactic
Plane rather  difficult.  Originally  (circa 2002) the  SAS task  {\em eposcorr}
corrected EPIC  source positions  for residual attitude  errors by  matching the
X-ray  source positions  with  the  USNO-A2.0.  However  in  the 2XMM pipeline,  the
cross-matches are  done with the  deeper USNO-B1.0 catalogue.   When applicable,
i.e., if  the field stellar  density is not  too high to confuse  the statistical
cross-identification process,  {\em eposcorr} can significantly  reduce the 90\%
confidence error circle by decreasing  systematic errors and thus easing optical
identification. However, as noted above, the positions of most of the 2XMM sources could not be refined by {\em eposcorr}.  Even with this improvement, XMM-Newton  error circles are sufficiently large  that for Galactic
studies  it is often  the case  that several  optical candidates  have positions
consistent with the X-ray source, with none of these  candidates standing out as
the  main contender.   In  this  respect, the  excellent  positions provided  by
Chandra with a sky uncertainty area an order of magnitude  smaller than that of
XMM-Newton constitute a major advantage for Galactic Plane studies.

Nevertheless,  it  is possible  to  find probable  counterparts  to  at least  a
fraction   of  the  XMM-Newton   sources  found   in  low-latitude   fields   by
cross-correlation with major optical and  infrared catalogues.  Thus for all the
XGPS/2XMM sources we looked  for possible identifications in archival catalogues
using the  XCat-DB\footnote{http://xcatdb.u-strasbg.fr/2xmmi/home} developed in
Strasbourg \citep{motch2007}.

\subsection{Cross-correlation with USNO-B1.0 and 2MASS}\label{xcorsection}

In a sizable number of cases, the \xmm error circle contains a relatively bright
optical  object. This  is consistent  with the  fact known  from  previous ROSAT
surveys, namely that the majority of soft Galactic X-ray sources are active stars
\cite[e.g.][]{motch1997} which, apart from the faintest K and  M types, appear as
relatively   bright  optical   objects   even  at   the   flux  sensitivity   of
XMM-Newton. There is obviously a limiting magnitude at which the appearance of a
relatively bright optical object in the EPIC error circle has a high probability
of being a chance coincidence due to the increasing density of field objects with magnitude.  In order to properly estimate the probability of
identification of the X-ray source with a USNO-B1.0 \citep{monet2003} or 2MASS
\citep{cutri2003} entry, we used the likelihood ratio scheme especially 
developed by  \cite{pineau2008,pineau2010} in the framework  of the XMM-Newton  SSC for the
cross-identification  of  the  2XMMi  sources with  several  large  astronomical
catalogues. In a few words, the  process considers the widely used ratio between
the likelihood that  the optical and X-ray sources are  associated and thus have
the same  position, to the likelihood to find by chance an optical object as bright as the candidate
at  that angular  distance from the  X-ray source. This  likelihood ratio,
however, does not provide a direct measure of the value of the association since
it formally depends on  the probability that a given X-ray source  has to have a
counterpart in the archival  catalogue considered.  Eventually, this probability
depends on the  properties and relative fractions of  the various populations of
X-ray sources  entering the overall sample.   Most authors build  a histogram of
the associations likelihood ratios (LR)  and estimate the contributions of  spurious matches to this histogram by randomizing the X-ray  positions. This Monte Carlo process might be time consuming when applied to very large samples and the scheme used in
\cite{pineau2008,pineau2010} uses a geometrical approach instead. For a given bin of
likelihood  ratio the  probability of  identification is  then derived  from the
ratio of total observed (true + spurious) to estimated  spurious associations. In practice, it is necessary to  merge  the LR  histograms  of  several  XMM-Newton observations  at  similar Galactic  latitudes to  obtain  enough  statistics and fit in a reliable manner the relation between spurious match rate and likelihood ratio.   The   probabilities  of identifications with 2MASS and USNO-B1.0 entries were retrieved from the XCat-DB for each XGPS source having a corresponding entry in the 2XMM catalogue.

\begin{center}
\begin{figure*}
\begin{tabular}{cc}
\psfig{figure=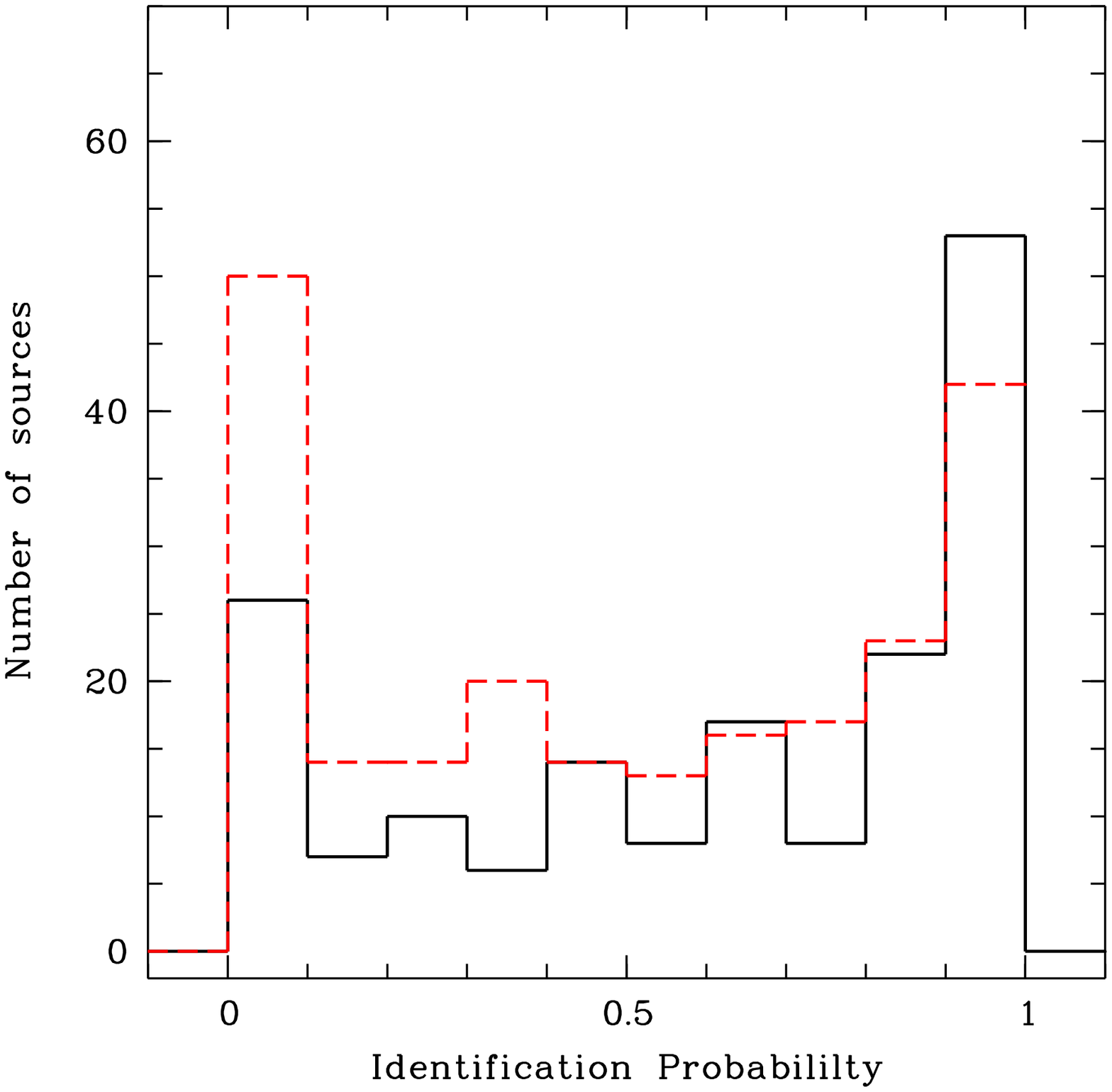,width=8.5cm,bbllx=1.0cm,bburx=21.5cm,bblly=1cm,bbury=21.5cm,angle=-90,clip=true} &
\psfig{figure=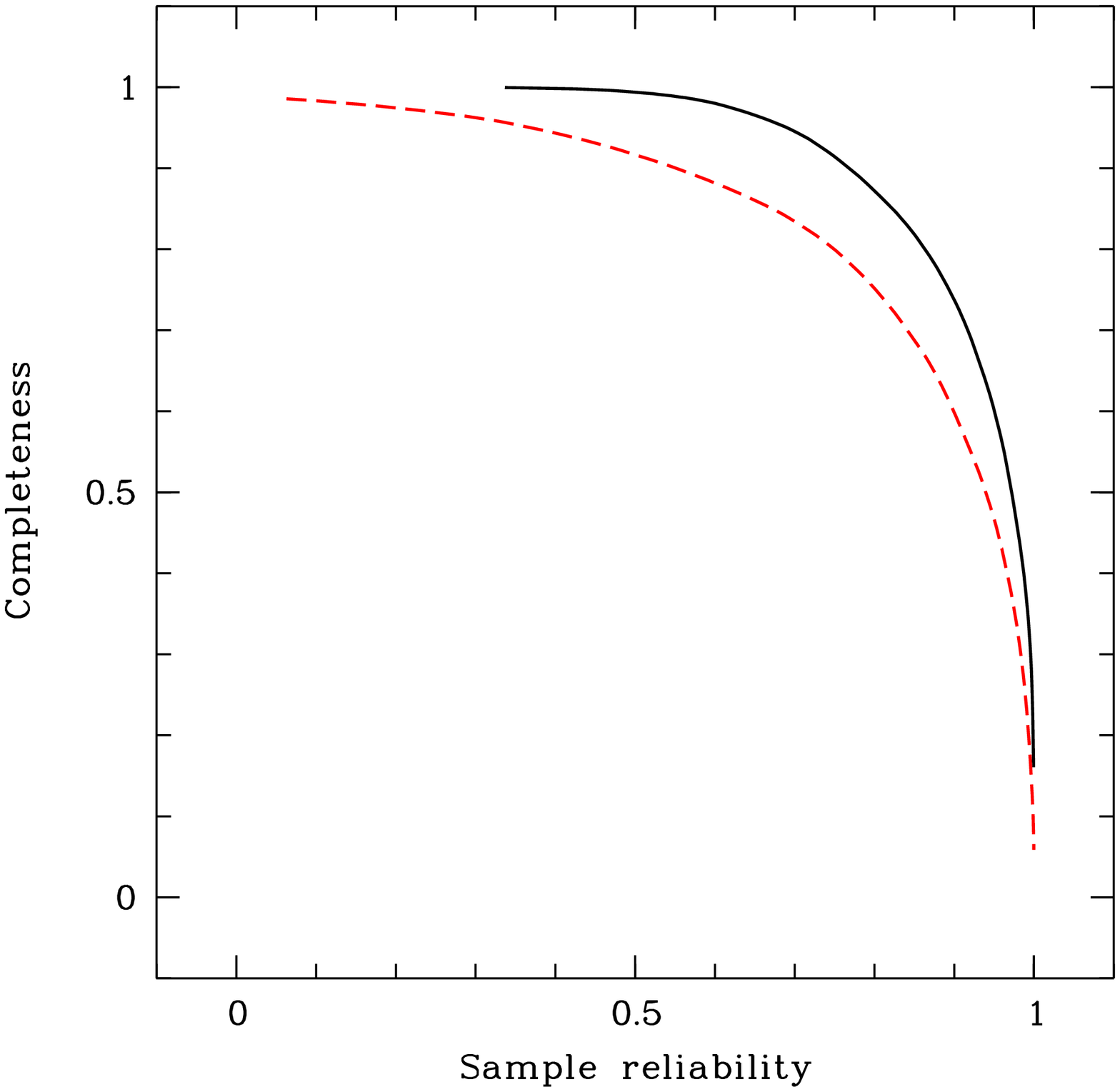,width=8.5cm,bbllx=1.0cm,bburx=21.5cm,bblly=1cm,bbury=21.5cm,angle=-90,clip=true}  \\
\end{tabular} 
\caption{Left: Histogram of the USNO-B1.0 (plain black) and 2MASS (dashed red) probabilities of  identification in the XGPS region. Right: relation between the integrated sample reliability versus completeness for USNO-B1.0 (plain black) and 2MASS (dashed red).}
\label{make_usno_2mass_histos}
\end{figure*}
\end{center}

The cross-correlation  process searches for possible candidate counterparts within a radius corresponding to 3.44 times the combined 1$\sigma$ positional error
(obtained by adding the X-ray and catalogue 1$\sigma$ errors in quadrature). This corresponds to the area in which 99.7\% of the true  matches (3 Gaussian $\sigma$) should be found. For the 2XMM catalogue, we use the combined intrinsic plus systematic\footnote{0.35\arcsec or 1.0\arcsec depending on the outcome of {\em eposcorr} task \citep{watson2009}} positional error as listed in the column labelled {\em poserr} which has a mean value of 1.8\arcsec\ (1$\,\sigma$) for the sample of 269 2XMM/XGPS sources. Typical positional errors on USNO-B1.0 and 2MASS entries are much lower, of the order of a few tenth of an arcsecond \citep{monet2003,cutri2003}. Over the whole XGPS area,  a total of 171 and  223  2XMM X-ray  sources  have  a   match  with  a  USNO-B1.0  and  2MASS entry
respectively. Although the number of matches appears significantly larger in the
near infrared than at optical  wavelengths, the relatively high density of 2MASS
entries   in   the  Galactic   Plane   decreases   the   reliability  of   these
cross-identifications. The left panel of Fig. \ref{make_usno_2mass_histos} shows
that  the number of  high probability  ($>$ 90\%)  USNO-B1.0 identifications is 53
while only  42 2MASS entries have  as high probabilities  of identification.  
We define   the  sample   reliability   as  the   expected   fraction  of   correct
identifications among matches with an  individual identification probability above a  given threshold. Completeness  is then the fraction of  true identifications recovered  above that threshold divided by the total number of true identifications expected in the given survey.  Computed over
the entire  XGPS area,  the completeness versus  reliability curve shown  in the
right panel of Fig. \ref{make_usno_2mass_histos} reveals that the USNO-B1.0
 catalogue provides more reliable and more complete identifications
than the 2MASS catalogue. On average, matches with identification probabilities of less than 10\% are with USNO-B1.0 and 2MASS objects only one or two magnitudes brighter than the limiting magnitudes ($\sim$\,20 in R and $\sim$\,12.5 in K). In contrast, candidate counterparts with identification probability $\ga$\,90\% are typically either USNO-B1.0 entries brighter than R\,$\sim$\,15 or 2MASS entries with K ranging from 12 to 8 with a mean value of $\sim$\,10.

Therefore, thanks to a careful statistical handling of the cross-correlation process, we are able, in a large number of cases, to recover with high efficiency the actual counterpart of the X-ray source and quite significantly extend the identification rate.  This can be achieved in spite of  the poorer XMM-Newton positional accuracy as compared to that of Chandra for X-ray sources having bright enough optical counterparts.

We list in Table \ref{rcstat} values of the fraction of spurious identifications and completeness as function  of   the  threshold  set  on  the   individual  source  identification probability. At  the 90\% threshold generally used  in this work, the number of wrong identifications should be of the order of 1\%, and only 30\% to 40\% of the true matches are identified.

\begin{table}
\begin{center}
\caption{Fraction of spurious identifications and completeness for the 2XMM / USNO-B1.0 and 2XMM / 2MASS cross-correlation in the XGPS area as a function of the individual probability threshold. The spurious fraction and completeness values are given as a percentage.}
\begin{tabular}{rccccc}
\hline \hline
            & \multicolumn{5}{c}{2MASS ident. prob.} \\ \hline
            & 50\% & 60\%  &  70\%  & 80\%  & 90\% \\
Spurious Fr.& 21.4 & 13.6  &  08.8   & 04.3  & 01.3  \\
Completeness& 76.7 & 66.6  &  57.2  & 44.1  & 27.4 \\
\hline
            & \multicolumn{5}{c}{USNO-B1.0 ident. prob.} \\ \hline
            & 50\% & 60\%  &  70\%  & 80\%  & 90\% \\
Spurious Fr.& 18.9 & 12.8  &  08.8  & 04.3  & 01.2  \\
Completeness& 86.2 & 79.1  &  71.1  & 57.5  & 39.4 \\
	    
\hline
\end{tabular}
\label{rcstat}
\end{center}
\end{table}

\begin{table*}[ht]
\tabcolsep=3pt
\caption{XGPS/2XMM - 2MASS - GLIMPSE I Cross correlation.  We only list here 2MASS sources with an individual probability of identification with the corresponding 2XMM source above 80\% and identified with a GLIMPSE I source. The entries under the column labelled ``Class'' refer to the nature of the source as derived from our optical follow-up programme:  AC (active corona); WR (Wolf-Rayet star); HMXB (high-mass x-ray binary);  UNID (un-identified source). The measurements in the different wavebands
are given in magnitudes.}
\begin{tabular}{rrccrrrrrrrrrrrrrrr}
\hline \hline
XGPS                &Class               &2MASS name          &2MASS               &J                   &{\it err}                 &H                   &{\it err}                 &K                   &{\it err}                 &3.6$\mu$            &{\it err}                 &4.5$\mu$            &{\it err}                 &5.8$\mu$            &{\it err}                 &8.0$\mu$            &{\it err}                 \\   
id                  &                    &                    &Prob id             &                    &                    &                    &                    &                    &                    &                    &                    &                    &                    &                    &                    &                    &                    \\   
 \hline
   2&AC   &18284546-1117112& 0.97& 9.80& 0.02& 9.14& 0.02& 8.96& 0.02& 8.91& 0.04& 8.92& 0.05& 8.74& 0.04& 8.77& 0.03\\   
   6&AC   &18284771-1013354& 0.98& 7.48& 0.02& 7.22& 0.05& 7.13& 0.03& 7.23& 0.03& 7.25& 0.05& 7.18& 0.04& 7.14& 0.02\\   
  11&AC   &18250691-1204304& 0.89&10.39& 0.02& 9.38& 0.02& 9.03& 0.02& 8.80& 0.05& 8.80& 0.05& 8.68& 0.04& 8.64& 0.04\\   
  14&WR   &18311653-1009250& 0.98& 9.09& 0.03& 8.29& 0.04& 7.63& 0.02& 7.17& 0.04& 6.83& 0.06& 6.66& 0.03& 6.32& 0.03\\   
  16&AC   &18303575-1021039& 0.96&11.06& 0.06&10.32& 0.07&10.04& 0.05& 9.77& 0.05& 9.75& 0.07& 9.55& 0.05& 9.49& 0.07\\   
  20&AC   &18263956-1142150& 0.88&11.36& 0.02&10.81& 0.04&10.57& 0.03&10.31& 0.11&10.43& 0.10&10.31& 0.11&10.19& 0.07\\   
  21&AC   &18310376-0958153& 0.86&12.39& 0.03&11.73& 0.05&11.38& 0.05&11.08& 0.07&11.07& 0.09&10.97& 0.10&10.81& 0.13\\   
  23&AC   &18301701-0953272& 0.96&11.40& 0.03&10.60& 0.03&10.32& 0.02&10.09& 0.04&10.09& 0.06& 9.94& 0.05& 9.94& 0.04\\   
  30&AC   &18274009-1028124& 0.84&11.56& 0.04&10.83& 0.04&10.49& 0.04&10.28& 0.06&10.34& 0.06&10.25& 0.06&10.31& 0.08\\   
  36&HMXB?&18301593-1045384& 0.93&10.54& 0.02&10.10& 0.02& 9.67& 0.02& 9.01& 0.05& 8.77& 0.05& 8.55& 0.04& 8.26& 0.04\\   
  42&UNID &18283337-1026505& 0.95&11.41& 0.03&10.21& 0.02& 9.72& 0.02& 9.37& 0.04& 9.43& 0.05& 9.24& 0.05& 9.19& 0.05\\   
  46&AC   &18312487-1047478& 0.98& 8.32& 0.03& 8.19& 0.05& 8.11& 0.03& 8.02& 0.06& 8.02& 0.04& 8.06& 0.03& 8.05& 0.04\\   
  51&     &18253369-1214505& 0.95&11.27& 0.02&10.76& 0.02&10.59& 0.02&10.47& 0.07&10.42& 0.06&10.40& 0.09&     &     \\   
  59&     &18281589-1154333& 0.92&10.54& 0.02& 9.24& 0.03& 8.78& 0.03& 8.46& 0.04& 8.52& 0.04& 8.39& 0.04& 8.41& 0.03\\   
  60&AC   &18251158-1157257& 0.96&10.58& 0.02&10.24& 0.03&10.15& 0.03& 9.98& 0.06&10.00& 0.05& 9.94& 0.05&10.03& 0.05\\   
  61&     &18282189-1011279& 0.98& 8.98& 0.02& 8.63& 0.03& 8.51& 0.03& 8.45& 0.06& 8.42& 0.06& 8.47& 0.04& 8.43& 0.03\\   
  66&     &18284762-1155529& 0.91&11.92& 0.03&10.90& 0.03&10.51& 0.02&10.39& 0.06&10.28& 0.11&10.13& 0.06&10.15& 0.06\\   
  71&     &18292438-1029568& 0.92&12.19& 0.02&11.37& 0.02&11.15& 0.02&10.99& 0.05&10.95& 0.07&11.02& 0.09&11.14& 0.10\\   
  72&     &18321298-1006343& 0.94&11.81& 0.02&11.14& 0.03&10.89& 0.03&10.72& 0.04&10.59& 0.06&10.60& 0.06&10.70& 0.08\\   
  73&     &18273223-1133571& 0.98&12.03& 0.03&10.50& 0.03& 9.88& 0.03& 9.49& 0.05& 9.47& 0.05& 9.27& 0.05& 9.22& 0.04\\   
  75&     &18263832-1043417& 0.90&11.40& 0.02&10.90& 0.03&10.76& 0.02&10.43& 0.04&10.50& 0.06&10.52& 0.07&10.42& 0.05\\   
  82&     &18264265-1121246& 0.98& 6.11& 0.05& 5.81& 0.04& 5.65& 0.03& 5.48& 0.05& 6.05& 0.12& 5.39& 0.03& 5.31& 0.03\\   
  95&     &18263599-1146469& 0.97&10.32& 0.02& 9.56& 0.03& 9.27& 0.02& 9.06& 0.04& 9.17& 0.05& 9.06& 0.04& 9.01& 0.04\\   
 110&     &18272856-1137400& 0.92&11.86& 0.03&11.23& 0.04&11.02& 0.05&10.99& 0.06&10.94& 0.09&10.88& 0.09&11.27& 0.11\\   
 112&     &18254529-1131541& 0.95&10.57& 0.02&10.23& 0.04&10.00& 0.03& 9.95& 0.05& 9.94& 0.06& 9.94& 0.06& 9.89& 0.05\\   
 113&     &18291343-0957303& 0.95&10.82& 0.02&10.27& 0.03&10.08& 0.03& 9.98& 0.04&10.04& 0.07& 9.81& 0.06& 9.85& 0.04\\   
 117&     &18293962-1058346& 0.89&11.80& 0.02&11.14& 0.02&10.86& 0.02&10.49& 0.05&10.49& 0.13&10.29& 0.07&10.35& 0.08\\   
 119&     &18301584-1024593& 0.88&12.33& 0.04&11.40& 0.05&11.11& 0.04&10.82& 0.05&10.84& 0.08&10.72& 0.06&10.65& 0.06\\   
 122&     &18303260-0947301& 0.89&11.20& 0.03&10.60& 0.04&10.37& 0.03&10.25& 0.04&10.15& 0.04&10.11& 0.06&10.10& 0.04\\   
 125&     &18260807-1215145& 0.97&11.39& 0.02&10.68& 0.02&10.54& 0.02&10.40& 0.03&10.53& 0.06&10.38& 0.06&10.60& 0.16\\   
 130&     &18281914-1002074& 0.98& 7.12& 0.02& 5.97& 0.03& 5.49& 0.03& 5.31& 0.07& 6.03& 0.11& 5.27& 0.03& 5.23& 0.03\\   
 132&     &18301787-1027277& 0.92&10.39& 0.03&10.01& 0.03& 9.88& 0.02& 9.90& 0.07&     &     & 9.89& 0.10&10.52& 0.22\\   
 135&     &18272674-1120398& 0.93&12.34& 0.03&11.73& 0.05&11.52& 0.05&11.35& 0.07&11.28& 0.08&11.37& 0.12&11.33& 0.12\\   
 142&     &18274465-1139576& 0.98& 9.59& 0.02& 8.95& 0.03& 8.78& 0.02& 8.67& 0.05& 8.70& 0.05& 8.70& 0.04& 8.67& 0.03\\   
 156&     &18270828-1134567& 0.90&     &     &     &     & 9.19& 0.03& 8.51& 0.03& 8.63& 0.05& 8.26& 0.03& 8.37& 0.03\\   
 174&     &18283639-1036553& 0.85&12.18& 0.02&11.12& 0.02&10.70& 0.02&10.40& 0.04&10.51& 0.06&10.35& 0.08&10.39& 0.06\\   
 182&     &18274957-1025358& 0.92&10.32& 0.03& 9.89& 0.03& 9.58& 0.03& 9.57& 0.21& 9.23& 0.13& 9.40& 0.05& 9.25& 0.04\\   
 203&     &18262946-1100587& 0.95&10.60& 0.02& 9.70& 0.03& 9.43& 0.03& 9.20& 0.04& 9.23& 0.05& 9.16& 0.05& 9.08& 0.03\\   
 208&     &18283165-1151464& 0.80&     &     &     &     &11.11& 0.03&10.78& 0.07&10.86& 0.06&10.83& 0.10&11.06& 0.10\\   
 214&     &18303226-1051502& 0.98& 6.89& 0.02& 6.68& 0.03& 6.57& 0.02& 6.81& 0.04& 6.57& 0.05& 6.60& 0.04& 6.58& 0.02\\   
 224&     &18285835-1039133& 0.86&10.01& 0.02& 8.44& 0.06& 7.73& 0.03& 7.34& 0.03& 7.54& 0.05& 7.28& 0.04& 7.18& 0.03\\   
 249&     &18284536-1156528& 0.86&11.59& 0.02&11.24& 0.02&11.10& 0.02&11.08& 0.05&10.98& 0.06&10.91& 0.07&11.07& 0.09\\   
 254&     &18271801-1128218& 0.82&12.95& 0.05&12.07& 0.07&11.62& 0.03&11.05& 0.10&11.02& 0.10&10.97& 0.09&10.78& 0.09\\   
 276&     &18271464-1130273& 0.96&12.08& 0.03&11.21& 0.04&10.84& 0.03&10.49& 0.07&10.41& 0.09&10.45& 0.09&10.40& 0.07\\   
 283&     &18274118-1127156& 0.94&     &     &     &     &11.54& 0.05&11.26& 0.06&11.27& 0.10&11.18& 0.07&     &     \\   
 287&     &18280597-1137354& 0.82&11.97& 0.03&11.60& 0.04&11.35& 0.03&11.11& 0.13&10.99& 0.12&10.98& 0.09&     &     \\   
 305&     &18281180-1025424& 0.98& 8.36& 0.03& 6.93& 0.04& 5.63& 0.03& 4.50& 0.09&     &     & 3.54& 0.06&     &     \\   
 319&     &18294250-1100494& 0.84&11.33& 0.03& 9.91& 0.03& 9.34& 0.03& 8.91& 0.06& 8.86& 0.05& 8.77& 0.04& 8.75& 0.04\\   
 322&     &18291086-1042454& 0.93&10.93& 0.02& 9.74& 0.03& 9.31& 0.03& 8.99& 0.05& 9.00& 0.05& 8.96& 0.04& 8.82& 0.04\\   
 326&     &18273027-1135116& 0.83&10.91& 0.02&10.65& 0.03&10.56& 0.03&10.42& 0.06&10.41& 0.06&10.45& 0.06&10.40& 0.07\\   
 350&     &18260167-1220208& 0.88& 9.97& 0.02& 7.02& 0.03& 5.52& 0.02& 4.40& 0.07&     &     & 3.69& 0.03& 3.41& 0.05\\   
 383&     &18302135-1050371& 0.93&11.06& 0.02&10.56& 0.02&10.37& 0.02&10.27& 0.05&10.24& 0.06&10.17& 0.06&10.31& 0.06\\   
 405&     &18244642-1204260& 0.98& 7.82& 0.03& 7.43& 0.02& 7.22& 0.02& 7.05& 0.05& 7.01& 0.04& 6.89& 0.03& 6.88& 0.02\\   

\hline
\end{tabular}
\label{glimpsexcor}
\end{table*}

\subsection{Cross-correlation with Spitzer source lists}

We also explored  the content of the recently  released Spitzer source catalogue
from  the Galactic  Legacy Infrared  Mid-Plane Survey  Extraordinaire (GLIMPSE)
project \citep{benjamin2003}.  The survey provides images and source lists with
magnitudes  and fluxes  in the  four bands  of the  IRAC instrument,  centered at
approximately  3.6, 4.5, 5.8 and 8.0\,$\mu$.   GLIMPSE I  covers  the Galactic
longitudes from 10\degr\ to 65\degr\ with  a width of $\pm$ 1\degr\ in latitude
and  thus perfectly  overlaps with  the area  covered by  the XGPS  survey.  We
extracted    the    individual    source    lists   from    the    GLIMPSE    I
archive\footnote{http://data.spitzer.caltech.edu/popular/glimpse/
20070416\_enhanced\_v2/source\_lists/north/}  merging  the catalogues  covering
the range of Galactic longitude $l =  19.0$\degr\ to $22.15$\degr.  We did not compute the probabilities of identifications of EPIC sources with GLIMPSE entries using the
LR  based method  described above  but  instead simply  considered all  GLIMPSE
candidate counterparts in a radius of 7\arcsec ( $\geq$ 3 average $\sigma$) around the 2XMM X-ray position. We checked that beyond a radius of 7\arcsec, no more 2MASS identifications of 2XMM sources are found with probabilities larger than 80\%. GLIMPSE positions are only subject to small errors since 95\% of the Spitzer sources are within 0.2\arcsec\ from their 2MASS counterparts\footnote{GLIMPSE Quality Assurance v1.0}.  The Spitzer  IRAC flux  measurements were  used  to constrain  the spectral  energy distributions of some of the candidate   counterparts  and   provide   additional identification  diagnostics. However,  the GLIMPSE  source lists  provide 2MASS identifications  which  can  be  used  as proxies  to  compute  likelihoods  of associations using the 2XMM/2MASS cross-correlation. We list in Table  \ref{glimpsexcor} the 55 XGPS/2XMM sources having an identification probability above 80\% with a 2MASS + GLIMPSE entry, together with their identification class (see \S \ref{sid}). 

\section{Optical observations}\label{s3}

The optical observations presented here were  obtained in 2002 and 2003 with the
ESO  3.6m telescope  and the  ESO-VLT.  At  the ESO-3.6m  equipped  with EFOSC2,
observations took  place from July  9 to  12 in 2002  and from February  28 till
March 3 and from June 27 till July 2 in 2003. An additional run was performed in
2003 with  the VLT-Yepun and FORS2 on July 1st and 2nd.  Bright optical stars
were observed with grism \#14 on EFOSC2, while spectra of fainter candidates were
obtained using  grisms \#6 and \#1. At  the VLT, we used  grisms GRIS\_1400V for
the brightest objects  and GRIS\_300I or GRIS\_300V for the  faintest ones. Absolute flux calibration was  derived from the  observation of  standard stars
done every night.  Weather conditions were photometric in most cases.  Table
\ref{spectro}  shows the  details of  the instrumental  setups relevant  to this
work.

\begin{table*}[ht]
\begin{center}
\caption{Optical spectroscopic settings}
\begin{tabular}{rrcccl}\hline \hline
Telescope     & Instrument   & Spectral    &FWHM Spectral      &Slit & Grisms\\
              &              & range (\AA) & resolution (\AA ) &width (\arcsec) \\ 
\hline 
ESO-3.6m      & EFOSC2       &  3185-10940     & 59                       & 1.5  & Grism \#1 \\
ESO 3.6m      & EFOSC2       &  3860-8070      & 17.3                     & 1.5  & Grism \#6 \\
ESO 3.6m      & EFOSC2       &  3095-5085      & 7.1                      & 1.5  & Grism \#14 \\
ESO-UT4       & FORS2        &  3850-7500      & 12                       & 1.0  & GRIS\_300V \\
ESO-UT4       & FORS2        &  5900-10240     & 11                       & 1.0  & GRIS\_300I \\
ESO-UT4       & FORS2        &  4560-5860      & 2                        & 1.0  & GRIS\_1400V \\
\hline
\label{spectro} 
\end{tabular}
\end{center}
\end{table*}

All CCD  frames were  corrected for bias  and flat-fielded using  standard MIDAS
procedures. Arc-lamp  exposures of various atomic species were  used to derive a
wavelength calibration  generally accurate  to better than  5\% of  the spectral
resolution. One  dimensional spectra were extracted with  a procedure optimizing
the accumulation region in order to reach the best signal to noise ratio and using sky background cleaned from  cosmic-ray impacts. The requirement to  put two optical
candidates in the  slit, in order to minimize observing  time, prevented it from
being  aligned  along or  close  to  the  parallactic angle.  Consequently,  the
accuracy  of the  flux  calibration, whilst  sufficient to derive
relative  fluxes, may  still suffer  from moderate  spectral distortion  in some
instances.

Stellar spectral classification  can be  difficult  when  using  only low  to  moderate
spectral resolution, particularly in the case of faint stellar counterparts with
spectra of  commensurately low  signal to  noise ratio.  In  order to  perform a
classification with  a minimum of  bias, we designed  a MIDAS procedure  in which
flux calibrated  reference spectra  degraded to the  spectral resolution  of the
observations are fitted to the observed  spectrum by adjusting the mean flux and
interstellar  absorption.   We used  standard  spectra  from various  references
depending on the  resolution and wavelength range of  the observation.  The most
useful atlases were those of \cite{jacoby1984}, \cite{pickles1998}, STELIB
\citep{stelib2003} and a collection of spectra extracted from the NASA/JPL 
NStars    project\footnote{http://stellar.phys.appstate.edu/}.    The   spectral
fitting process was visually controlled  and checked by monitoring the behaviour
of  several  important  diagnostic  lines  \citep{turnshek1985,jaschek1987}.  We
estimate  that our spectral  classification is  on  average accurate  to one
subclass.

In addition, we obtained photometrically uncalibrated broad band imaging at the ESO 3.6m and ESO-VLT telescopes in the B, R and I filters for XGPS positions lacking relatively bright optical candidates. Standard MIDAS procedures allowed to correct these frames for bias and flat-field. These exposures were used to prioritize targets for subsequent spectroscopic observations. Some of the finding charts presented in this work are also based on these data. Short raw white light acquisition images are also used on occasion. Finally, many of the charts are based on MAMA scans of the R band plate SRC734S extracted from the second epoch survey of the ESO/SERC Southern Sky Atlas. All images were astrometrically corrected using reference stars in the GSC 2.2 catalogue. We estimate that the resulting positions are good to $\sim$ 0.3\arcsec. This error is quadratically added to XMM-Newton astrometric errors when plotting the 90\% confidence error circle.  

\section{Optical identification of a bright source sub-sample}\label{sid}

As mentioned earlier,  the basic strategy of the  optical spectroscopic campaign
was to start with the brightest XGPS  sources in the broad band and then to work
down  the XGPS  list to  fainter count  rates. However, even  at the low Galactic
latitude ($b$  $\sim$ 0\degree) covered by the XGPS, the integrated 
Galactic interstellar absorption along the line of sight  is not  large  enough
(\nh  $\sim$  $10^{23}$\,cm$^{-2}$) to  efficiently screen hard  X-rays above
$\sim$  2\,keV. We thus  expected that a  rather large number of  the XGPS
sources might actually  be background AGN.  

In order not to waste valuable telescope  time, we did not investigate the 
content of the error circles of sources void of optical candidates bright 
enough to obtain optical spectra in a reasonable amount of time and exhibiting a
MOS hardness ratio\footnote{HR = (H$-$S)/(H+S) with H and S the count rates in the 2.0--6.0\,keV and 0.4$-$2.0\,keV bands respectively \citep{hands2004}} consistent with that expected for  an extragalactic source undergoing the total Galactic absorption in the source direction. The expected MOS hardness ratios of extragalactic sources were computed assuming a power-law spectrum of photon index equal to 1.7. As it was, we used the finding charts provided by the XMM-Newton pipeline\footnote{SSC provided finding charts are based on images extracted from the Aladin image server at CDS \citep{bonnarel2000}} which in the region of the Galactic Plane covered by the XGPS are based on the same R band second epoch survey of the ESO/SERC Southern Sky Atlas as used for some of the finding charts shown in Fig. 33. We did not consider for optical follow-up X-ray sources having no detectable objects in their error circles implying a limiting magnitude of R $\sim$ 21.5\footnote{http://www.roe.ac.uk/ifa/wfau/ukstu/platelib.html}. This corresponds to the faintest targets reachable with the instrument used at the 3.6m telescope. Evidently, optically faint Galactic X-ray sources undergoing large foreground or internal absorption would also be discarded by this selection criterion.

We used the cross-correlations of EPIC source positions with over 170 astronomical catalogues and the finding charts available in the pipeline data produced by the Survey Science Centre to investigate the content of the EPIC error circles\footnote{The SSC pipeline lists all archival catalogue entries with positions consistent within 3$\sigma$ with the EPIC source. However, at this stage, no probability of identification is computed. These cross-correlations are available from the XCat-DB. See \cite{watson2009} for a complete description}. 

We have computed the expected total Galactic absorption in the direction of each XGPS source from  the IR maps provided by COBE/DIRBE  and IRAS/ISSA \citep{schlegel1998}.  We compared these values to those  obtained by  summing the  HI \citep{dickey1990}  and  CO
\citep{dame2001} column densities in  each direction. Values were interpolated 
over the 4 nearest pixels (of size 2\farcm37$^{2}$) for the COBE/DIRBE maps, whereas the HI and CO maps were not interpolated and had pixel sizes of 0\fdg5$^2$ and 0\fdg121$^{2}$ respectively.  We assumed \nh\ = N{\tiny 
HI} + 2$\times$N{\tiny  H2} with N{\tiny  H2} =  2.7 10$^{20}$ $\times$ Wco 
where Wco  is the intensity of the CO  1-0 rotational transition. 
The  two determinations of the total column density agree well. However, the radio  determined  \nh\ value clearly saturates at Log(\nh) $\geq$ 23.1. We thus
retained the \nh\ values derived from the  IR maps  as  providing a  more 
realistic estimate  of  the total  Galactic absorption.   

In the event  our programme of optical spectroscopic  follow-up extended down to
the 63rd source in the count-rate ordered preliminary XGPS list (XGPS-69) to a limiting MOS broadband  count rate  of 4.2  ct/ks.  However, of these  63 sources, 12 where discarded due to the optical candidate / MOS hardness-ratio criterion outlined above and 8 due to lack of observing time at the ESO 3.6m telescope or due to very offset XGPS positions.

This leaves us with a total of 43 XGPS sources which were investigated at the
telescope.  Of these, we failed to identify a likely counterpart in 13 cases due to the lack of spectral signatures of known classes of X-ray emitters in the observed candidates or because the optical candidates turned out to be too faint for the instrumental setting used at the given telescope. 
Further details of these 43 sources are given later - in Table \ref{xgps_tab}. Possible identifications are discussed in sections 5.1 to 5.4, while some of the unidentified sources for which we have the most constraining information are presented in section 5.5. Optical spectra
for all identified sources are available as on-line material (Fig. 32). Finding charts for all of the 43 investigated sources are either shown in \S 5 or in the on-line material section (Fig. 33).

As a check for unforeseen biases  we have investigated some of the characteristics
of  our ``telescope'' sample  in relation  to the  whole XGPS  catalogue. For
example, Fig. \ref{plot_xgps_MOSfluxdistrib} shows the histogram  of the broad-band (0.4--6.0\,keV)  MOS count rate for sources scheduled for optical follow-up compared to that of the entire final XGPS sample of 424 sources. 
The difference between the two histograms is explained by the fact we ignored 20 sources. In addition, as mentioned in \S 2 (see also \S \ref{sms}), the preliminary version of the XGPS catalogue used at the telescope was missing a few bright sources.

We plot in  Fig. \ref{plot_nh_distribution}  the histogram  of the integrated 
Galactic \nh\ for  the entire  XGPS set of 424 sources and for  the sub-sample
of 43 bright  sources investigated optically.  The distribution  shows that  the
broad band  bright sample  of XGPS  sources selected  for optical  follow-up 
does not exhibit any bias towards obscured or clear Galactic  directions. 

\begin{figure}
\psfig{figure=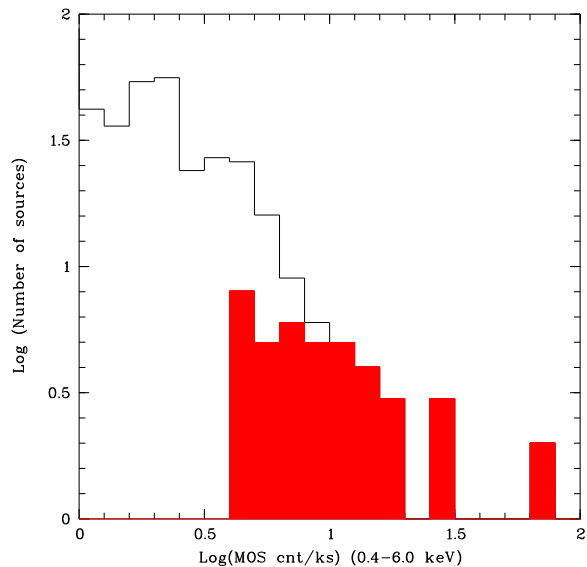,width=8.5cm,bbllx=1.0cm,bburx=21.5cm,bblly=1cm,bbury=21.5cm,angle=-90,clip=true}    
\caption{Histogram of the MOS broad-band flux for the entire XGPS sample and for the sub-sample scheduled for identification at the telescope (red histogram).} 
\label{plot_xgps_MOSfluxdistrib}
\end{figure}

\begin{figure}
\psfig{figure=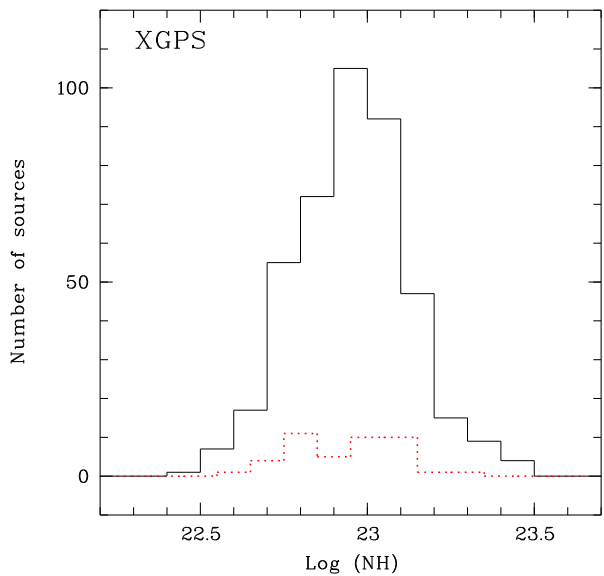,width=8.5cm,bbllx=1.0cm,bburx=21.5cm,bblly=1cm,bbury=21.5cm,angle=-90,clip=true}    
\caption{Distribution of integrated \nh\ for the entire set of XGPS sources (black solid line) and for the sources investigated optically (red dashed line).} 
\label{plot_nh_distribution}
\end{figure}

Table \ref{xgps_stat}  provides a  break-down by identification  type for the 43
sources studied at the telescope.  
In practice, the faintest identified XGPS source is XGPS-65 with a  broad band
MOS count rate of  4.5 ct/ks. 

\begin{table}[ht]
\begin{center}
\caption{Identification statistics for the sub-sample of XGPS sources selected for follow-up optical observations. AC: Active Coronae, WR: Wolf-Rayet star.}
\begin{tabular}{lr}\hline \hline
Class     & Number   \\ \hline
AC       & 16 \\
Possible AC & 2 \\
B stars   & 2 \\
WR       & 1 \\
CV       & 3 \\
LMXB candidates   & 2 \\
HMXB candidates      & 4 \\
Unidentified     & 13 \\
\hline
Total  & 43 \\
\hline
\end{tabular}
\label{xgps_stat} 
\end{center}
\end{table}

\subsection{Stellar identifications}

\subsubsection{Normal stars}

Our identification of an X-ray source  with an active corona (AC), in the main, rests on the detection  of spectroscopic evidences such as Balmer  emission or Ca H\&K re-emission.  However, when  the  optical  spectrum of  the  candidate star  was lacking such  features, the  identification was then  based on a  probability of association with the USNO-B1.0 or 2MASS object above 90\%. Note that in this statistical identification process, we did not apply any constraint on X-ray hardness ratio. 

Most coordinates of active coronae are derived from their GSC2 entry and thus reflect their position near epoch 1988. However, whenever a TYCHO entry was available we used the TYCHO coordinates corrected for proper motion at epoch J2000, i.e. at a time very close to X-ray observations. A close inspection of finding charts in Fig. 33 reveals the relatively high proper motion of XGPS-6 whose 2001 X-ray position is overplotted on a plate acquired in 1982 and of that of the counterpart of XGPS-21 whose 2003 position is well inside the error circle.

A total of 16 (or 18 including the two dubious cases) of the 30 sources having a confirmed or a tentative identification fall into the AC category. Stars located across all  regions of the HR
diagram have been  detected in X-rays \citep[see e.g.][]{vaiana1981,guillout1999}
barring A-type stars and cool M-type giants (although even in these cases 
exceptions do exist).

For the  hottest O,B stars,  shocks forming in  unstable winds heat  the ambient
plasma to million degree temperatures, while for late-type stars, X-ray emission
results  from magnetic  activity  in the  outer  stellar atmospheres  due to  an
interplay between rotation and subsurface convection. In the latter case whether
the coronae  are energised by  flares, waves or  currents is still  debated.  If
the counterpart is a late B-type or an A-type star then  very often the X-ray emission is ascribed to an unresolved, active, binary companion \citep[see for instance][]{neff2008}.
 
In our spectroscopically identified sample we find 2 K stars,  3 G stars and one late B and one late A star ( in these two later cases, we assumed that X-ray emission arises from a late-type stellar companion). We  also have 9 counterparts  which are Me stars  in the range  of M1e to M6e. Source XGPS-16 is identified with a likely physical pair of  a M4 and a M5 emission line stars, based on their comparable photometric distances. The X-ray brightest coronal source, XGPS-2, was also detected in the ASCA Galactic Plane Survey \citep[AX J182846-1116,][]{sugizaki2001}. We also identify two sources XGPS-19 and XGPS-56 with early B type stars. 

Unfortunately, the absence of reliable optical spectrophotometric flux calibration resulting from the observing procedure prevents us from deriving accurate enough reddening corrected X-ray to optical flux ratios. As a consequence, we cannot sensitively search our stellar sample for possible anomalous X-ray emission which could be the signature of neutron star or black hole accreting binaries in quiescence for instance. However, neglecting all reddening effects, the $log(f_x/f_r)$ distribution of our active coronae appears fully consistent with that of ChaMPlane stellar identifications in the direction of the Galactic bulge \citep{koenig2008}.

Despite their  fully convective interior  structure, the X-ray activity  of very
low mass  M dwarfs shows  similar trends to  more massive stars, for  example an
ultra-cool M9 star has been detected in X-rays at a quasi-quiescent level
\citep{robrade2009}.  Although F, G and K-type stars display higher X-ray
luminosity than M  dwarfs, only the 'tip of the iceberg'  of the stellar coronal
population has been sampled here.  In terms  of the likely composition of X-ray source
catalogues, it  is probable  that M dwarfs will overwhelmingly outnumber other
stellar categories at faint X-ray fluxes.

The high fraction ($\sim 50\%$) of Me stars found among active coronae  in the XGPS  is at variance with  the results  of the  ROSAT Galactic  Plane Survey  \citep{motch1997}, which recorded only  19\% of M-type identifications. It is, however, known  from X-ray stellar population models \citep{guillout1996}  that the contribution of M stars rises  with increasing  sensitivity. However,  considering the  small  number of stars  in each  class,  it is  difficult  to make a detailed comparison with the stellar-content inferred for published X-ray surveys covering
other segments of the Galaxy and with different flux sensitivities. 

\subsubsection{XGPS-14: an X-ray selected Wolf-Rayet star}

\begin{figure}
\psfig{figure=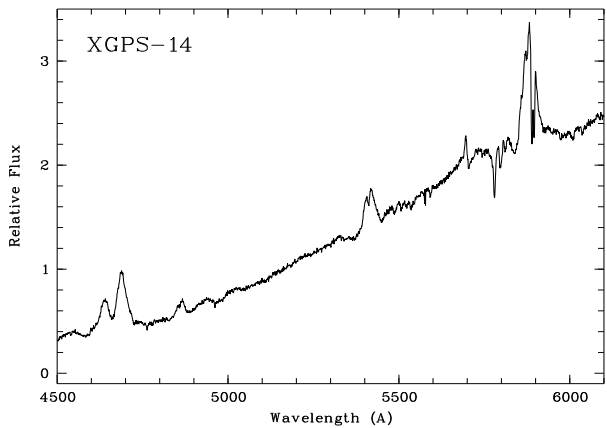,width=8.5cm,height=6.0cm,angle=-90,bbllx=0.5cm,bburx=20cm,bblly=1.5cm,bbury=27.5cm,clip=true}   
\caption{The WHT optical spectrum of the WN8 counterpart of XGPS-14.}
\label{figspecxgps14}
\end{figure}

This X-ray source has previously been detected in the ASCA Galactic Plane Survey
\citep{sugizaki2001} as  entry AX~J183116-1008, but  has not been  identified at
other  wavelengths. Our improved  position shows  it to  be coincident  with the
southern  component  of a  highly  reddened visual  binary  at  a separation  of
15\arcsec\ which consists of USNO-A2.0 0750-13397443 and  USNO-A2.0 0750-13397376,
the former being the actual optical counterpart of the X-ray source.
According   to   the  2MASS \citep{cutri2003} and DENIS 
database\footnote{http://cdsweb.u-strasbg.fr/denis.html}, the observed 
optical/IR  magnitudes  of  the  counterpart  are  as  follows:  B=17.2;  R=13.4
I=11.63$\pm0.03$; J=9.09$\pm0.03$; H=8.29$\pm0.04$;  K=7.63$\pm$0.02, where the B
and R magnitudes are deduced from photographic plates \citep[USNO-A2.0 Catalogue, ][]{monet1998} and should accordingly be treated with caution. XGPS-14 is also identified with the GLIMPSE source SSTGLMA G021.6064-00.1970 (see Tab. \ref{glimpsexcor}).

A first optical spectrum shown in Fig. 32 was obtained at the ESO-3.6m in 2003 with EFOSC2 and Grism\#6 (see Tab. \ref{spectro}). Fig.\ref{figspecxgps14} shows the second, higher resolution spectrum (FWHM resolution of $\sim$ 2\AA). It was acquired on the 18/05/2005 with the ISIS spectrograph attached on the William Herschel Telescope. On this occasion, the R300B grating was used in the blue arm with the EEV12 CCD. 
Our  optical  spectra  confirm  the  highly  reddened nature  of  the  star  and,
furthermore, reveal  the presence  of strong broad  emission lines typical  of  a
Wolf-Rayet  star.  The presence of HeI,  HeII, and nitrogen emission lines  and the lack of  strong carbon lines defines  a nitrogen-sequence  WN star.  Drawing on  the  spectral classification criteria    described   by   \cite{smith1996}, the emission    line  ratio HeI$\lambda5876$/HeII$\lambda5411$ $>$  1 (where the  $\lambda5876$ component is furthermore diminished by interstellar Na I absorption) suggests a spectral type later than WN7, whereas the presence of NIV ions
and the ratio NIV$\lambda4057$/NIII$\lambda4640$ ($\sim$ 0.3) constrains the spectral type to WN8. This  classification is confirmed  by comparison with  the \cite{vreux1983}
near-IR-spectral-atlas of WR stars. The  higher resolution WHT spectrum shows
that the  HeII emission  lines do  have a two-component  structure or  a central
absorption component  which might  indicate (but do  not necessarily  imply) the
presence of an early O type close binary companion.
Given the  intrinsic colours of  an early-type star,  which in the IR  displays 
the Rayleigh-Jeans tail of a  black body spectrum (here we use an O9  type star), 
we estimate the reddening E(B-V) = 3.05$\pm0.15$  towards XGPS-14  by fitting 
the  B to  K band photometry.  A  similar reddening  is  derived from  the  
slope  of the  optical spectrum only. We  note that the same colour excess is found
for  the northern  component of  the visual  binary, which  appears intrinsically
redder and which  on the basis of its IR colours \citep{koornneef1983}  
is consistent with a late F-type supergiant.

The analysis  of the EPIC  pn and MOS  observations of XGPS-14  is unfortunately
limited  by the  short  usable exposure  time  of the  source,  typically just a  few
thousand  seconds  at  a large off-axis  angle.   Nevertheless  the  results  are
sufficiently interesting to be reported here. The source is in the field of view
of  the EPIC cameras  in six  observations, 0135741401,  0135744301, 0135744501,
0135746601,   0135741601  and  0135746301.   However,  apart   for  observations
0135741601  and 0135746301, the  high flaring  background or  a position  at the
extreme  edge  of  the  field  of  view  prevent  the  extraction  of  an  X-ray
spectrum. Owing to the position of XGPS-14  close to the edge of the field of view,
we measured the  background contribution in a circular area  located on the same
CCD and at slightly smaller off-axis angle. The small number of photons collected led us  to use the Cash
statistics. Fitting together  or separately the EPIC pn and MOS  data of the two
observations  yield consistent  results. Models based on an absorbed thin
thermal {\em mekal} spectrum or an absorbed power-law continuum are acceptable, 
although the former gives a  slightly  better  fit.  Using observation  0135746301  
only,  which displays the  best signal-to-noise  ratio, we find  
$\rm kT$ = 2.14$_{-0.52}^{+1.09}$\,keV and $\Gamma$ = 2.91$_{-0.86}^{+0.96}$ for  
the thin thermal temperature and the photon spectral index
respectively. In both cases, the X-ray column densities
N$_{\rm  X}$   =  5.21$_{-1.25}^{+1.38}$  and   5.34$_{-1.68}^{+1.99}$$\times$$10^{22}$ H~atoms cm$^{-2}$, respectively, are three times larger than the optical/IR derived value of \nh\,=\,1.7\,$\times$\,10$^{22}$\,atoms\,cm$^{-2}$, 
(assuming the  canonical  relation  N$_{\rm  X}$=\,5.5\,$\times$\,10$^{21}$\,E(B-V)
atoms\,cm$^{-2}$ of \cite{predehl1995} corrected for A$_{\rm V}$/E(B-V)=3.1)).

Using these spectral  models we derive observed/unabsorbed 0.2--10\,keV fluxes of
$\sim 8.2 \times 10^{-13} / 4.4 \times 10^{-12}$\ergscm\ for the thin thermal {\em
mekal}  model. The  temperature is  significantly higher  than the  more typical
values of $<1$\,keV observed from stellar-wind shocks in early (O and WR ) type
stars,  but  agrees  with  the  values  expected  in  colliding  winds  binaries
\citep{skinner2007}, which also tend to be  more X-ray luminous than single O and
WR stars.

A reliable  estimate of the distance to  XGPS-14 may be derived  in at least
two  ways. Either  the  distance modulus  can  be established  by comparing  the
unabsorbed magnitude  with the  absolute magnitude of  the stellar type,  or the
distance can be estimated by  comparing the observed reddening with a previously
determined   relation   between  reddening/absorption   and   distance  in   the
line-of-sight of  the object in  question.  The latter  may be derived  from the
absorption maps of \cite{marshall2006}. The distance determinations from the two
methods are  in agreement  if we assume that the absolute magnitude M$_{\rm V}$ of a Galactic WN8  star lies in the range $-6$ to $-7.2$, as suggested  by \cite{abbott1987} and \cite{Hamann2006}. The distance so-inferred using the optical/IR reddening/absorption is $\sim$ 5 kpc.  At this distance, the unabsorbed X-ray luminosity is  $\sim$\,1.3\,$\times$\,10$^{34}$\ergs. This is more than 3 orders of magnitude brighter than observed from single WN8 stars \citep{oskinova2005}, but only slightly higher than the luminosities of several WR + O star colliding wind binaries \citep{skinner2007}.  The high X-ray output also leaves open the possibility of a WR + compact object binary that can reach up to several 10$^{38}$\ergs.
Such systems are in vogue again since the discovery of the WR counterparts of IC 10 X-1 and NGC300 X-1 \citep{carpano2007,prestwich2007}.
In  fact,  we believe  that XGPS-14 is among the very rare cases of an X-ray selected  WR star. These are rare objects since only $\sim$ 300 WR stars are known in the Galaxy. In a recent paper \cite{shara2009} discuss the use of different near-infrared photometric techniques that allow them to discover 41 new WR stars. They fail, however, to detect any late WN (7-9) stars. 

\subsection{Cataclysmic variables}

Three cataclysmic variables were positively identified in the XGPS survey; 
their properties are described in the following paragraphs.

\subsubsection{XGPS-9}

\begin{figure}
\centerline{\psfig{figure=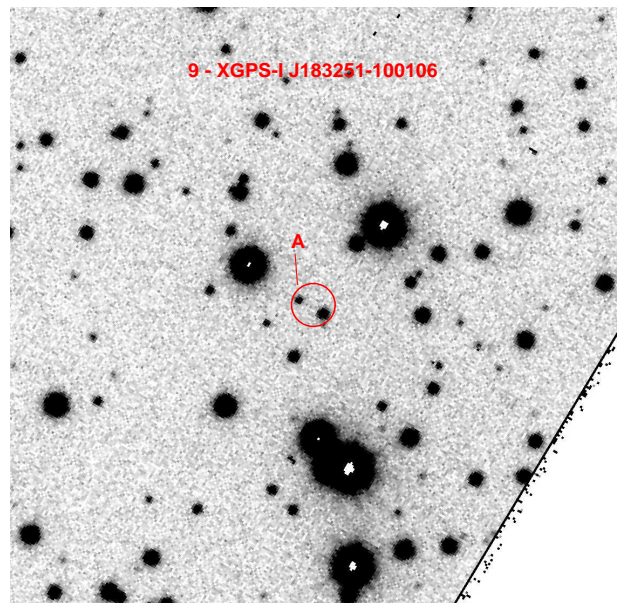,width=8cm,clip=true}}
\caption{A 2 min long ESO-VLT V band exposure showing the field of
  XGPS-9. The cataclysmic variable (object A) is the fainter of the two objects located in the 2XMM error circle. North is at the top and East to the left. The field of view shown is approximately 1\arcmin$\times$1\arcmin.} 
\label{f:xgps9_fc}
\end{figure}
 
\begin{figure}[t]
\psfig{figure=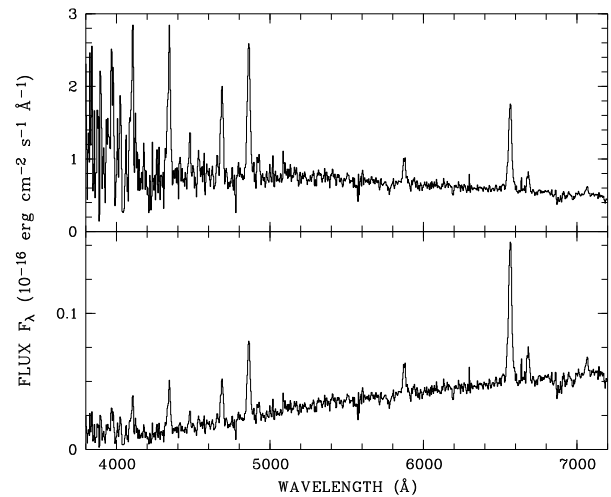,width=8.5cm,bbllx=30pt,bburx=560pt,bblly=30pt,bbury=650pt,angle=-90,clip=true}
\caption{Original (lower panel) and de-reddened (upper panel) identification spectra of XGPS-9. The upper spectrum was corrected for a reddening corresponding to the best X-ray \nh\ (8$\,\times\,10^{21}$\,cm$^{-2}$).}
\label{f:xgps9_osp}
\end{figure}

\begin{figure}[t]
\resizebox{\hsize}{!}{\includegraphics[clip=]{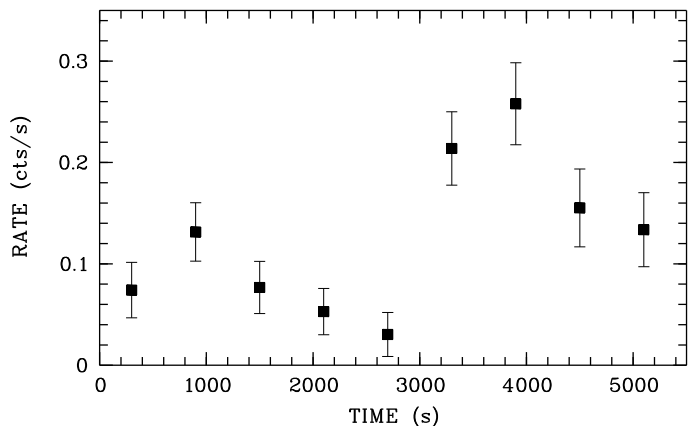}}
\caption{EPIC pn X-ray light curve of XGPS-9 with bin size 600\,s.}
\label{f:xgps9_xlc}
\end{figure}

The optical finding  chart for XGPS-9 is shown in Fig. \ref{f:xgps9_fc}.
The optical spectrum of object A, the fainter of the two objects present in the
X-ray error circle was acquired at the ESO-VLT  using  FORS2 and grism GRIS\_300V.  
The observed and the de-reddened (see below) optical spectra are shown in 
Fig.~\ref{f:xgps9_osp}.  These spectra show the typical  features of a  
magnetic cataclysmic  variable, namely  strong H, HeI and HeII emission, superposed  on a blue continuum.   We note that the  spectrum does not show signs of 
the secondary star nor does it display evidence of cyclotron lines, which are sometimes 
seen in polar systems. If the orbital period were known, the size and mass 
of the secondary star could be determined and based on the non-detection of
the secondary, a lower distance limit could be estimated.

The emission  lines have an observed  width of 21\,\AA\  (FWHM), somewhat larger
than the formal FHWM spectral resolution of 12\,\AA\ provided by grism 
GRIS\_300V (see Table \ref{spectro}).

Due to its large off-axis angle, the source  was not in the field of view of the
optical monitor (OM) on XMM-Newton, an independent brightness determination of the 
counterpart is thus not possible. ESO-VLT imaging observations unfortunately lack proper photometric calibration. However, the counterpart is certainly very faint since the flux calibrated spectrum indicates V $\sim$ 23.3. 

To create an X-ray spectrum, photon events were extracted using a circular 
aperture of radius $\sim$ one arcmin. The resulting count rate spectrum was binned
with a minimum of 20 photons per bin to allow the application of the
\chisqr\ statistic in the model fitting. 

A  one-component thermal plasma  spectrum absorbed  by cold  interstellar matter
provided a satisfactory fit to the data. The temperature was not well constrained
and was  thus fixed at 20\,keV\footnote{Unconstrained thermal  plasma model fits
of  magnetic CVs  tend  to reveal  unphysical  high plasma  temperatures due  to
reflection    and/or    warm    absorption,   see    e.g.~\cite{schwarzetal2009,
staudeetal2008}}.  The column  density of  the best  fit is  
\nh\ $\sim 8\times10^{21}$\,cm$^{-2}$ and the 0.2--12\,keV unabsorbed  
flux of the source was determined to 
be \Fx $= 1.4 \times 10^{-12}$\,ergs\,cm$^{-2}$\,s$^{-1}$.
Taking  the fitted  value  of the  column  density at  face  value, the  optical
spectrum was de-reddened (see above) using the calibrations in
\cite{predehl1995} (for $A_V$) and \cite{cardellietal1989} (for
$A_{\lambda}/A_V$) and  assuming $R=3.1$ for  Galactic dust.   

The  X-ray light curve  shown in Fig.~\ref{f:xgps9_xlc}  
was created  with task  {\it epiclccorr} (with dead-time and vignetting 
corrections applied) and has time  bins of 600\,s. 
There is evidence of variability of up to  100\% on a timescale as  short as a
few 100\,s.  Unfortunately, an estimate of any  periodic behavior (orbital period 
or spin period of the white dwarf) is not possible given the rather short X-ray 
observation and the low count rate.

Taking all facets together,  the source XGPS-9  is almost certainly  a magnetic
cataclysmic variable. A final classification, AM  Her type or DQ Her type (polar
or intermediate polar), i.e.~with synchronously or asynchronously rotating white
dwarf, would need measurements of both  the orbital period and the spin period of
the white  dwarf.  The absence of  a soft X-ray spectral  component, a prominent
defining  feature  for   polars  in  the  ROSAT  era  does   not  rule  out  the
classification of XGPS-9 as a  polar. Follow-up observations of definite polars
with XMM-Newton revealed a $\sim$30\% fraction without a soft component
\citep{ramsaycropper2004}. 
To  the best  of our  knowledge XGPS-9  is the  first magnetic  CV to  have been
uncovered  in  the 2XMM/2XMMi  catalogue  as a  result  of  a dedicated  optical
identification programme. However, previously two rather bright eclipsing polars
were  discovered  amongst  the  2XMM   sources  due  to  their  prominent  X-ray
variability \citep{vogeletal2007,ramsayetal2009}. 

\subsubsection{XGPS-24 and XGPS-65}

We  show in Fig.~\ref{f:xgps24_fc}  and  Fig.~\ref{f:xgps65_fc} the  
finding charts  for sources  XGPS-24 and XGPS-65,  respectively.  The VLT spectra obtained  with FORS2 and grism GRIS\_300V for the respective counterparts are reproduced in Figs.~\ref{f:xgps24_osp} and \ref{f:xgps65_osp}. The optical spectra of the counterparts of sources \#24 (object B) and \#65 USNO-B1.0  0789-0404378 are very similar.  Both   show  Hydrogen  Balmer   ($\alpha,  \beta,   \gamma$)  and   He\,{\sc  I} ($\lambda\lambda5875, 6678$) emission lines on a reddened smooth continuum.
The absence of He\,{\sc II} emission suggests a classification as non-magnetic
cataclysmic variables, i.e. disk CVs whose X-ray emission likely originates
from the boundary layer between the disk and the accreting white
dwarf. A further division according to CV subclass is not possible/reliable
given the sparse observational data base. 

However, following \cite{patterson1984} we estimate the absolute brightness
$M_{\rm V}$ of the accretion disk from the observed equivalent width of the
H$\beta$ line, $EW(H\beta) = 0.3M^2_{\rm V} + exp(0.55(M_{\rm V}-4))$. 
Using the same observable the X-ray to optical flux ratio may be
inferred with the empirical relation of \cite{pattersonraymond1985}:
$\log(F_{\rm X}/F_{\rm V}) = -2.21 + 1.45 \log(EW)$. Measurements and results
are compiled in Table\,\ref{t:eqw}. 

\begin{table}
\tabcolsep=3pt
\caption{\Hbet\ equivalent widths (in \AA) and derived quantities for the non-magnetic CVs XGPS-24 and -65.}
\begin{tabular}{cccccccc}
XGPS & B & V & EW(H$\beta$) & $M_{\rm V}$ & $\log({F_{\rm X}/F_{\rm V})}$
& $A_{\rm V}$ & $D$/kpc \\
\hline
24 & 24.3 & 22.8 & $-25\pm5$ & $7.6\pm0.6$ & $-0.18$ & $1-5$ & $7-1.1$ \\
65 & 23.4 & 22.0 & $-9\pm1$  &   $5\pm0.3$ & $-0.82$ & $1-5$ & $16-2.5$ \\
\hline
\end{tabular}
\label{t:eqw}
\end{table}

Many  caveats   have  to   be  made  when   interpreting  the  numbers in 
Table \ref{t:eqw}. Patterson's EW-$M_{\rm  V}$ relation displays a large  scatter by 1.5 mag above EW $\sim  15$\,\AA, and it is even less certain at  lower widths. The apparent
B and V magnitudes were  derived from the  observed spectra by  folding those
through  model filter  curves and are probably accurate to
about 30\%. The  biggest   uncertainty  is  the  rather  unconstrained   $A_{\rm  V}$.  Some constraints  on the  absorbing  column density  can  be derived  from the  X-ray
spectra. They are compatible with a one-component thermal spectrum with $kT \geq
5$\,keV. Fixing the  temperature at 5\,keV, \nh$ \sim (0.2-2.4)\times
10^{22}$\,cm$^{-2}$ for  XGPS-24 and $\sim  (0.8-2.4)\times 10^{22}$\,cm$^{-2}$ for
XGPS-65,   implying   $A_{\rm V} =1-13$ using the calibration of
\cite{predehl1995}.

Alternatively, $E(B-V)$ and $A_{\rm V}$ can be estimated using the observed 
$B-V$ colour of the systems as derived from the spectrophotometry and an assumed
intrinsic colour. Assuming $(B-V)_0 \simeq 0.3$ \citep{tylenda1981} and $R_{\rm V}
= 3.1$ for Galactic dust, $A_{\rm V} \simeq 4$ for the two systems.  
In Table~\ref{t:eqw} distance limits are derived for the range $A_{\rm V} = 1
- 5$ using $M_{\rm  V}$ derived from Patterson's EW-$M_{\rm  V}$ relation.

The X-ray spectral parameters yield an unabsorbed 0.2--12\,keV spectral flux of
a few times $10^{-13}$\,ergs\,cm$^{-2}$\,s$^{-1}$, which at the lower limit
distance estimates of 1.1 and 2.5 kpc listed in Table \ref{t:eqw} give luminosities of
$(1-5) \times 10^{31}$\,erg\,s$^{-1}$. Values in that range are expected for 
nova-like CVs or dwarf novae in quiescence and we propose such an identification 
for the two systems. 

\begin{figure}
\centerline{\psfig{figure=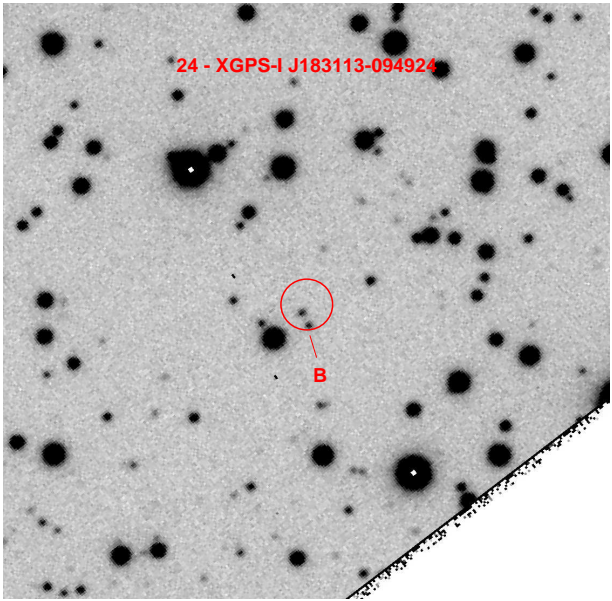,width=8cm,clip=true}}
\caption{A 2 min long ESO-VLT V band exposure showing the field of
  XGPS-24. The CV counterpart of the X-ray source (object B) is the object located at the South-West. North is at the top and East to the left. The field of view shown is approximately 1\arcmin$\times$1\arcmin.}
\label{f:xgps24_fc}
\end{figure}

\begin{figure}
\centerline{\psfig{figure=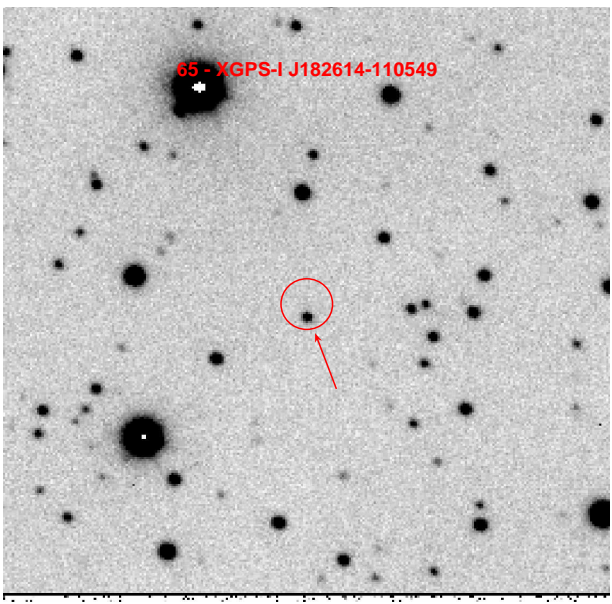,width=8cm,clip=true}}
\caption{A 2 min long ESO-VLT V band exposure showing the field of
  XGPS-65. The CV counterpart of the X-ray source, USNO-B1.0 0789-0404378, is marked. North is at the top and East to the left. The field of view shown is approximately 1\arcmin$\times$1\arcmin.}
\label{f:xgps65_fc}
\end{figure}

\begin{figure}[t]
\resizebox{\hsize}{!}{\includegraphics[clip=]{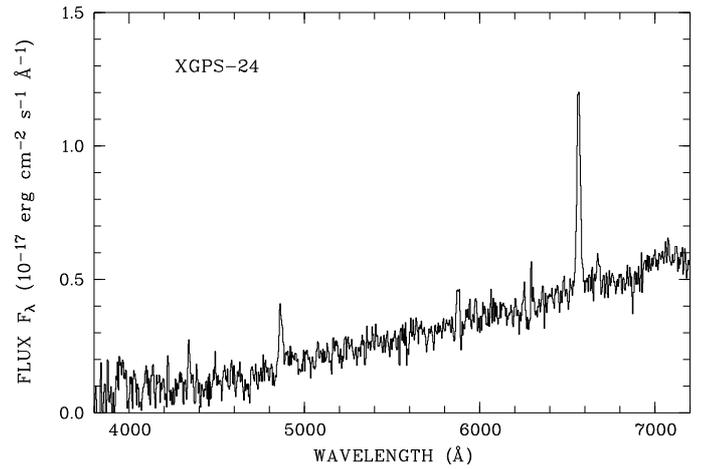}}
\caption{Identification spectrum of XGPS-24 (i.e., of object B).}
\label{f:xgps24_osp}
\end{figure}

\begin{figure}[t]
\resizebox{\hsize}{!}{\includegraphics[clip=]{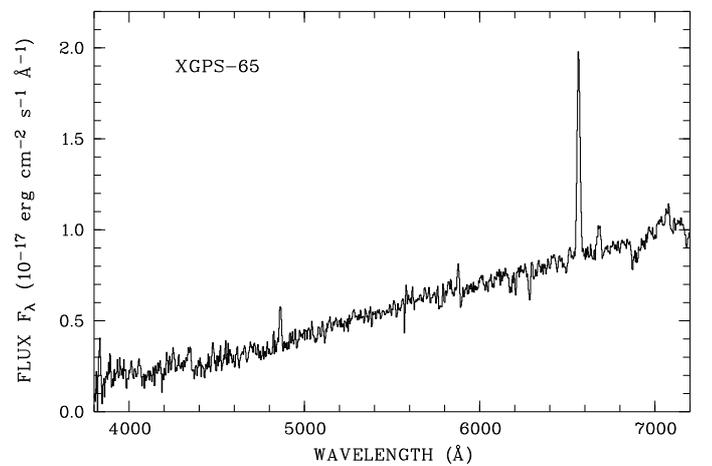}}
\caption{Identification spectrum of XGPS-65 (i.e., of USNO-B1.0 0789-0404378).}
\label{f:xgps65_osp}
\end{figure}

\subsection{Candidate High-Mass X-ray Binaries}

Einstein and ROSAT all-sky survey observations of OB stars \citep{pallavicini1981,berghoefer1997} have established a rather well defined relation between X-ray and stellar bolometric luminosities, \Lx\ $\sim$ 10$^{-7}\,\times\,$\Lbol , with maximum X-ray (0.1--2.4\,keV) luminosities of the order of a few 10$^{32}$\ergs\ for the hottest O-type stars. X-ray spectra are adequately described by multi-temperature thin thermal emissions. When fitted with a single temperature component, the mean kT is $\sim$ 0.2--0.6\,keV for O stars and kT $\sim$ 1\,keV for B stars \cite[see e.g.][for a recent survey of XMM-Newton observations]{naze2009}.  
The XGPS sources listed in this section display properties departing from those of normal early type stars, either based on their large X-ray luminosities in excess of several 10$^{33}$\ergs\ or on the basis of an unusually hard X-ray spectrum. Extreme wind collisions such as proposed for the X-ray selected WR star XGPS-14, may explain the observed properties to some extent. However, the high \Lx\ and/or the X-ray hardness combined with the absence of WR signatures and evidences of Balmer emission suggest the presence of a compact object accreting from a circumstellar disc as in ``classical'' Be/X-ray systems \cite[see ][for a recent review]{negueruela2007} or from the massive stellar wind of a supergiant star \citep{corbet1986}. Our sample includes four tentative identifications of X-ray  binaries with high mass  primaries. We outline below  the supporting  evidence for  each  of these sources.

\subsubsection{XGPS-3}
 
The XMM-Newton source position coincides with a radio source detected in the 1.4
GHz survey of \cite{zoonematkermani1990} and the 5GHz VLA survey of the Galactic
Plane \citep{becker1994}.  It also matches  with the ASCA Galactic  Plane Survey
detection AX J1832.1-0938 \citep{sugizaki2001}.   The X-ray source is identified
with GSC2 S300302241761 / 2MASS 18320893-0939058 with a probability greater than
98\%  (see optical finding chart in Fig. 33). Our  optical spectrum (Fig. \ref{xgps3spec}) shows that this relatively  bright star (R  = 16.80
$\pm$  0.45) exhibits  a  very strong  \Halp\  emission line  (EW of  $-$104\,\AA)
superposed  on a  highly reddened  continuum and  is indeed  the  likely optical
counterpart  of the X-ray  source.  \cite{stephenson1992}  also reports  a strong
near infrared excess and identify this star with IRAS 18293-0941. The source is
also present  in the DENIS survey  (I = 12.325$\pm$0.020).  Its 2MASS magnitudes
are  very  bright  with  J  =  7.05$\pm$0.007,  H =  5.49  $\pm$0.015  and  K  =
4.34$\pm$0.027.   We  show  in  Fig. \ref{Xgps3SpitzerCh3}  the  5.8$\mu$  image
acquired by  the IRAC  instrument on board  Spitzer, which illustrates  the large
infrared  brightness of  the counterpart.  Amazingly, this  bright source  is not
found in  both the  "reliable" and  "less reliable" versions  of the  final 2007
GLIMPSE I  catalogue available at  IPAC. This absence  might be due to  the fact
that saturation did not allow clean extraction of the infrared parameters.

\begin{figure}
\centerline{\psfig{figure=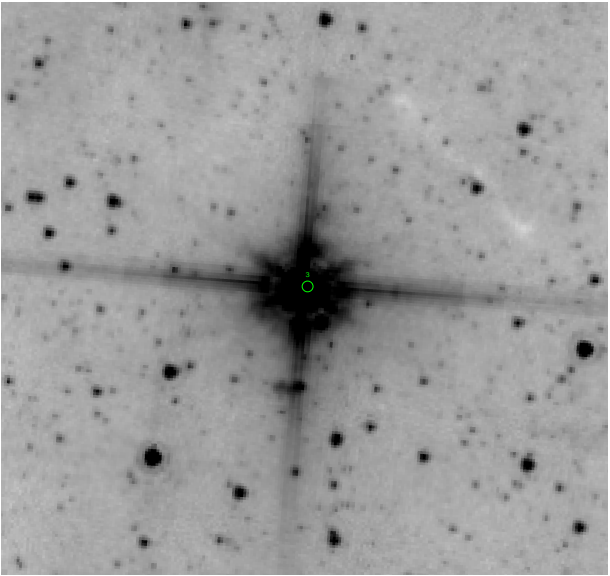,width=8.cm}} 
\caption{The 5.8$\mu$ image of the field of XGPS-3 as seen by Spitzer. North is at the top and East to the left. The field approximately covers 4$\times$4} arcmin.   
\label{Xgps3SpitzerCh3}
\end{figure}

\begin{figure}
\psfig{figure=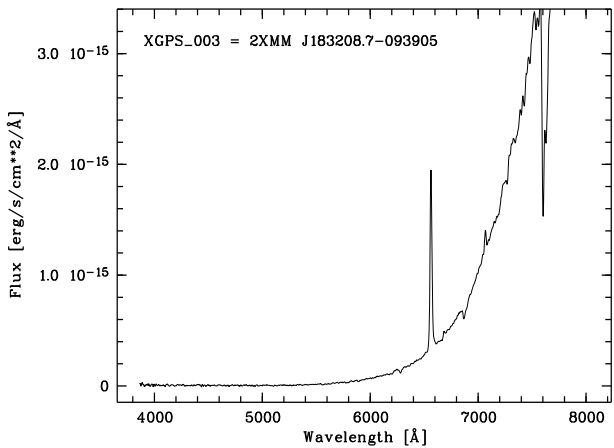,width=8.5cm,height=6.0cm                
,bbllx=1.5cm,bburx=18.5cm,bblly=16cm,bbury=27.5cm,clip=true}  
\caption{The observed optical spectrum of XGPS-3.}
\label{xgps3spec}
\end{figure}

In addition to the prominent \Halp\  line, our spectrum exhibits HeI emission lines
at $\lambda$6678 and  $\lambda$7065 \AA\ (Fig. \ref{xgps3spec}).
The absence of  any visible  photospheric absorption  features makes
spectral classification extremely difficult. Furthermore, since our discovery spectrum was acquired with EFOSC2 and grism \#6, the wavelength range does not cover the Paschen line region. Model spectra of B to K giant stars
can reasonably well account for the slope of our optical spectrum assuming a large
E(B-V) in  the range of  5 to  7. Later spectral  types would display  TiO bands
which  are  not  detected  in  the   red  part  of  our  spectrum.  We  plot  in
Fig. \ref{sed_xgps3} the overall optical  to infrared energy distribution of the
star, using our own flux calibrated spectrum, and GSC 2.2, DENIS, 2MASS and IRAS
photometry,   together    with   a   theoretical   model   for    a   O6I   star
\citep{castelli2004}.  In fact good fits are obtained with model energy distributions
corresponding to super-giant O6  to K0 stars.  However, in  all cases, the  12 and 25  $\mu$ IRAS
fluxes seem  an order of magnitude  above model expectations and might be indicative of a cool dusty circumstellar envelope.  Assuming that the
underlying spectrum is  that of an early type OB star  implies E(B-V) $\sim$ 7.0
while a G0I star would still  require E(B-V) $\sim$ 6.0.  Such a high reddening,
if  not intrinsic  to the  source  indicates astonishingly  large distances  and
accordingly  a  very high  luminosity. To find the distance corresponding to this reddening we  used  both  the  3-D extinction  model  of
\cite{drimmel2003}\footnote{In general, we prefer to use Marshall's determinations which are not sensitive to variations of dust temperature along the line of sight.} based on FIR  COBE/DIRBE data and that of \cite{marshall2006}
relying  on  a modelling  of  the giant  stellar  population  detected at  large
distances by 2MASS.  Both models predict that E(B-V) $\sim$ 7 is reached between
8 and 9 kpc.   At this distance the absolute K magnitude  of the star reaches an
exceptional value  of K$_0$  = $-$12.4.  Interestingly,  \cite{lockman1996} note
the presence of  a nearby HII region [LPH96] 022.162-0.157,  only 33 arcsec away
from XGPS-3. The HII region has two velocity components which, using the velocity
distance  relation of  \cite{anderson2009}, imply  a distance of either 4.8 kpc or 9 kpc, the latter  being  close  to  that estimated  from  the reddening.

\begin{figure}
\psfig{figure=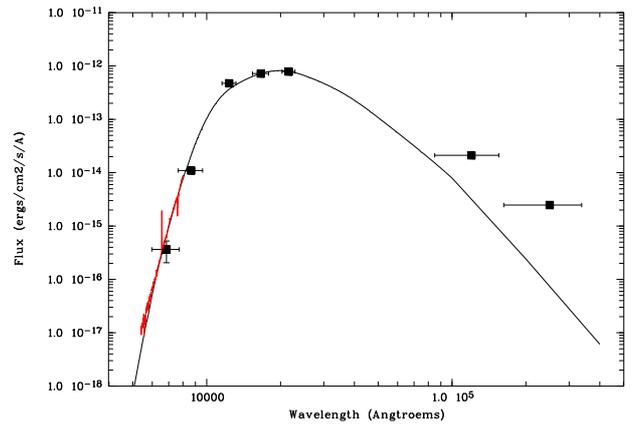,width=8.5cm,angle=-90,clip=true}    
\caption{The spectral energy distribution of the optical counterpart of XGPS-3. Data points show the 2MASS and IRAS photometry, while the optical spectrum is plotted. The continuous line represents the theoretical energy distribution of a O6I star with E(B-V) = 6.9.} 
\label{sed_xgps3}
\end{figure}

We   reprocessed  the   XMM-Newton   data  of  observation
0135741701. Only the MOS detectors were on  at the time of this observation with
both medium filters deployed. XGPS-3 is  detected with a count rate of 0.13
ct/s. However, the  modest exposure time of 6.9\,ks and  a strong vignetting due
to the large off-axis angle of the  source yields only about 300 photons in each
MOS camera. We  thus grouped the X-ray  spectra with a minimum of  15 counts per
channel and  used the XSPEC  C statistics for  fitting.  A power-law  model (see
Fig.\ref{xgps3mosspec}) provides  a good  fit to the  MOS spectra with  a photon
index of 1.45$^{+0.42}_{-0.40}$ and a  strong photoelectric absorption of \nh\ =
2.69$^{+0.76}_{-0.67}$$\times$10$^{22}$cm$^{-2}$ (90\% confidence level). A thin  thermal ({\em mekal}) energy distribution also provides a good fit to the observed spectra with  kT = 4.2 $\pm$\,0.6\,keV, when  \nh\ is fixed at the value inferred from the optical/infrared reddening (E(B-V) = 7 or \nh $\sim$ 3.8$\times$10$^{22}$). Assuming a power-law model, the observed 0.2--10\,keV flux is 3.2$\times$ 10$ ^{-12}$\ergscm\ equivalent to an unabsorbed 0.2--10\,keV flux of 5.6$\times$10$ ^{-12}$\ergscm. The corresponding X-ray luminosity is \Lx\ = $6.7\,\times\,10^{34}$\,(d/10\,kpc)$^{2}$\ergs , corrected for interstellar absorption. It may be noted that the flux ($1.97 \times 10^{-12}$\ergscm ; 0.7--10\,keV) and spectral shape (photon index = 3.67$^{+0.76}_{-0.69}$ and \nh = 5.48$^{+1.87}_{-1.47} \times 10^{22}$\,atoms cm$^{-2}$) observed by ASCA are not consistent with the XMM-Newton values and provide evidence of long-term flux and spectral variability. No significant flux variability is detected during the $\sim$ 2 hour XMM-Newton observation.
  
\begin{figure}
\psfig{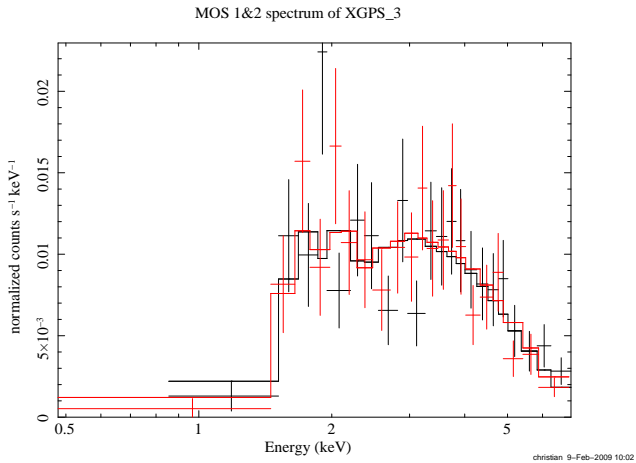}    
\caption{The MOS 1 (black) and MOS 2 (red) spectra of XGPS-3 fitted with a power law of photon index 1.45 and \nh = 2.7$\times$10$^{22}$cm$^{-2}$.}
\label{xgps3mosspec}
\end{figure}

Taken at  face value, the large  reddening and inferred distance  imply that the
optical counterpart is  a rare super-giant star with  an extreme luminosity. The
overall observational picture presented by XGPS-3 is clearly reminiscent of that
of $\eta$ Carina, which displays  similar optical and X-ray luminosity and X-ray
spectral shape  \citep{davidson1997,chlebowski1984}.  By analogy  with the model
favoured to explain the absorbed  high-energy emission from $\eta$ Carina, we may
speculate that  wind collision in a  binary system may  be at the origin  of the
copious X-ray emission. For XGPS-3, however, we would expect a different viewing
condition  to  the  X-ray  source,  since the  photoelectric  X-ray  absorption,
is consistent (at least for the thermal model) with that  seen  in  the
optical/infrared range. The detection of radio emission also fits well in this picture since $\eta$ Carina itself is a strong variable radio emitter \citep[][and references therein]{duncan2003}.
Alternatively, assuming that the star is a more common dwarf or giant hot star,
then the presence of strong intrinsic X-ray absorption and optical reddening would suggest that a cocoon of absorbing matter surrounds the source. 

However, the large inferred X-ray luminosity is also consistent with a scenario in which a compact object, a neutron star or a black hole, accretes matter from the dense wind of an early type companion star. 
   
\subsubsection{XGPS-10} 

This source is  located in the field designated XGPS-2 (nb recall that the XGPS field names are not related to the source  numbering  system  employed  here). Although the source position and finding charts produced by an early (2001) version of the standard analysis are available, the observation data unfortunately suffer  from an  ODF description problem  which prevents  it from being processed  by the  current pipeline. XGPS-10  is thus  not in the  2XMM or 2XMMi catalogues. However, at the telescope we used the position and associated errors computed by the 2001 pipeline (RA = 18 27 45.49 , DEC=-12 06 06.7 and a 90\% error radius of 2.3\arcsec) which is fully consistent with that listed in the XGPS-I catalogue while providing a smaller error radius. The XGPS MOS-based hardness ratio of this source is 0.56, indicative of a hard spectrum but not the degree of absorption expected for a background extragalactic object (assuming a $\Gamma$ = 1.7 power law energy distribution).
The  two   faint  candidates   A  and   B,  are   shown  in
Fig. \ref{xgps10}.  The third faintest object to the North [of candidate A] 
was not spectroscopically observed. Candidate B shows \Halp\
in absorption and  no evidence for Paschen lines, suggesting  a rather late type
star.   Because of  the relatively  large  reddening, the  G band  could not  be
measured. The absence of strong Mg  "b" feature suggests a possible F-G spectral
type.  However, candidate A might  exhibit a double-peaked \Halp\ line in emission
and  shows  high-order Paschen  absorption  lines (see Fig. \ref{fig_spec10}) consistent  with an  early  B spectral type or with a $\sim$ mid  F type star \citep{fremat1996}. XGPS-10 has no match with 2MASS
stars within 5 arcsec. There is however a GLIMPSE source at RA = 18 27 45.44 and
DEC  = -12  06  06.3 detected  only in  the  3.6$\mu$ and  4.5$\mu$ bands.   Its
position is  consistent with  either candidate  A or the  fainter object to the north. Assuming  that  the  GLIMPSE source  is  indeed  candidate  A  yields an optical  to infrared energy distribution suggesting E(B-V) $\sim$  3.9 and E(B-V) $\sim$ 3.2 for a B0V and a F2V star respectively. Fitting the spectral distribution to the flux calibrated optical spectrum alone yields a slightly smaller value of the reddening of E(B-V) $\sim$ 3 and E(B-V) $\sim$ 2.2 for a B0V and a F2V star respectively. The estimated V band magnitude of candidate A (V $\sim$ 23.7) would imply distances of the order of 10\,kpc assuming an early
type star. Such a large distance is in fact consistent with a total Av of 12.1 derived by 
\cite{marshall2006}, since in this direction the line of sight appears much
clearer than  towards source  XGPS-3. The unabsorbed 0.2--10\,keV X-ray
luminosity is of the order \Lx\ $\sim$ 3.7$\times$10$^{34}$ (d/10 kpc)$^{2}$\ergs, assuming a spectral shape similar to that of XGPS-3. Overall, the observational picture could be consistent with a Be/X-ray system in quiescence. 
However, considering the  need to confirm the Balmer emission  of candidate A we
regard  the  identification of  XGPS-10  with  a  high-massive X-ray  binary as
uncertain. 

\begin{figure}
\psfig{figure=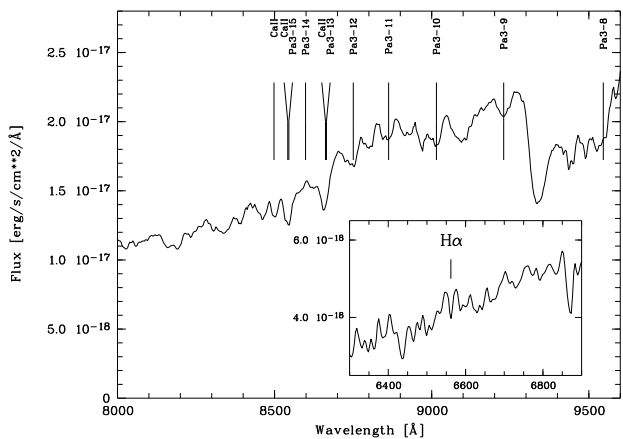,width=8.5cm,angle=-90,bbllx=1.0cm,bburx=19.5cm,bblly=2cm,bbury=26.5cm,clip=true}    
\caption{Red part of the ESO-VLT FORS2 spectrum of candidate A for XGPS-10 displaying evidences of Paschen absorption lines. The CaII triplet is mostly of interstellar origin (spectrum filtered with a 10\AA\ window). Insert: the \Halp\ region  (spectrum filtered with a 3.3\AA\ Gaussian).} \label{fig_spec10}
\end{figure}

\begin{figure}
\centerline{\psfig{figure=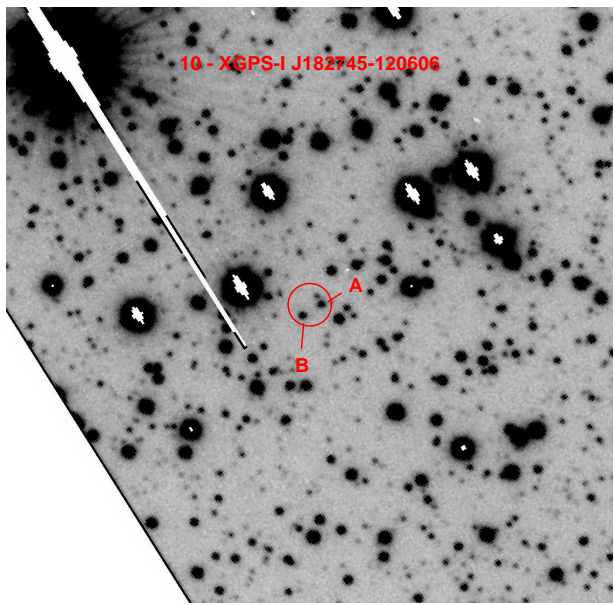,width=8cm,clip=true}}
\caption{A 20s long ESO-VLT white light exposure showing the field of XGPS-10 and the position of candidates A and B. North is at the top and East to the left. The field approximately covers 1$\times$1 arcmin.  The position and 90\% confidence radius shown here are those produced by the 2001 version of the XMM-Newton pipeline.}
\label{xgps10}
\end{figure}

\subsubsection{XGPS-15} 

Two relatively bright stars occupy the error circle (see Fig.\ref{xgps15}). We tentatively identify the X-ray source  with the object located on  the Eastern side, which  is also 2MASS 18281430-1037288  and identical to the GLIMPSE source SSTGLMA
G020.8457+00.2476. The formal probability of identification of XGPS-15 with the 2MASS entry is 55\% and therefore the association cannot be proven solely on the basis of positional coincidence.  
Our  optical  spectrum   reveals  a  heavily  reddened  flux
distribution.  It  shows  rather   clear  evidence  for  \Halp\  emission  (see
Fig. \ref{plotHalphaXGPS15})  with an equivalent  width of $\sim$  $-$1.9\AA\ and,
again, high order Paschen lines  in absorption with depths consistent with those
expected from an early B0V type  star.  In contrast, the optical spectrum of the
western  star, which  is the  brightest  in the  optical, exhibits  a much  less
reddened continuum and \Halp\ in absorption. The slope of the optical spectrum 
of the candidate 2MASS identification indicates E(B-V) $\sim$ 5.0, assuming an  early B-type stellar  continuum. The  optical and GLIMPSE energy  distribution are indeed  
well represented by  a \Teff\,30,000\,K stellar  atmosphere 
model \citep{castelli2004}  subject to E(B-V)  $\sim$ 5.0 (see Fig.\ref{sed_xgps15}).

\begin{figure}
\centerline{\psfig{figure=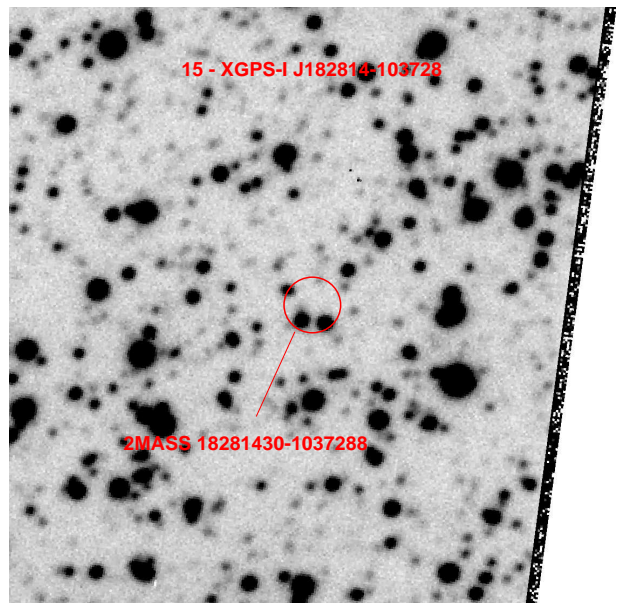,width=8cm,clip=true}}
\caption{A 30s long ESO-VLT I band exposure showing the field of XGPS-15. The proposed optical counterpart is the 2MASS source. North is at the top and East to the left. The field approximately covers 1$\times$1 arcmin.}
\label{xgps15}
\end{figure}

\begin{figure}
\psfig{figure=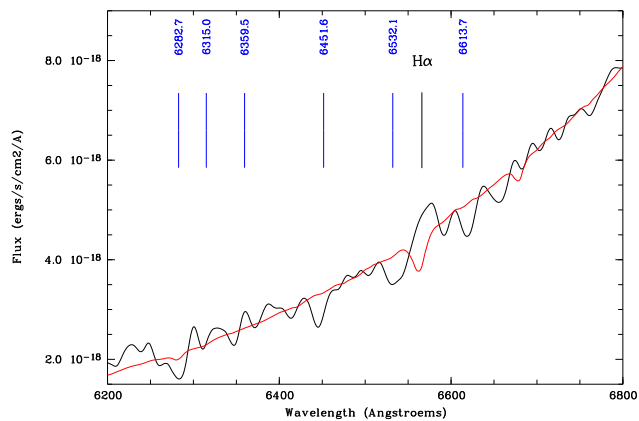,width=8.5cm,angle=-90,clip=true}    
\caption{Black line; combined GRIS\_300I and GRIS\_300V optical spectrum of  2MASS 18281430-1037288, the possible counterpart of XGPS-15, obtained at ESO-UT4 and smoothed with a $\sigma$ = 5\AA\ Gaussian filter. Red line; the best fit spectrum of a reddened B0V star. Blue labels show the positions of the strongest interstellar absorption bands.} \label{plotHalphaXGPS15}
\end{figure}

\begin{figure}
\psfig{figure=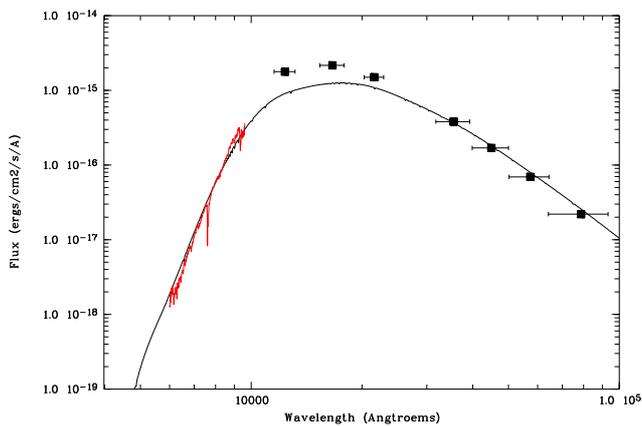,width=8.5cm,angle=-90,clip=true}    
\caption{The spectral energy distribution of the possible optical counterpart of XGPS-15. Data points show the 2MASS and GLIMPSE photometry, while the optical spectrum is plotted in red. The continuous line represents a theoretical energy distribution of a B0V star with E(B-V) = 5.0.} 
\label{sed_xgps15}
\end{figure}

The fact that the 2MASS photometry  does not fit the model suggests the presence
of an interloper within the 2MASS  photometric radius or the existence of variability. With an estimated R
$\sim$ 22.2  derived from the optical spectrum,  a B0V star would  be located at
$\sim$ 5\,kpc  for a  total absorption of  Av =  15.8.  However, this reddening  is
reached at a  distance of $\sim$ 10\,kpc in  that direction \citep{marshall2006}
for which  the absolute  K magnitude of  $\sim$ -5.3  would then indicate a  giant or
super-giant star.

The large  off-axis  position  and  relative  faintness of  the  source  limits  
the constraints that can be derived from the shape of the X-ray energy distribution. A power law adequately fits the combined data from the three EPIC cameras ($\chi^2_r $ = 1.27) and gives \nh\ = 7.8$^{+5.9}_{-3.4} \times 10^{22}$\,cm$^{-2}$ with a photon index of  2.4$^{+1.5}_{-1.0}$.  The corresponding unabsorbed X-ray luminosity is
\Lx $\sim$ 5.4$\times$10$^{34}$ ($d$/10\,kpc)$^{2}$\ergs (0.2--12\,keV). In  the direction  of XGPS-15,  the total Galactic absorption is $\sim  6.4\ \times 10^{22}$\,cm$^{-2}$ and therefore the X-ray spectrum could  be consistent with an extragalactic origin if the early type star is simply a chance interloper..  

If the identification of XGPS-15 with 2MASS
18281430-1037288 were confirmed, the reddening of the optical 
candidate (E(B-V)  $\sim$ 5.0) implies \nh $\sim$ 2.8\,10$^{22}$\,cm$^{-2}$ \citep{predehl1995} and  would  suggest the  presence  of an  additional  local absorption  in the  X-ray domain, as is sometimes seen in the super-giant high-mass X-ray binaries discovered by INTEGRAL  \citep[see e.g.][]{walter2006}.  An identification with  a giant star could be  consistent with the absence of strong infrared excess in the Spitzer bands. An accreting Be/X-ray system would probably exhibit a noticeable excess of infrared emission from the circumstellar disc.

\subsubsection{XGPS-36}

\begin{figure}
\centerline{\psfig{figure=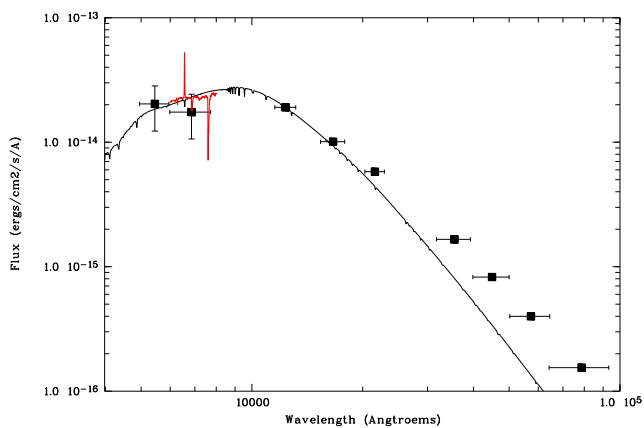,width=8.5cm,angle=-90,clip=true}}   
\caption{The spectral energy distribution of the optical counterpart of the Be/X-ray source XGPS-36. Data points show the GSC2.2, 2MASS and GLIMPSE photometry, while the optical spectrum is plotted in red. The continuous line represents a theoretical energy distribution of a B1V star with E(B-V) = 1.5.} 
\label{sed_xgps36}
\end{figure}

A bright optical object, GSC2 S300302371, occupies the error circle (see finding chart in Fig. 33). Its optical spectrum (see Fig. 32) was unfortunately not acquired under the best conditions. The continuum shape of the flux-calibrated spectrum is distorted as the slit was not aligned to the parallactic angle, and the observation was carried out under non-photometric conditions. Many diffuse interstellar
bands are visible. The equivalent width of the $\lambda$5780 and $\lambda$6613
\AA \  lines are  consistent with E(B-V)  $\sim$ 1.5  \citep{herbig1975}. Balmer
absorption  lines are  also clearly  visible blue-ward  of \Hbet .  Several HeI
absorption lines  are seen  with equivalent widths  suggesting an early  B type star.
The \Halp\ line  is strongly in emission with an equivalent  width of 27\AA. The
overall spectral energy distribution (SED) based on the optical, 2MASS and GLIMPSE data (see Tab. \ref{glimpsexcor}) is shown in Fig. \ref{sed_xgps36}. Assuming E(B-V) = 1.5, the B to H band flux distribution fits well that of a
\Teff\,25,000\,K model \citep{castelli2004}  corresponding to a B1V  star. However, the
SED shows evidence for an infrared excess starting red-wards of 2$\mu$ which may
indicate the presence of a large  circumstellar envelope. A B1V star would need to
be  at  $\sim$ 2.6\,kpc  to  account  for the [K-band] 2MASS  observed magnitude  of
9.67. Interestingly,  the 3-D  absorption map  of  \cite{marshall2006} indicates
that  the  reddening  of E(B-V)  =  1.5  is  reached  at about  2.5\,kpc,  fully
consistent with  the photometric distance.   

The 2XMM EPIC pn hardness ratios (see \S \ref{section6}) reveal a
rather hard  X-ray spectrum with,  for EPIC  pn, HR2 =  0.882 $\pm$ 0.430;  HR3 =
0.613  $\pm$ 0.211  and  HR4 =  -0.039 $\pm$  0.196. Assuming a power-law energy distribution with a photon index of 1.9 and a column density fixed at the optically derived E(B-V), the observed  flux of  2.57\,$\pm$\,0.34$\times$10$^{-13}$\ergscm\  yields an unabsorbed X-ray
luminosity  of  3.4$\times$10$^{32}$\ ($d$/3\,kpc)$^{2}$\ergs  (0.2--12\,keV).  The
overall  observational picture, spectral  type, equivalent  width of  the \Halp\
line,  X-ray hardness  ratios and  luminosity,  clearly suggest  that XGPS-36  is
another    member   of    the    growing   class    of   $\gamma$-Cas    analogs, a group of early Be stars exhibiting very hard thermal X-ray emission (kT $\sim$ 10\,keV) and X-ray luminosities of a few 10$^{32}$\ergs\ \citep{lopes2006,motch2007g}.

\subsection{Candidate low-mass X-ray binaries}

\subsubsection{XGPS-1} 

This source is coincident in position with SAX J1828.5-1037
\citep{cornellisse2002} which is known to exhibit X-ray bursts. It belongs to 
the growing class of very faint X-ray transients
\citep{sakano2005,king2006} whose exact nature remains uncertain. 
An  optical spectrum  of GSC2  S300302034199, the  bright star  located slightly
outside the 2XMM 90\% confidence error circle (see Fig.
\ref{trans}) shows that the object is a normal G-type star and thus 
unrelated to  the X-ray  source. The  faint object in  the error  circle located
to the north-west of  GSC2 S300302034199  has  I $\sim$  21.6. The  absence of a bright
optical counterpart  is consistent with a  low-mass X-ray binary type
and a large interstellar absorption.

\begin{figure}
\centerline{\psfig{figure=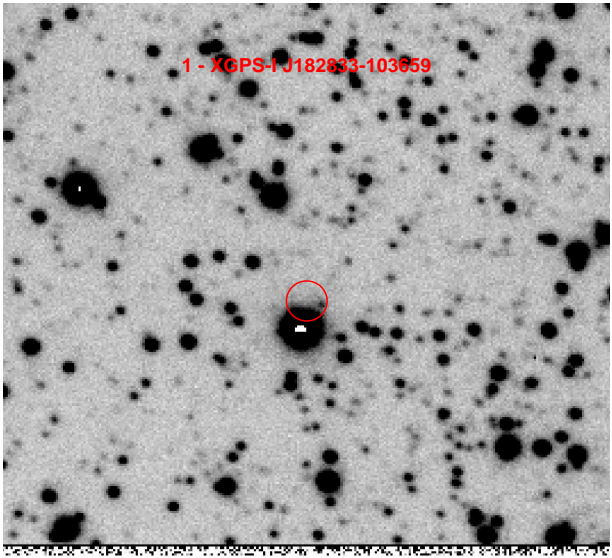,width=8cm,clip=true}}
\caption{A 30 second ESO-VLT I band exposure showing the 2XMM position of the 
bright X-ray transient [here designated XGPS-1]. North is at the top and East 
to the left. The field approximately covers 1$\times$1 arcmin.}
\label{trans}
\end{figure}

\subsubsection{XGPS-25}

XGPS-25,  alias 2XMM  J182854.6-112656, is  identified with  a  relatively bright
object (R $\sim$ 17.05) detected in the USNO and GSC2 R band and 2MASS and DENIS
infrared surveys but with no GLIMPSE counterpart (see optical finding chart in Fig. 33).  A radio source detected in the
Parkes-MIT-NRAO (PMN) catalogue \citep[][source catalogue for the tropical survey 
(-29\degr $<$  Delta $<$  -9.5\degr) at 4850 MHz]{griffith1994} is  found 20.2\arcsec\
away from the X-ray  source and owing to the error on  the radio position, could
still be associated with the 2XMM source. The optical spectrum exhibits a wealth
of interstellar bands  suggesting a rather strong reddening.  In addition to the
\Halp , \Hbet\ and $\lambda$6678  HeI emission lines, there is clear evidence for
the $\lambda$4686 HeII and  $\lambda\lambda$4630-4660 CIII-NIII  Bowen complex
high  excitation  lines.  The  equivalent  widths  remain  modest  ranging  from
$-$4.3\AA\  for  \Halp\ to  $\sim$  $-$1.1\AA\ and  $-$0.9\AA\  for  \Hbet\ and  HeII
respectively. The optical candidate is only  detected in the DENIS I and J bands, with only non-constraining upper limits available in the 2MASS H and K bands.  Although the
flux derived  from the GSC2.2 R band  appears fully consistent with  that of our
calibrated  spectrum, the  I  and J  band  seem about  $\sim$  0.4 mag  brighter
than the extrapolation toward longer wavelengths of the optical spectrum, suggesting some long term variability. The optical spectrum and the I
and J  flux can be independently fitted by a  B0V star energy distribution  undergoing E(B-V) $\sim$  1.5.  
XGPS-25 was only detected in observation 0051940601 which, unfortunately, suffers from a large background flare. The combined EPIC spectrum gathers 1.7\,ks and 3.8\,ks of usable pn and MOS exposure time. Owing to the small number of counts we used the C statistics in {\em XSPEC}. The energy distribution of XGPS-25 is extremely hard with a power-law photon index of 1.07$^{+0.32}_{-0.32}$ or a thin thermal temperature in excess of 10\,keV with \nh\ ranging from 1 to 8$\times$10$^{21}$\,cm$^{-2}$. The upper range of \nh\ is compatible with the E(B-V) estimated from the slope of the optical flux calibrated spectrum. In the direction of the source, E(B-V)  = 1.5 is reached in the distance range  of $\sim$ 2 to 4\,kpc \citep{marshall2006}, yielding M$_{\rm R}$ $\sim$ 0.6 - 2.1, while the unabsorbed X-ray luminosity  would be $\sim 4.3\times$10$^{32}$\ergs\ at 3\,kpc. 

\begin{figure}
\psfig{figure=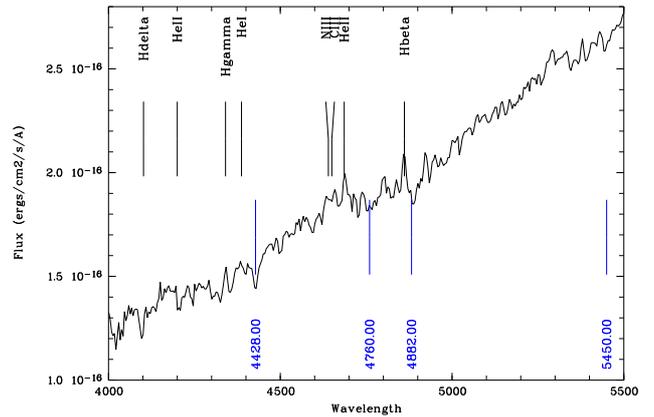,width=8.5cm,angle=-90,clip=true}    
\caption{Blue part of the optical spectrum of the counterpart of XGPS-25 showing the pronounced interstellar absorption lines (marked at the bottom) and the HeII and CIII-NIII Bowen emission lines.} 
\label{plot_heII_Xgps25}
\end{figure}

XGPS-25  is clearly  not a high -mass system  and the
determination of  its true nature  requires further observations.  The X-ray luminosity and spectral shape are consistent with those expected from a cataclysmic variable of the intermediate polar type \citep[see e.g.][]{patterson1994}. However, the likely high optical brightness of the counterpart seems at first glance inconsistent with a CV nature \citep{warner1987}. Although some old novae and novae-like systems do exhibit the N{\sc III}-C{\sc III} complex in emission \citep{williams1983} as well as some AM Her systems \citep{schachter1991}, the high relative strength of the Bowen complex compared to He{\sc II}\,4686 is at odds with what is observed in CVs.  In addition, the X-ray emission of novae-like systems arising from shocks in the boundary layer between the accretion disc and the white dwarf usually have soft thermal energy distributions at variance with that observed in XGPS-25. Although some
symbiotic binaries emit hard  X-ray  spectra \citep[see
e.g.][]{muerset1997,kennea2009}, the absence of M giant signature in the optical
spectrum and the low equivalent width of the emission lines seem to exclude this hypothesis. A possibility would  be that of a transient low-mass  X-ray binary  in the  quiescent state. Alternatively, XGPS-25  could be an X-ray  bright LMXRB whose  X-ray emission is mostly shielded  by an accretion disc  seen at high  inclination, such as for the accretion disc coronae (ADC) sources 4U 1822-37 and 4U 2129+47 \citep{white1982}. ADC sources exhibit a wide range of optical and X-ray properties. In the case of  4U 2129+47 for instance, a type I X-ray burst suggests that only 0.2\% of the central X-ray luminosity is observed \citep{garcia1987}. The amount of visible scattered X-ray emission will depend both on the optical depth and extent of the ADC and on the inclination of the accretion disc which determines the fraction of the ADC shielded by its edge \citep{white1982}. In this framework, the modest \Lx\ emitted by XGPS-25 may indicate a low central X-ray luminosity and/or a small visible scattered fraction. The LMXRB scenario is also supported by the relatively small EW of the optical lines which appear comparable to those exhibited by many LMXBs \cite[see e.g.][]{motch1989}.

\subsection{Notes on other unidentified XGPS sources}

In this section, we report on the 7 unidentified sources for which we could obtain significant spectroscopic information, useful for further optical identification work, and from which constraints on the possible nature of the X-ray emitter can be derived. The remaining six unidentified sources either have too faint optical candidates, offset XGPS positions with respect to 2XMM coordinates or very large error circles leading to an impracticable number of possible candidates.

\subsubsection{XGPS-5} 

This source has MOS and EPIC pn hardness  ratios consistent with an extragalactic  origin. The only bright star visible close to the 2XMM error circle (marked on Fig. \ref{xgps5}) displays the Mgb band,  \Halp\ absorption and weak high order Paschen lines. The overall spectrum  is consistent  with  a late  G  star undergoing  a  mild reddening  of E(B-V)$\sim$ 1.5  and therefore  is unlikely to be the  optical counterpart of  the hard X-ray source.

\begin{figure}
\centerline{\psfig{figure=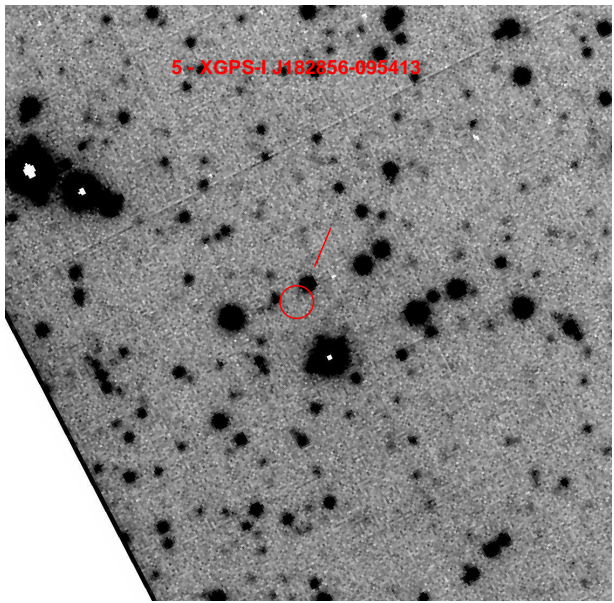,width=8cm,clip=true}}
\caption{A 20 second ESO-VLT white band exposure showing the 2XMM position of source XGPS-5. North is at the top and East to the left. The field of view shown is approximately 1\arcmin$\times$1\arcmin.}
\label{xgps5}
\end{figure}

\subsubsection{XGPS-7}

The R $\sim$ 18.8 star USNO-B1.0 0782-0453438 lies slightly outside of the 2.3\arcsec\
90\% confidence  error circle (see optical finding chart in Fig. 33).  The identification probability  is  less than 2\%, based on positional coincidence.  USNO-B1.0 0782-0453438 displays
\Halp\ in absorption and evidence for high-order Paschen
 lines  in  absorption.  There  is  a  GLIMPSE  source  G019.5207+00.3343  only
1.2\arcsec\ away  from the X-ray position  and probably coincident  with a faint
optical object exhibiting a reddened featureless spectrum. The overall SED can be
fitted equally  well by  an O9V  or a F6V  star undergoing  a high  reddening of
E(B-V) $\sim$ 5 and $\sim$ 4 respectively.

\subsubsection{XGPS-8} 

This source remains unidentified down to very faint optical limits. XGPS-8  is 
probably identical to the ASCA source  AX J1830.6-1002 \citep{sugizaki2001}  located only 18.6\arcsec\ away from the XMM position. XGPS-8 is also probably detected by the
SPI  instrument on  board  INTEGRAL \citep{bouchet2008}  up  to 100\,keV and  by
IBIS/ISGRI up to 40\,keV \citep{bird2007}, although the identification of the SPI
detection  with the  ASCA source  is  quoted as  tentative since  more than  one
possible identification could exist.  Interestingly XGPS-8 also coincides with a
radio  source,  GPSR5 21.631-0.007,  detected  in the  5GHz  VLA  survey of  the
Galactic Plane conducted by \cite{becker1994}. This source is also present in the NVSS
catalogue. The XMM EPIC spectrum is not consistent with that of an extragalactic
object  since it  displays a  significant  flux down  to 0.5\,keV  and seems  to
consist of  two distinct components peaking  at $\sim$ 1\,keV  and 5\,keV. Three
faint  stars are visible  in the 90\% confidence  XMM error
circle (see  Fig. \ref{xgps8}). We have  collected spectra of the brightest object
located at the south of the error circle  and of the faint one to the north both marked on Fig. \ref{xgps8}. The
two  spectra are  rather noisy  and none  of these  two candidates  displays any
convincing evidence  for Balmer or  Paschen emission or absorption  lines. The
third object located at the west of the error circle is probably coincident with
2MASS   18303813-1002463,  DENIS   J183038.2-100246  and   the   GLIMPSE  source
G021.6316-00.0061. The  probability of identification  of XGPS-8 with the 2MASS  source is $\sim$ 40\%.  The 1.2$\mu$ to  4.5$\mu$ SED does  not allow us to meaningfully 
constrain the spectral type, but suggests a reddening of the order of E(B-V)
$\sim$ 4, which might not be consistent  with the presence of a soft component in
the X-ray spectrum.

\begin{figure}
\centerline{\psfig{figure=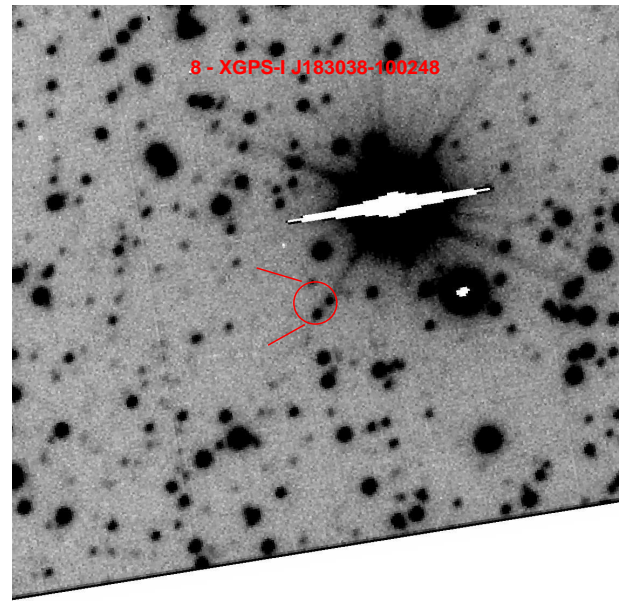,width=8cm,clip=true}}
\caption{A 20 second ESO-VLT white band exposure showing the 2XMM position of source XGPS-8. North is at the top and East to the left. The field of view shown is approximately 1\arcmin$\times$1\arcmin.}
\label{xgps8}
\end{figure}

\subsubsection{XGPS-28} 

This source  was  detected  three  times  by  XMM-Newton  (in
observations  0051940601,  0104460401  and  0135747001) and  could  be  slightly
variable. Its MOS hardness ratio suggests a likely Galactic nature. We  observed the relatively faint  star at the East  edge of the
error  circle which  is  probably identified  with  2MASS 18282782-1117500 (see optical finding chart in Fig. 33).  The
spectrum is highly reddened and displays  \Halp\ and high order Paschen lines in
absorption. The error circle contains a relatively bright GLIMPSE source G020.2759-00.1138 with a
3.6$\mu$ mag of 11.66$\pm$0.07. The Spitzer source has no obvious 2MASS or optical counterpart.

\subsubsection{XGPS-31} 

XGPS-31 has EPIC pn and MOS hardness ratios suggesting a Galactic origin. Its rather large error circle of  3.0\arcsec\ (90\% confidence level) {overlaps with}  three relatively bright objects. We  obtained optical  spectra of  the two bright  stars located  on the Eastern edge  of the error circle  (marked in Fig. \ref{xgps31}) but  failed to acquire spectral information  on the object located  at the Western edge. Neither of these two objects exhibit spectral signatures of X-ray activity such as Balmer emission lines. 

\begin{figure}
\centerline{\psfig{figure=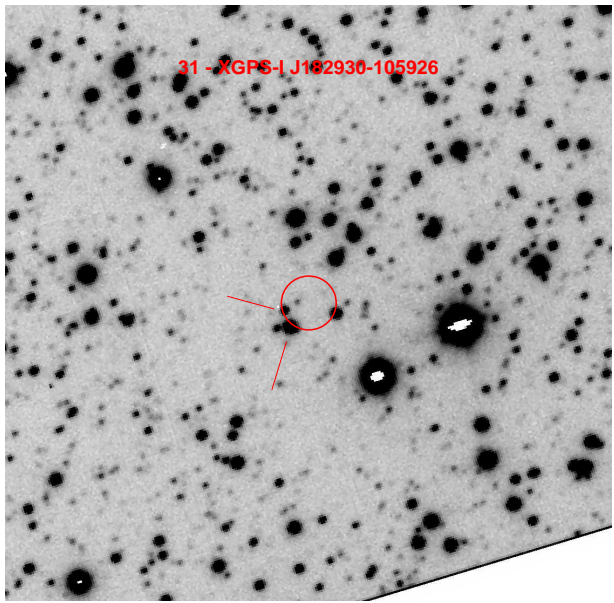,width=8cm,clip=true}}
\caption{A 20 second ESO-VLT white band exposure showing the 2XMM position of source XGPS-31. North is at the top and East to the left. The field of view shown is approximately 1\arcmin$\times$1\arcmin.}
\label{xgps31}
\end{figure}

\subsubsection{XGPS-42}\label{s42}

The position of 2XMM J182833.3-102650 coincides with GLIMPSE G021.0390+00.2608, USNO-B1.0 0795-0405657 and 2MASS 18283337-1026505. In spite of the rather large 90\% confidence radius of 2.7\arcsec, the individual probabilities of identification of the X-ray source with the USNO-B1.0 and 2MASS entries are 94\% and 95\% respectively and thus qualify this relatively bright object (R $\sim$ 16.2) as a likely candidate (see Fig. \ref{xgps42}). The optical spectrum reveals a reddened continuum with \Halp\ in absorption and no evidence for Paschen lines. In the pipeline processing used for the 2XMM catalogue production, the {\em eposcorr} task had failed to correct X-ray positions in observation 0135740901 for remaining boresight errors. However, in the 2002 pipeline processing we actually used at the telescope, the {\em eposcorr} task had been able to find a likely shift and translation improving the attitude and moved the position of the source towards candidate "A", the fainter object South-East of USNO-B1.0 0795-0405657 but still located close to the 2XMM 90\% confidence circle. Candidate A might be associated with the GLIMPSE source G021.0385+00.2599. Our ESO-VLT optical spectrum of object A reveals again a reddened continuum with \Halp\ absorption. However, deep Pashen lines are clearly detected in absorption and hint at an early type star. On this basis, we consider candidate A is a more likely counterpart of XGPS-42 than the brighter USNO-B1.0 star which fails to show spectroscopic evidence of strong coronal activity.

\begin{figure}
\centerline{\psfig{figure=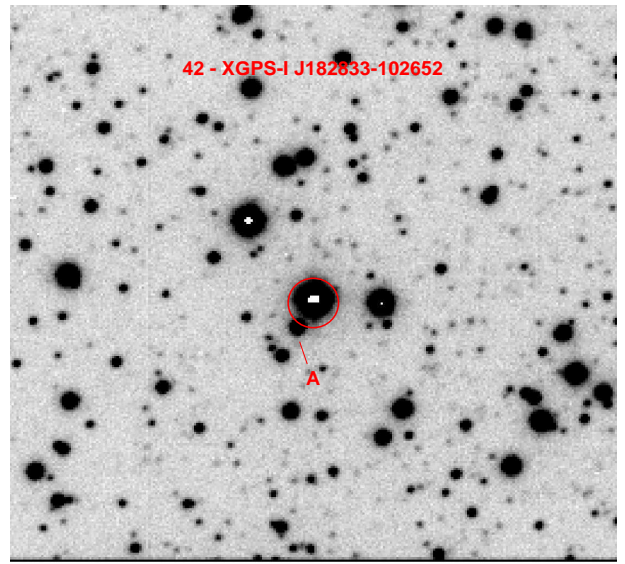,width=8cm,clip=true}}
\caption{A 20 second ESO-VLT white band exposure showing the 2XMM position of source XGPS-42. North is at the top and East to the left. The field of view shown is approximately 1\arcmin$\times$1\arcmin.}
\label{xgps42}
\end{figure}

\subsubsection{XGPS-69}

The relatively bright star (R\,$\sim$\,17), located somewhat outside of the south-west edge of the 90\% confidence error circle is USNO-B1.0 0785-0414570 (highlighted in Fig. 33). Its probability to be associated with the X-ray source if of only 22\%. Our optical spectrum shows Balmer, Paschen absorption lines and a continuum consistent with a F star reddened by E(B-V)\,$\sim$1.1.

\subsubsection{Tentative identification of several bright XGPS sources missing in the preliminary catalogue}\label{sms}

A  small  number  of  relatively  bright  X-ray  sources present in the
final catalogue were actually missing  in  the preliminary  XGPS list  we
used  at the  time of  the observations.  These sources would have been
bright enough in the broad X-ray band to enter our sample of sources
scheduled for optical follow-up. We thus discuss below their possible
identifications:

XGPS-I J183216-102303:  Ranks as the fourth brightest in  the broad
(0.4-6.0\,keV) band. It is also known  as 2XMM J183216.1-102301 and has a bright
USNO-B1.0  (R  =  11.80)  and  2MASS  (K  =  9.2)  match  with  a  probability  of
identification of  99\% and 91\%  respectively.  XGPS-I J183216-102303  was also
detected by  ROSAT as  2RXP J183216.1-102255.  The  overall picture  suggests an
identification with  a bright active  corona. The  MOS hardness ratio of 0.51 given in \cite{hands2004} suggests a hard spectrum possibly indicative of an RS CVn system. 

XGPS-I J182830-114516: Ranks at the twelfth  position in the broad band.  As for
the previous  case the source has a  bright (R = 12.8)  optical counterpart with a
high probability of association (86\%). The  soft MOS hardness ratio of $-$0.82 and the EPIC pn hardness ratios of HR2\,=\,$-$0.319$\pm$0.067 and HR3\,= \,$-$0.625$\pm$0.096 clearly indicate an active corona.

XGPS-I  J183053-101936: Stands  among the  faintest of  our sample  at  the 66th
position. It is clearly  identified with an active corona on the  basis of a highly
significant  match with  2MASS 18305320-1019370 (K = 9.4; 95\%)   and  USNO-B1.0
0750-13374636 (R  = 14.4;  92\%) and of a soft MOS hardness ratio of -0.08.

XGPS-I J183057-103452  (35th), J183156-102447 (60th) have no  match with USNO-B1.0
nor 2MASS and no tentative identification in Simbad.

Accordingly,  we assigned  a  stellar nature  to the first  three sources  and include 
these alongside the 43 objects studied at the telescope when
discussing the nature of the XGPS source content.
 
\section{Properties and characteristics of the XGPS source sample}\label{section6}

We  show in Fig.  \ref{plot_2XMM_HR_distribution} the  distribution of  the full
sample  of  XGPS/2XMM sources in an X-ray  two-colour  diagram  based on  the
hardness ratios  deriving from  the EPIC pn  camera. The XMM-Newton EPIC hardness ratios are based on five energy bands which expressed in keV units are 0.2--0.5, 0.5--1, 1--2, 2--4.5 and 4.5--12.0. Hardness ratio $i$ is then defined as; 
\begin{equation}
\rm HR_{i} = \frac{C_{i+1} - C_{i}}{C_{i+1} + C_{i}}
\end{equation} 
with $C_{i}$ the count rate in band {i} corrected for vignetting. Soft sources exhibit negative hardness ratios. 

The diagram overplots the source nature (active corona, X-ray binary, cataclysmic variable or unidentified) as derived from our optical spectroscopic campaign for a subset of bright sources. Because of the limit of 0.3 set on the hardness ratio errors, only a fraction of the 269 XGPS/2XMM sources appear in the diagrams. In addition, those sources  with likely infrared or optical counterparts based on the probability of identification with 2MASS or USNO-B1.0 are highlighted. 

As expected, the sources with spectroscopically confirmed stellar identifications occupy the region with  pn\_HR3 $<$ 0.0 indicative  of an  underlying soft X-ray
spectrum, although  in some  cases with non-negligible  soft X-ray  absorption (as
measured  by pn\_HR2). It is  also   remarkable  that  most  of   the  XGPS  entries having  a likely identification with a bright USNO-B1.0 or 2MASS candidate (here with a probability of association larger than 90\%) are also found in the soft region implying that these are also  very likely active coronae. The handful of spectroscopically identified active coronae without high probability USNO-B1.0 or 2MASS associations are faint Me stars. In these cases, the low  identification probabilities reflect the high surface density of objects as faint as the M star counterparts.  

Interestingly, a small number of hard X-ray sources also have high probability matches with relatively bright optical or infrared counterparts. At the probability threshold of 90\% chosen here, the sample ``purity'' or integrated probability of identification is about 99\%, leaving only 1\% of spurious matches (see Tab. \ref{rcstat}). We note that this probability threshold also implies that we miss about two thirds of all true associations. The total number of XGPS/2XMM sources shown in Fig. \ref{plot_2XMM_HR_distribution} matching a USNO-B1.0 or a 2MASS entry and having errors on pn\_HR2 and pn\_HR3 hardness ratios smaller than 0.3, is of 38. Therefore, at most one of these high likelihood associations could be spurious.  Since the probability of a spurious match does not depend on source hardness, the frequency of wrong identifications will be larger for hard X-ray sources which on the average have fainter optical counterparts and are less likely to have a bright stellar counterpart than soft X-ray sources. This is possibly the case of source XGPS-42 (see \S \ref{s42}) for which the bright USNO-B1.0 counterpart has no spectroscopic evidence suggesting it is the true counterpart of the X-ray source, whereas the much fainter candidate A exhibits the spectral signatures of an early type star and on this basis can be considered as a serious contender. XGPS-42 (pn\_HR2 = 0.74$\pm$0.14; pn\_HR3 = 0.18$\pm$ 0.14; pn\_HR4 = -0.02$\pm$ 0.14) appears as the hardest undidentified source having a high probability of association with a bright optical object. However, the majority of the associations between a hard X-ray source and a relatively bright optical or infrared object, are still expected  to be true identifications  in agreement with the fact that the $\gamma$-Cas like object XGPS-36 and the Wolf-Rayet associated with XGPS-14 are found amongst them.

\begin{figure*}
\begin{tabular}{cc}
\psfig{figure=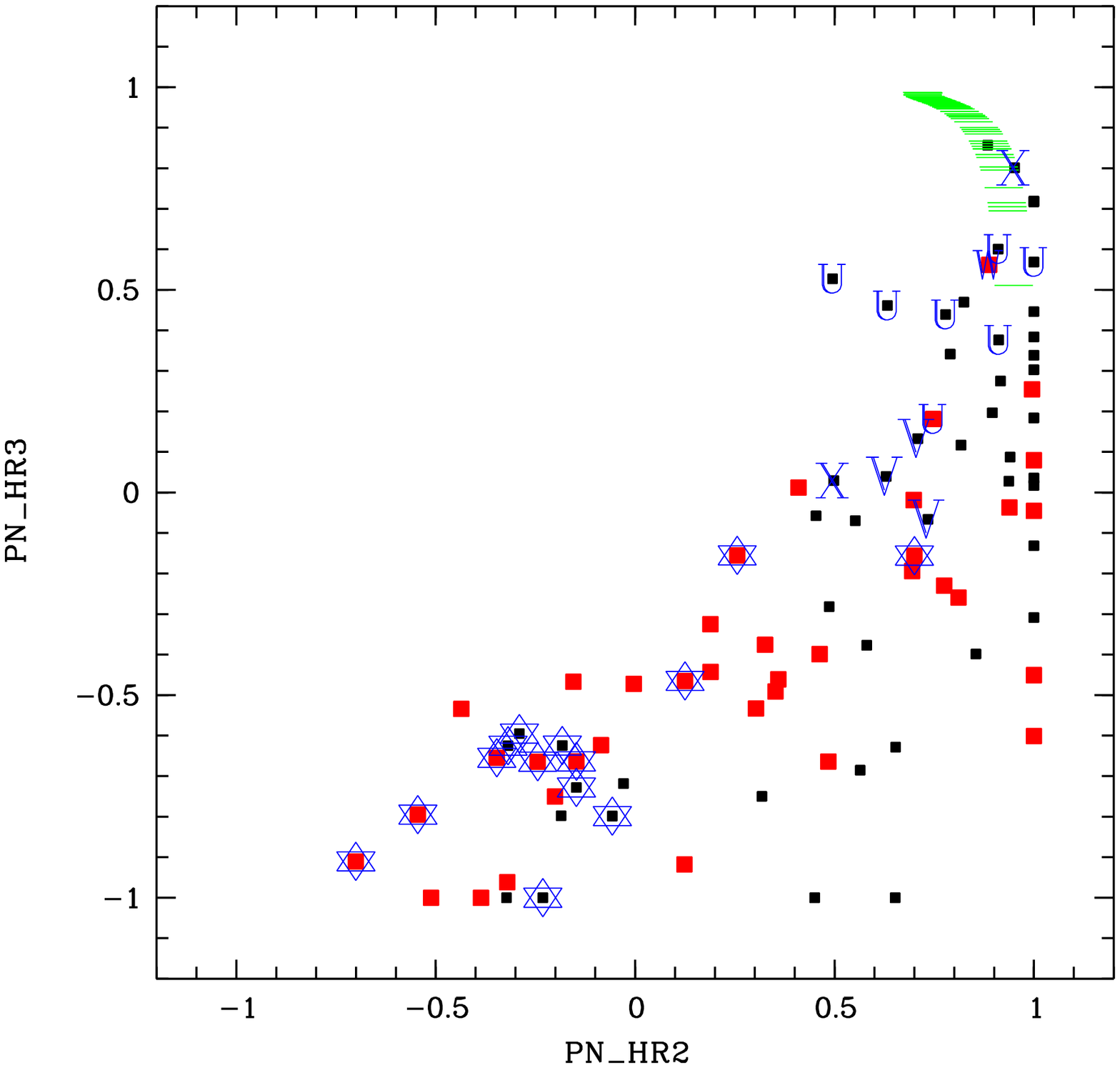,width=8.5cm           
,bbllx=1.0cm,bburx=21.5cm,bblly=1cm,bbury=21.5cm,angle=-90,clip=true}    
& 
\psfig{figure=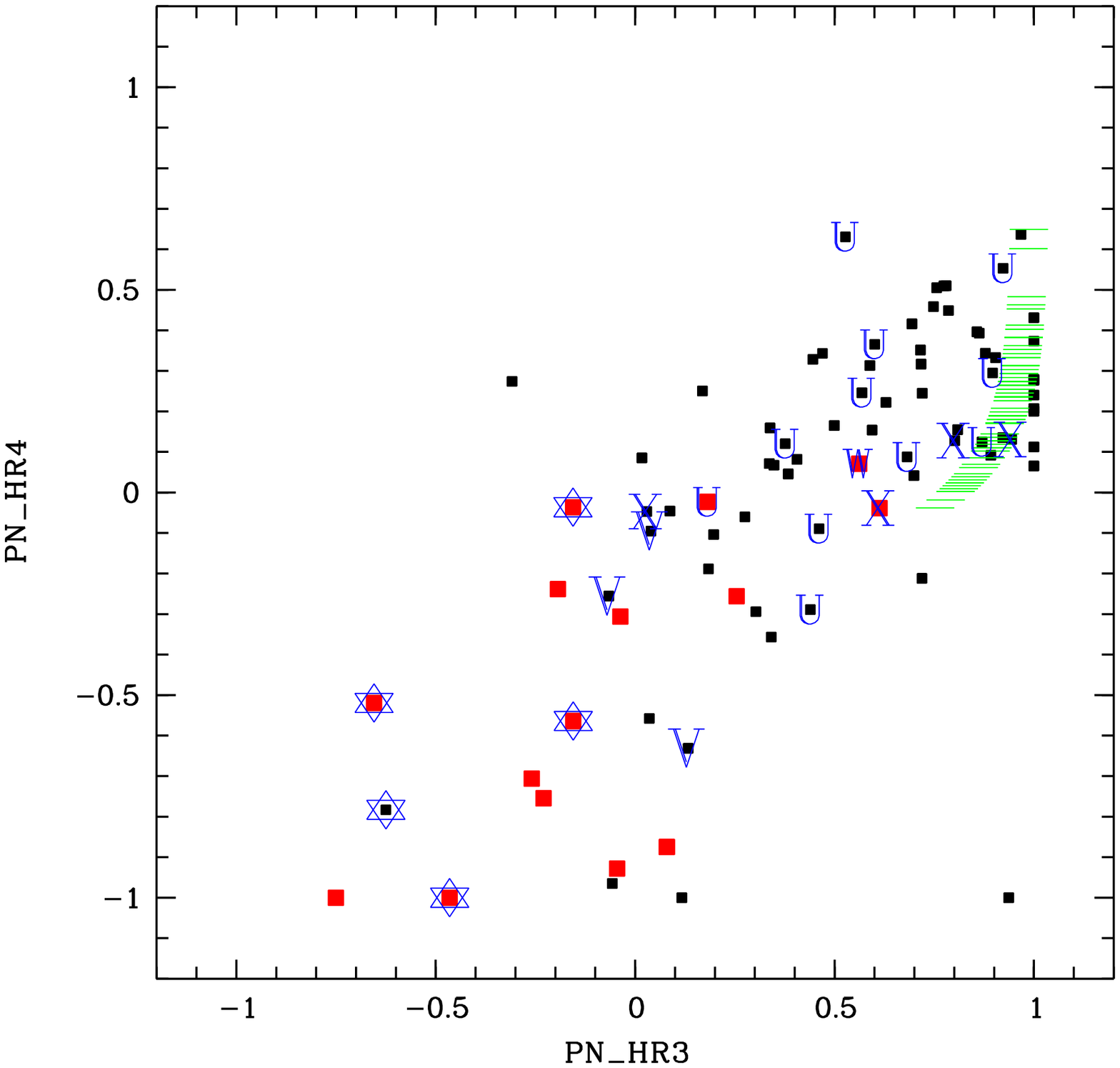,width=8.5cm,angle=-90              
,bbllx=1.0cm,bburx=21.5cm,bblly=1cm,bbury=21.5cm,clip=true}    
\\
\psfig{figure=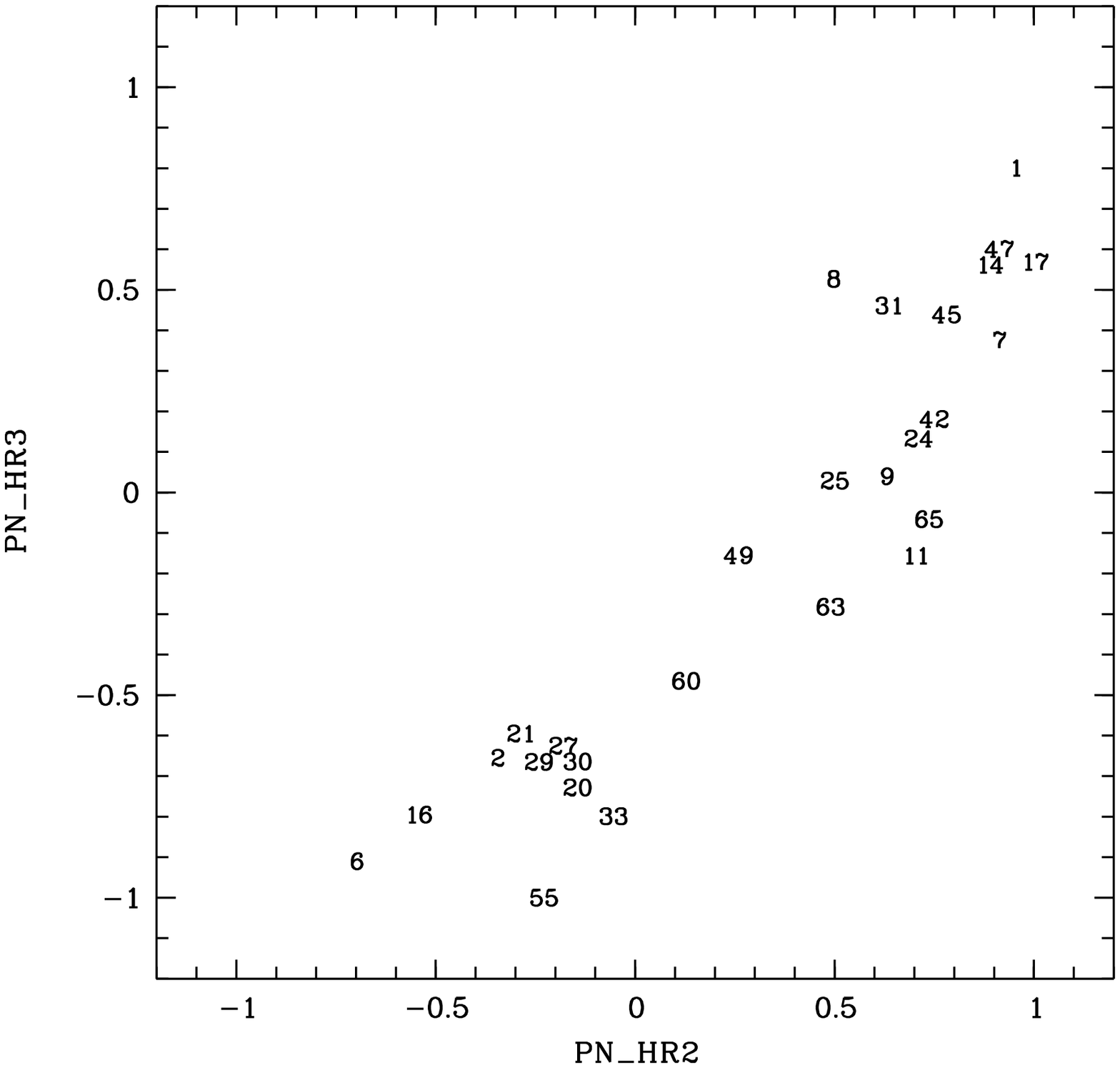,width=8.5cm           
,bbllx=1.0cm,bburx=21.5cm,bblly=1cm,bbury=21.5cm,angle=-90,clip=true}    
& 
\psfig{figure=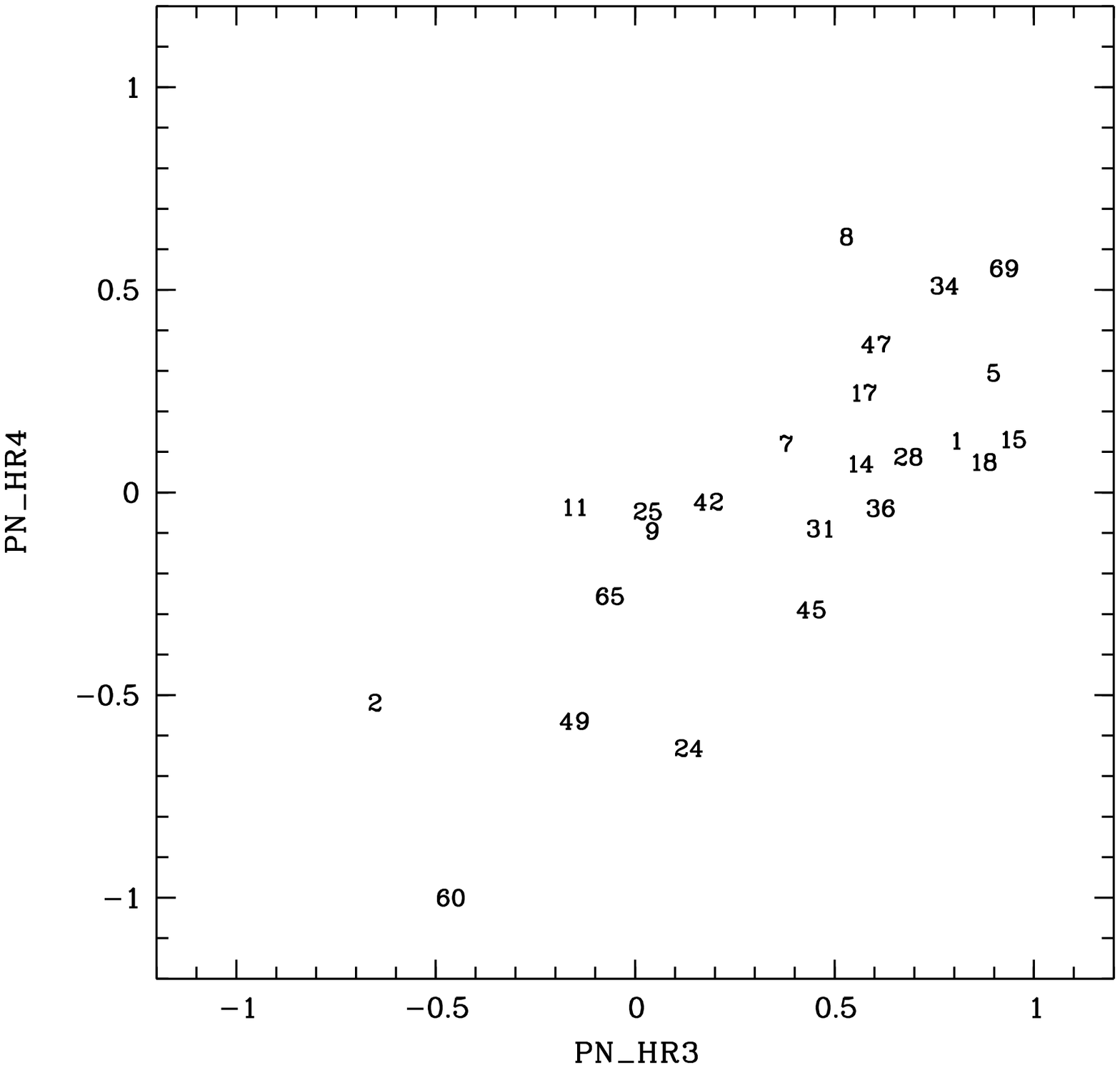,width=8.5cm,angle=-90              
,bbllx=1.0cm,bburx=21.5cm,bblly=1cm,bbury=21.5cm,clip=true}    
\\
\end{tabular}
\caption{The distribution of the XGPS/2XMM sources in the EPIC pn hardness ratio diagrams. For clarity, we show only those sources with hardness ratio errors less than 0.3. Upper panel: Larger red squares; sources having an optical or infrared counterpart with matching probability above 90\%. Small black squares; the remaining XGPS sources. The overplotted
characters (in blue) show the sources observed at the telescope. Stars - identified active coronae;  V - cataclysmic 
variables; X - high-mass and low-mass X-ray binary candidates; W - Wolf-Rayet;
U - sources unidentified at the telescope.  The green strips illustrate the
locus occupied by extragalactic sources, computed assuming a power-law spectrum of photon index 1.7 and the total Galactic absorption on the line of sight.  Lower panel: corresponding id numbers of the XGPS sources observed at the telescope.}
\label{plot_2XMM_HR_distribution}
\end{figure*}

The identified or candidate low-mass and high-mass X-ray  binaries and cataclysmic variables preferentially exhibit harder X-ray spectra, some approaching that expected for extragalactic sources. However, CVs in general display spectra significantly softer than those of background AGN. 
Only one of the two XGPS sources identified with an early type B star (XGPS-56, not plotted on Fig. \ref{plot_2XMM_HR_distribution} for clarity) has an EPIC pn measurement. Due to the low count rates in the soft band HR2 is not defined. Its HR3 and HR4 ratios of 0.92$\pm$0.11 and 0.14$\pm$0.16 respectively clearly indicate a hard source. Although early-type stars have thin thermal spectra of temperature comparable to that of active coronae, their somewhat higher average X-ray luminosity (\Lx $\sim$ 10$^{31-32}$\ergs) allow their detection at large distances in spite of the large hardening due to significant photoelectric interstellar absorption. Source XGPS-14, identified with a likely wind colliding Wolf-Rayet star also appears as a hard X-ray source. 
 
It can  also be  seen that  several of the  sources we  failed to  identify have
hardness ratios significantly softer than those expected  from a typical AGN absorbed by the full
Galactic line-of-sight column  density. The nature of these  Galactic sources is
hard  to  guess. In  the  absence  of a  2MASS  or  GLIMPSE  infrared source  an
identification with a  massive X-ray binary or a  $\gamma$-Cas like analog seems
unlikely. Cataclysmic  variables would make  viable candidates since  we know
from our identification  of the relatively X-ray bright source, XGPS-9 with a V $\sim$
23 object, that CVs may  turn out to be hard to identify in the
large XMM-Newton error  boxes and even beyond the spectroscopic  reach of the VLT,
at least  in the  optical band.   However,  some of  these unidentified
sources may belong to a new group of moderate- to low-luminosity X-ray
sources such as wind-accreting neutron stars \citep{pfahl2002} or pre low-mass X-ray binaries \citep{willems2003}. If  this putative  population has a  space density which  rises sharply
with declining  galacto-centric radius, then  the X-ray surveys of  the Galactic
Plane carried out  prior to XMM-Newton and Chandra may  contain very few relatively
nearby archetypes of this population.

With an assumed luminosity of the CVs of \Lx\,$\sim\,10^{31}$\ergs\ 
and a limiting flux of $\rm F_{\rm lim}\,\sim\,2\times
10^{-14}$\,erg\,cm$^{-2}$\,s$^{-1}$, the maximum distance for  a CV to
be included in  our sample, ignoring absorption,  is about 2\,kpc (absorption
with  a column density of  \nh$\,=\,10^{22}$\,cm$^{-2}$ only gives a 10\%
flux reduction  in  the $2--10$\,keV band).   The  volume probed
per XMM-pointing is  $\sim 1.1\times 10^5$\,pc$^3$. The  CV space
density  is not well known,  with recent observational  constraints
derived  from  the ROSAT  Bright Survey \citep{schwopeetal2002} and
the ROSAT NEP Survey \citep{pretoriusetal2007} for non-magnetic CVs giving 
$\rho \sim 1.5 \times 10^{-6} \dots 1.1 \times
10^{-5}$\,pc$^{-3}$. Both  values are nevertheless compatible with
each other, because  the  RBS  only probes  the  intrinsically  bright
part of  the  parent population, while  the NEP  also probes  the
faint end  of the luminosity function at  $\log({\rm L}_{\rm X}) <
30.5$\,erg\,s$^{-1}$, which currently is very poorly
understood\footnote{see the discussion  by Pretorius  in the
conference {\it  Wild Stars  in the  Old West} published on-line only 
http://www.noao.edu/meetings/wildstars2/talks/tuesday/pretorius.pdf}.
In 17 XGPS pointings   we   might   thus   expect   approximately  3
(RBS-type) \Lx\ $   \sim 10^{31}$\,erg\,s$^{-1}$  non-magnetic CVs.
Our survey area is  too  small to uncover  the   likely  more
abundant low-luminosity  CVs  at  \Lx\ $  = 10^{30}$\,erg\,s$^{-1}$.
In 17 XGPS pointings  we are  surveying a  volume of $6\times 
10^4$\,pc$^{3}$ and we expect to have one at maximum in our sample.

\subsection{Log N - Log S curves}

We  constructed \lnls\  curves using the final XGPS source list limited to count rates larger than 0.5 cts/ks and 1.5 cts/ks in the soft and hard bands respectively.  This relatively high threshold minimizes the effects of possible uncertainties in the area correction (see Fig. 8 in \cite{hands2004}) while keeping the flux range adapted to our optically identified sample of bright sources. The total number of sources entering the \lnls\ curves is of 212 for the soft band and of 129 for the hard band. We apply the count  to flux  conversion factor  of 2.6$\times$10$^{-14}$\ergscm / MOS count ks$^{-1}$ for the hard band spectra given in \cite{hands2004}. This factor roughly corresponds to an ``average'' spectrum in the context of both Galactic and extragalactic sources (\nh\ =  1$\times$10$^{22}$\,cm$^{-2}$ with a power law of photon index 1.7). 

Since the stellar component is likely to dominate source counts in the soft band, we assumed a thin thermal spectrum with kT\,$\sim$\,0.5\,keV absorbed by \nh\,$\sim$\,10$^{21}$\,cm$^{-2}$. These spectral parameters are typical of active stars \citep{guedel2009} observed at distances of a few hundred parsecs. The resulting flux conversion factor is 5.2$\times$10$^{-15}$\ergscm / MOS count ks$^{-1}$ in the 0.4--2\,keV band. 

As shown in Fig. 15 of \cite{hands2004}, the 2--10\,keV \lnls\ XGPS curve nicely merges at high fluxes with the ASCA  \lnls\ \citep{sugizaki2001} and at faint  fluxes with the results of a deep Chandra observation of a specific field at $l$ = 28.5\degr\
\citep{ebisawa2001,ebisawa2005}  (see Fig. \ref{lnls_Xgps}). 
We expect a non-negligible contribution from extragalactic sources in  this hard band despite the 
large Galactic foreground absorption. In order to estimate this fraction, we considered both  the  \lnls  \ relation  of \cite{campana2001} and that recently obtained by \cite{mateos2008} (using the three-powerlaw description). 

In the  region covered  by the  XGPS, the  mean total  Galactic column
density  of $\sim$  9$\times$10$^{22}$ cm$^{-2}$  is high  enough to  reduce the
normalisation of the \lnls\ curve for extragalactic sources by a factor of about two. Consequently, in the flux range covered by the XGPS (between 5$\times$10$^{-14}$ and 1$\times$10$^{-12}$\ergscm ; 2--10\,keV), the mean number of extragalactic sources per square degree is only 43\% of what would be observed at high galactic latitude. 

The X-ray bright part of the \lnls\ curve of \cite{campana2001}, most relevant for our study, is based on the results of the ASCA GIS medium sensitivity survey of \cite{dellaCeca2001} and \cite{ueda2005}. At low flux, \cite{campana2001} use data from several deep Chandra surveys. On the other hand, the \lnls\ curve of \cite{mateos2008} is only based on XMM-Newton measurements and being free of relative calibration uncertainties may be more directly comparable with our work. 

\begin{figure}
\psfig{figure=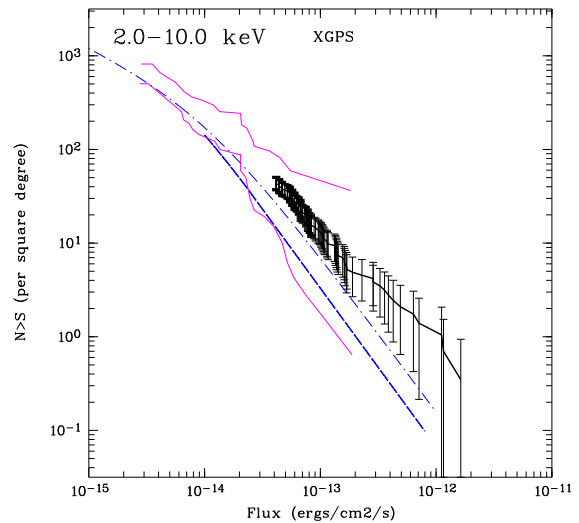,height=8cm,bbllx=1.0cm,bburx=21.8cm,bblly=1cm,bbury=22.5cm,angle=-90,clip=true}          
\caption{The \lnls~curve of XGPS sources in the hard band. Error bars represent the one sigma counting statistics. At faint fluxes, the 90\% confidence region of the \lnls~curve of \cite{ebisawa2001} is shown in magenta. The dash-dotted blue curve represents the expected extragalactic contribution assuming a mean \nh\ of 9.0\,$\times$ 10$^{22}$\,cm$^{-2}$. Thin line: after \cite{campana2001}. Thick line: after \cite{mateos2008}.} 
\label{lnls_Xgps}
\end{figure}

It can be seen in Fig. \ref{lnls_Xgps} that the curve based on ASCA is systematically higher by a factor of 1.2 to 2.0 than that derived from XMM-Newton. This large discrepancy is also exemplified in Fig. 6 of \cite{mateos2008}.  According to these authors, the reasons explaining the higher normalisation of the ASCA \lnls\ remain unknown, since cross calibration uncertainties are unlikely to account for more than a 10\% difference. We may argue, however, that since the curves are based on serendipituous detections in the field of view of observations devoted to bright targets, they are necessarily difficult to correct for in the flux range covered by the targets. Such a bias could explain the difference between the two \lnls\ curves and would then argue in favour of using the ASCA relation, since the targets of the ASCA pointings were in general much brighter than those selected for the XMM-Newton observations. Depending on the assumed \lnls\ curve, up to 50\% (ASCA) or 30\% \citep{mateos2008} of the low-latitude source population at \Fx\ $\ga$ 5\,10$^{-14}$\ergscm\ could be of extragalactic origin. At this threshold, the fraction of XGPS/2XMM sources with EPIC pn\_HR3 and pn\_HR4 consistent within 2\,$\sigma$ with that expected from an extragalactic source is $\sim$ 50\,$\pm$\,8\%. Given that this must represent an upper limit on the fraction of extragalactic sources contaminating the sample (since distant galactic sources
will have hardness ratios largely indistinguishable from extragalactic objects) it strongly suggests that at least half of the hard sources are indeed of Galactic origin. 

At a higher flux of \Fx $\sim$ 10$^{-12}$\ergscm, the relative contribution of AGNs decreases to $\sim$ 10\% in agreement with results from the ASCA Galactic Plane Survey \citep{sugizaki2001}, albeit with a close to 100\% upper limit owing to the large  uncertainty  caused  by small number statistics. The exact contribution of extragalactic and Galactic sources is also made somewhat uncertain by the intrinsic scatter of the count to flux conversion  resulting from the variety of photo-electric absorptions and intrinsic spectral shapes.
  
We investigated how the likely distribution of the intrinsic spectra and photoelectric absorptions of Galactic X-ray sources would affect the hard X-ray \lnls\ curves. For that purpose, we computed the distribution of the 2--6\,keV count to 2--10\,keV flux conversion factors assuming powerlaw indices evenly distributed between 1.3 and 2.1 and $log$(\nh) in the range of 21.5 to 22.5. Assuming a constant ISM density and a Galactic radius of 14\,kpc, a $log$(\nh) of 22.5 is reached at $\sim$ 6.5 kpc. We assumed that this value was representative of the limiting distance of our survey. Our most absorbed source  XGPS-3 (and probably most luminous apart form XGPS-1) also exhibits $log$(\nh) $\sim$ 22.5. This ``flat'' distribution of spectral parameters yields a rms scatter of 15\% on the conversion factor. With the \lnls\ slopes found here, the corresponding errors on the integrated surface densities of Galactic sources remain of the same order and become comparable to counting errors for the total Galactic + extragalactic curve near our faint flux limit. The effect should thus be relatively small and in the absence of good knowledge of the true \nh\ and spectral shapes exhibited by hard Galactic sources one cannot estimate more precisely how the expected scatter of spectral parameters impacts the Galactic \lnls\ curves. 

We also tried to estimate the consequences of assuming a common mean count to flux conversion factor for both Galactic and extragalactic sources. For instance, removing the ASCA \lnls\ seen through the mean survey \nh\ of 9$\times$10$^{22}$ cm$^{-2}$ from the hard MOS count distribution rises the expected number of hard Galactic sources brighter than 1$\times$10$^{-13}$\ergscm\ and 5$\times$10$^{-14}$\ergscm\ to values comparable to those obtained assuming the Mateos extragalactic \lnls\ and the count to flux conversion factor of \cite{hands2004}. However, the size of the effect will also depend on ISM clumpiness and on the intrinsic scatter of the AGN spectral parameters.

In the soft  0.4--2.0\,keV band, we have available the ROSAT X-ray extragalactic \lnls  \  curve reported  by  \cite{hasinger1998}  and at faint  flux  levels that in \cite{campana2001}. However, the effect of
the integrated Galactic photoelectric absorption is so large in this case that  the expected number of AGN detectable at the XGPS flux level is several orders of magnitude below the observed soft \lnls\ curve. We thus expect virtually no extragalactic ``contamination'' in the 0.4--2.0\,keV band. 

In order  to  illustrate  the  relative contributions  of  the  various  classes of identified  source  to the  total  Galactic  population, we  plot  in
Fig. \ref{egsub_lnls_Xgps} the  soft and hard band \lnls\  curves {\em corrected
for the contribution of AGNs} using for the hard band the \lnls\ curve of ASCA. 
We define here as stars  all   sources  with  a   positive  or  tentative   stellar  spectroscopic identification i.e. with classes AC or AC?? in Tab. \ref{xgps_tab}.
We also consider those XGPS sources having an individual probability of identification with a USNO-B1.0 or a 2MASS entry  greater than 90\% (corresponding to a $\sim$ 1\% level of spurious associations) and 50\%. At this later threshold, we expect the sample to be $\sim$ 80\% complete but with 20\% of spurious matches in the sample. In other words, although we do not know which sources are actually identified with a USNO-B1.0 or a 2MASS entry, the total number of matches over the whole sample will be approximately correct. 
We also note that at the flux limit used here, only 85\% of the XGPS sources entered in the \lnls\ curve have a corresponding entry in the 2XMM catalogue and were thus searched for a possible candidate USNO-B1.0 or 2MASS identification. The true fraction of statistically plus spectroscopically identified sources should thus be slightly higher than shown on Fig. \ref{egsub_lnls_Xgps} by the dashed lines. The fact that the source density of bright 2.0--10\,keV X-ray sources identified by spectroscopic or statistical means appears comparable to that expected for the Galactic population only, suggests that many of the hard XGPS sources discarded from the optical follow-up campaign on the basis of their too faint optical counterparts are indeed of extragalactic nature.

We used the maximum likelihood methods of \cite{crawford1970} and \cite{murdoch1973} to constrain the slopes of the 'Galactic only' \lnls\ curves. The hard X-ray slope appears consistent with Euclidean ($\alpha$ = $-$1.46\,$\pm$0.14) and is clearly at variance with that obtained from the ASCA Galactic Plane Survey \citep[$\alpha$ = $-$0.79\,$\pm$\,0.07;][]{sugizaki2001} in the flux range of 10$^{-10}$ to 10$^{-12}$\ergscm . This steepening was also noted in \cite{hands2004}. Such a quasi-Euclidean distribution suggests that the hard sources detected in the XGPS flux regime are relatively homogeneously distributed and do not suffer too much from absorption effects. On the other hand, the soft X-ray \lnls\ is found to be less steep with $\alpha$ = $-$1.26\,$\pm$0.10. This value is consistent with those reported for other shallow or medium deep soft X-ray Galactic surveys \citep[$\alpha$ = $-$1.10\,$\pm$\,0.26;][]{hertz1984} and \citep[$\alpha$ = $-$1.05\,$\pm$\,0.13;][]{motch1997}. 

We list in Tab. \ref{id_stat} the identification statistics based on spectroscopic work for various classes of source and also positional coincidence with bright USNO-B1.0 and 2MASS entries. The table also gives estimates of the true number of Galactic sources by removing the expected extragalactic contribution using the extragalactic \lnls\ curves. In this table, the two tentative stellar identifications (XGPS-34 and XGPS-63) are included in the group of active coronae.

In the  soft band, we spectroscopically identify  all sources down to  a flux of
2.7$\times$10$^{-14}$\ergscm\ with  $\sim$ 17\%  representing XRBs  candidates and  CVs. Adding USNO-B1.0  and  2MASS  90\%  probability  matches allows  us to  build  a  completely identified sample down to 2.3$\times$10$^{-14}$\ergscm.  At \Fx\ = 1$\times$10$^{-14}$\ergscm, the fraction of sources identified by spectroscopic or statistical means is still high, of  the order  of $\sim$ 75\%.  At this flux, XRB candidates and CVs contribute only $\sim$ 10\% of the total number of sources.

The census of optical identifications in the hard X-ray  band remains good down
to a flux level of $\sim$ 1$\times$10$^{-13}$\ergscm\ at which we identify one third of the total number of sources and $\sim$ 70\% of the expected number of Galactic sources (assuming the extragalactic ASCA \lnls). Above  \Fx\ = 2$\times$10$^{-13}$\ergscm , we identify about half of the total number of sources. The relative fraction of solar-type stars and XRB candidates plus CVs are reversed compared to what is observed in the
soft  band, with  typically 10\%  of  the expected  number of  non-extragalactic
sources being  identified with a  stellar corona and  $\sim$ 40\% with  a XRB
candidate or a CV at \Fx\ $\geq$ 1$\times$10$^{-13}$\ergscm .

\begin{table}[ht]
\centering
\caption{Source identification statistics as a function of flux threshold.}
\smallskip
\begin{minipage}{8.3cm}
\centering
\begin{tabular}{lcccc}\hline \hline
 & \multicolumn{4}{c}{0.4--2.0\,keV}\\
 \hline
Sample /\ Flux limit (cgs)  &         10$^{-13}$& 3\,10$^{-14}$ &10$^{-14}$  & 5\,10$^{-15}$\\
\hline
All  sources                   &   1  &15 & 52 &122 \\
All - extragalactic           &   1  &15 & 52 &122 \\
Active coronae \footnote{The three bright XGPS sources discussed in \S \ref{sms} were added to the spectroscopic sample}           &   1  &12 & 17 & 18\\
Accreting Candidates	      &   0  & 2 &  5 &  6\\
All ids                &   1  &15  &23 & 25\\
All ids + stat ids (90\%)&   1 & 15 & 39 & 55\\
All ids + stat ids (50\%) &  1 & 15 & 44 & 72\\
\hline
\hline
 &\multicolumn{4}{c}{2.0--10.0\,keV}\\
\hline
Sample /\ Flux limit (cgs) &    10$^{-12}$ & 3\,10$^{-13}$ & 10$^{-13}$ & 5\,10$^{-14}$ \\
\hline
All sources                   &  3  & 10 & 36 & 102\\
All - extragalactic  \footnote{ASCA \lnls}         &   3 &  7 & 17 & 48 \\
All - extragalactic  \footnote{\lnls \ from \cite{mateos2008}}         & 3   &  9 &  27 &  73\\
Active coronae   	      &   1 &   1 &  2&   5\\
Accreting Candidates	      & 1   &  4 &  7  & 8 \\
All ids                       & 2   & 6 & 11 & 15 \\
All ids + stat ids (90\%)     & 2   & 6 & 12 & 23\\
All ids + stat ids (50\%)     & 3   & 7 & 17 & 37\\
\hline
\end{tabular}\par
   \vspace{-0.75\skip\footins}
   \renewcommand{\footnoterule}{}
  \end{minipage}
\label{id_stat} 
\end{table}

\begin{figure*}
\begin{center}
\begin{tabular}{cc}
\psfig{figure=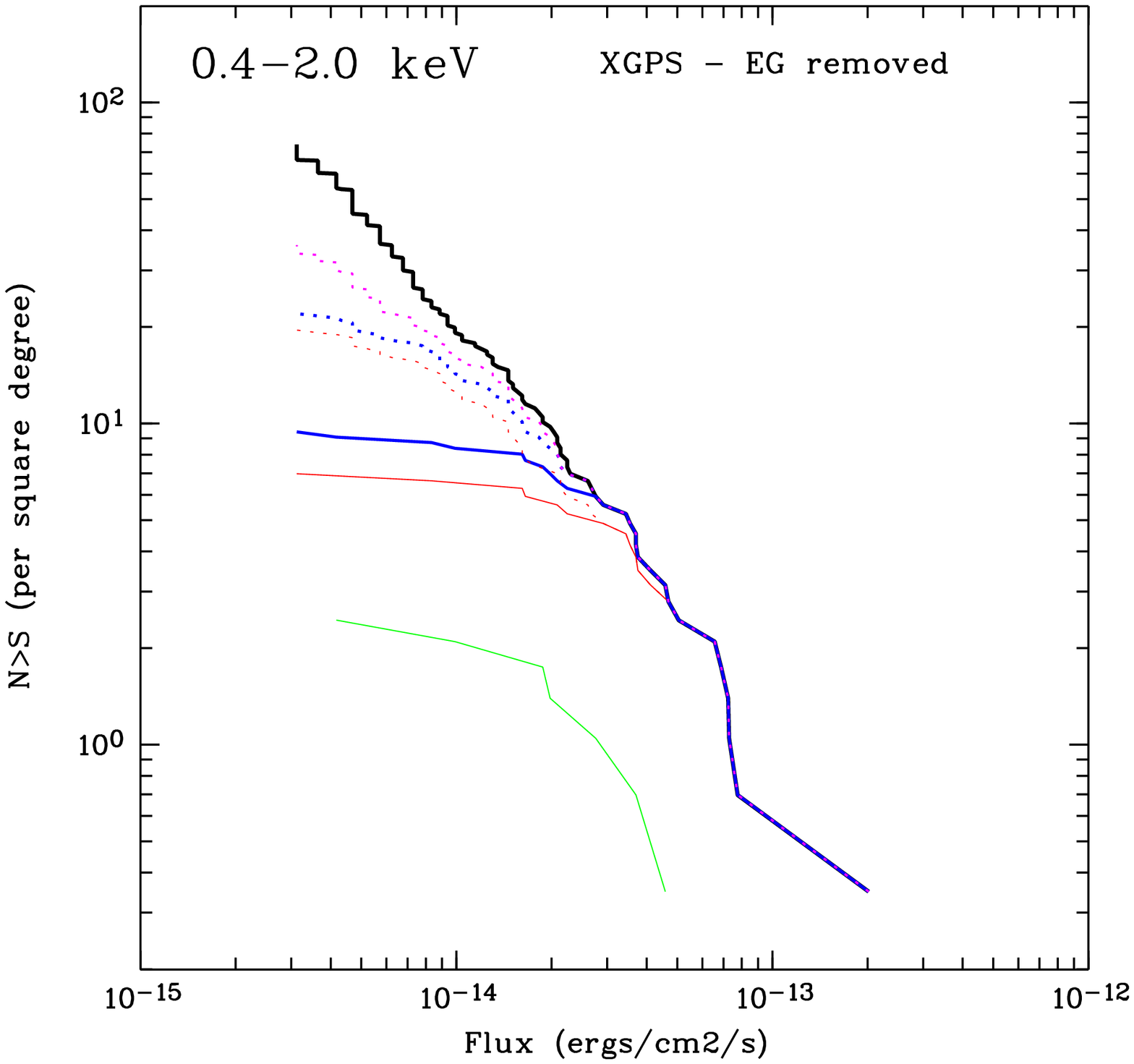,height=9cm,angle=-90              
,bbllx=1.0cm,bburx=21.8cm,bblly=2.5cm,bbury=22.5cm,clip=true}    
& 
\psfig{figure=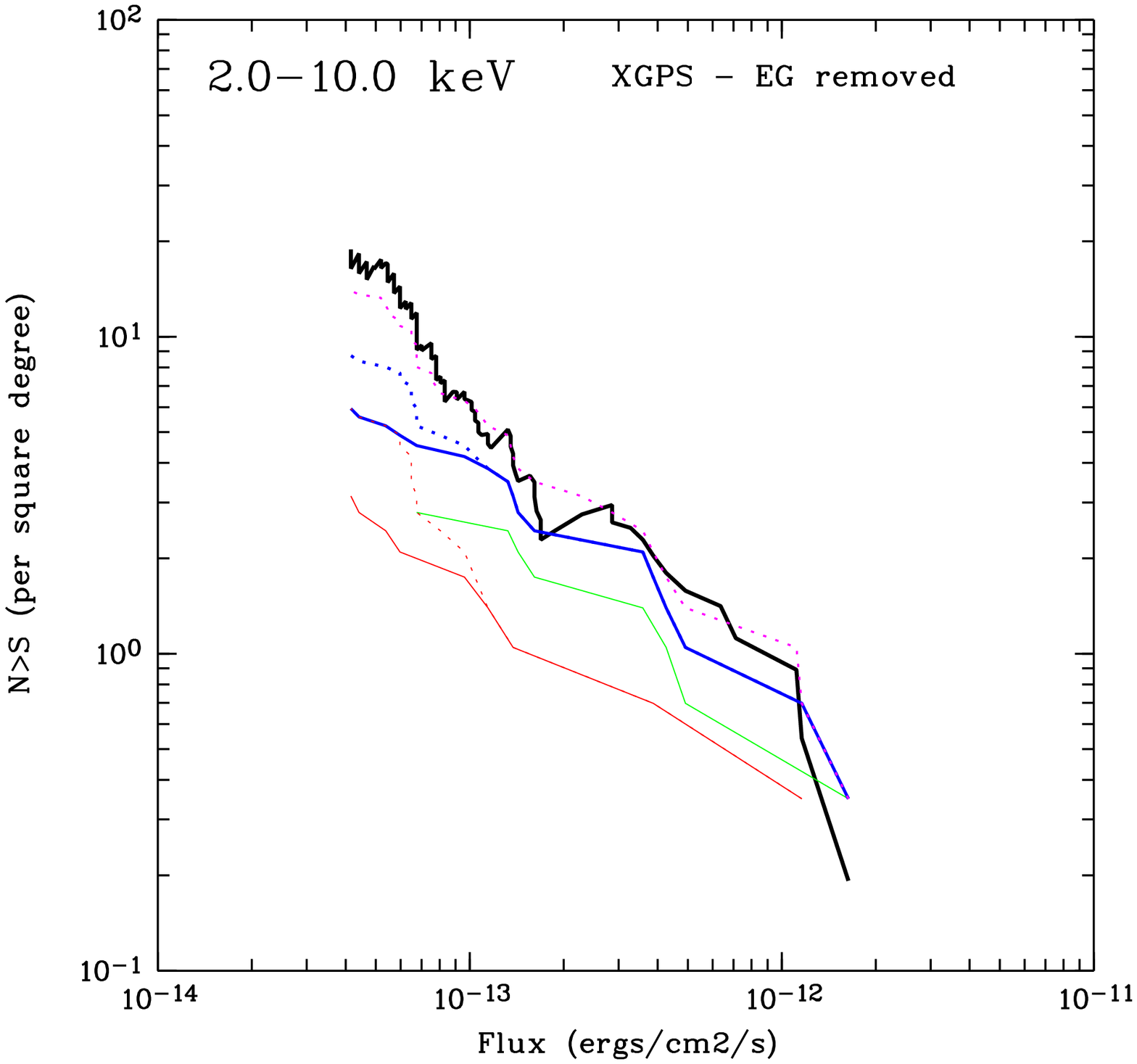,height=9cm,angle=-90              
,bbllx=1.0cm,bburx=21.8cm,bblly=2.5cm,bbury=22.5cm,clip=true}    
\\
\end{tabular}
\caption{Soft and hard XGPS \lnls\ curves after subtracting the ASCA extragalactic contribution (EG)
(thick black lines). The contribution of different categories of source are also shown 
as follows: identified stars (red, see text); accreting candidates (green); 
and all identified (blue). The dashed lines show the result of 
also including matches with USNO-B1.0 and 2MASS which have probabilities larger than 90\% to the assumed stellar (red) and total (blue) identifications. 
The upper magenta dashed line shows the sum of all spectroscopically identified sources added to sources with an individual probability $>$ 50\% to be positively associated 
with a USNO-B1.0 or 2MASS entry.}
\label{egsub_lnls_Xgps}
\end{center}
\end{figure*}

\section{Discussion and Conclusions}

We have successfully identified at the telescope a large number of the brightest X-ray sources detected in the XGPS survey. In addition, we were able to increase the number of identifications by cross-correlating X-ray positions with the USNO-B1.0 and 2MASS catalogues thanks to a carefully calibrated statistical method. The bulk of these statistical identifications with relatively bright optical or near infrared objects are likely stellar coronae based on their soft X-ray spectra. However, our optical campaign has demonstrated that some of them, in particular those exhibiting the hardest X-ray energy distributions, might be massive stars exhibiting particularly hard X-ray emission. Our optical spectroscopic follow-up observations have also confirmed the variety of the astrophysical objects contributing to the low to medium X-ray luminosity population of Galactic X-ray sources. We spectroscopically identify a total of 16 solar-type active stars, mostly among the softest X-ray sources with a high fraction of late Me stars. Furthermore, two early B type stars appear as the probable counterparts of some of our bright XGPS sources.
We have discovered a total of three new cataclysmic variables with optical magnitudes in the range of $\sim$ 23-22 and 0.5--10\,keV fluxes from 0.7 to 4.7$\times$10$^{-13}$\ergscm . Four sources can qualify as massive X-ray binary candidates on the basis of hard X-ray emission and of the presence in the error circle of an early type star exhibiting \Halp\ emission. The optical counterpart of XGPS-3 is likely to be an hyper-luminous star possibly as bright as $\eta$ Carinae. We also report the discovery of an X-ray selected Wolf-Rayet star. The brightest source of our sample, XGPS-1 is a known transient Low-Mass X-ray Binary exhibiting a bursting behaviour. The bright optical magnitude and unusual Bowen C{\sc III}-N{\sc III} emission of XGPS-25 might be the signature of a Low-Lass X-ray Binary in quiescence, although the observational picture could also be consistent with a rare type of CV. The main findings of our work can be summarized as follows:

\begin{enumerate}

\item In the 2--10\,keV flux range from 10$^{-12}$ to 10$^{-13}$\ergscm, we positively identify a large fraction of the hard X-ray sources  with Galactic  objects  at a  rate  consistent with  that expected for the Galactic contribution only. We note, however, that the  diversity of the observed X-ray spectra makes the count to flux conversion somewhat uncertain. Moreover, different normalisations of the extragalactic \lnls\ curves are proposed in the literature. Taken together, these two issues do not allow a precise estimate to be made of the fraction of Galactic sources. However, based on hardness ratio considerations, we find that about half of the hard XGPS sources are likely Galactic objects, in rough agreement with the number estimated by subtracting the extragalactic ASCA \lnls\ curve from the relation obtained using the whole XGPS sample. At the faintest hard X-ray flux considered in our study of \Fx\ = 5$\times$10$^{-14}$\ergscm , we still identify spectroscopically or statistically between 30\% and 50\% of the 'Galactic fraction' of the hard X-ray sources. Since most of the identifications at the faint flux end rely on coincidence in position with a bright optical or infrared candidate, the nature of this optically bright population remains somewhat uncertain. However, drawing on the spectroscopic identifications of the optically bright X-ray brightest objects, we expect them for the most part to be made of particularly active late type stars and of early type stars. 

\item The soft X-ray band is completely dominated by coronally emitting stars with CVs contributing only a small percentage of the identifications. We identify all sources down to \Fx$\sim$3$\times$10$^{-14}$\ergscm . With a mean integrated Galactic \nh\ of $\sim$ 9$\times$10$ ^{22}$ cm$^{-2}$ the background of extragalactic sources is completely screened out by line-of-sight absorption. Although the small number statistics does not allow us to draw robust conclusions, it seems that a larger number of faint Me stars is present in the XGPS survey than was identified in the ROSAT Galactic Plane Survey. Such an evolution is in rough agreement with the prediction of X-ray stellar population models.

\item A few relatively bright active coronae do contribute significantly above 2\,keV. This probably indicates unusually high temperatures for the hottest of the two thin thermal components generally needed to represent the X-ray spectra of active stars \citep[see a recent review in][]{guedel2009}. The observation of open clusters and field stars of different ages has established clear relations between stellar age, X-ray luminosity and temperature. For instance, the young stars in Orion (age $<$ 1\,Myr) require on average $kT_1\,\sim$ 0.8\,keV and  $kT_2\,\sim$ 2.9\,keV, while those of the $\sim$ 160\,Myr old stars in NGC 2516 have already dropped to $kT_1\,\sim$ 0.5\,keV and $kT_2\,\sim$ 1.7\,keV  \citep[see][
and references therein]{sung2008}. In contrast, the population of X-ray bright stars identified in the high galactic latitude XMM-Newton Bright Serendipitous Survey by \cite{xbss2007} is well characterized by $kT_1\,\sim$ 0.3\,keV and $kT_2\,\sim$ 1.0\,keV, indicating a significantly older age on average. The hard X-ray coronae unveiled in the XGPS are thus likely to be very young stars in agreement with the predictions of X-ray stellar populations models. However, a fraction of these hard X-ray emitting stars may also be active binaries, mainly of the RS CVn type. Close binaries in which rotation and coronal activity is maintained by tidal synchronisation do amount to about one third of the ROSAT/Tycho-2 sample of stars \citep{guillout2009}.

\item The hard X-ray source population has a strong component made of  early-type  stars   exhibiting  enhanced  emission   from  a   hard  X-ray component. The physical mechanism responsible for the hard X-ray excess is unclear but could be related to wind collisions in the case of the Wolf-Rayet star counterpart of XGPS-14 and for the likely hyper-luminous star associated with XGPS-3, which shares several similarities with $\eta$ Carinae. XGPS-36 is another example of an early Be star containing a well developed circumstellar disc and exhibiting thin thermal X-ray emission with typical temperatures of $\sim$ 10\,keV, much hotter than usually measured for normal OB stars in which shocks in the high-velocity radiation-pressure driven wind are responsible for the X-ray emission. The origin of the relatively modest X-ray emission (\Lx\ $\sim$ 10$^{32-33}$\ergs ) remains a matter of lively debate. Two distinct mechanisms have been proposed, mainly in the context of $\gamma$-Cas, namely accretion on a companion star, most probably a white dwarf, or magnetic interaction between the early-type star and its decretion disc. XGPS-36 is the eighth identified member of the growing  class of $\gamma$-Cas analogs, and the fourth found so far in XMM-Newton observations of the Galactic Plane.  

\end{enumerate}

Our optical spectroscopic campaign has successfully unveiled the nature of the Galactic sources in the hard X-ray flux interval between 10$^{-12}$ and 10$^{-13}$\ergscm. This \Fx\ range corresponds to the faintest flux limit covered by the ASCA Galactic Plane Survey but still remains between one and two orders of magnitudes above the sensitivity reached by the deepest Chandra pointings. Our identified sample of hard X-ray sources mainly consists of massive stars, possible X-ray binary candidates and CVs with X-ray luminosities in the range of \Lx $\sim$ 10$^{32}$ to 10$^{34}$\ergs\ and located within a few kpc from the Sun. At first glance, the nature of the ``local'' population of hard X-ray sources does not differ from that inferred from infrared follow-up imaging of deep Chandra pointings in the direction of the Galactic Centre \citep[see e.g.][and references therein]{mauerhan2009}. However, the relative numbers of the various classes might well be quite different. Our work also illustrates the difficulty of optically identifying sources in this \Fx\ range, especially the cataclysmic variables as well as the importance of having arcsecond positions so as to alleviate the confusion issue. 

\begin{acknowledgements}
We thank an anonymous referee for useful comments which helped to improve the quality of this paper. We are grateful to O. Herent for carrying out some of the observations presented in this work. This work has been supported in part by the DLR (Deutsches Zentrum f\"ur Luft-und Raumfahrt) under grants 50 OX 0201 and 50 OX 0801. I.N. is supported by the Spanish Ministerio de Ciencia e Innovaci\'on under grants AYA2008-06166-C03-03 and CSD2006-70. This publication makes use of data products from the Two Micron All Sky Survey, which is a joint project of the University of Massachusetts and the Infrared Processing and Analysis Center/California Institute of Technology, funded by the National Aeronautics and Space Administration and the National Science Foundation. The DENIS project has been partly funded by the SCIENCE and the HCM plans of the European Commission under grants CT920791 and CT940627.
It is supported by INSU, MEN and CNRS in France, by the State of Baden-W\"urttemberg 
in Germany, by DGICYT in Spain, by CNR in Italy, by FFwFBWF in Austria, by FAPESP in Brazil, by OTKA grants F-4239 and F-013990 in Hungary, and by the ESO C\&EE grant A-04-046. Jean Claude Renault from IAP was the Project manager.  Observations were  
carried out thanks to the contribution of numerous students and young 
scientists from all involved institutes, under the supervision of  P. Fouqu\'e,  
survey astronomer resident in Chile. The WHT is operated on the island of La
Palma by the Isaac Newton Group in the Spanish Observatorio del Roque
de los Muchachos of the Instituto de Astrof\'{\i}sica de Canarias. 
The observation presented here was taken as part of the ING service programme (proposal SW2005A06). This research has made use of Aladin, of the VizieR catalogue access tool and of Simbad at CDS, Strasbourg, France.

\end{acknowledgements}

\bibliography{xgps}

\newpage
\begin{landscape}
\voffset=7truecm
\centering
\renewcommand{\footnoterule}{}  
\pagestyle{empty}
\begin{table*}
\scriptsize
\tabcolsep=5pt
\caption{Optical identifications of the brightest XGPS sources. All coordinates are given in J2000. For each XGPS-id source, we list the corresponding XGPS-I entry in \cite{hands2004}, the matching 2XMM entry and its combined multi-detection coordinates (when available) and the position of the optical identification. The 90\% confidence error radius, $r_{90}$, is that of the 2XMM entry, while a constant value of 4\arcsec\ is assumed for XGPS-I only entries \citep{hands2004}. $d$~x-o is the offset of the proposed optical counterpart from the 2XMM (or XGPS-I) position. MOS count rates refer to the broad (0.4--6.0\,keV) band and the hardness ratios are those of the original XGPS processing. In the Class column , AC stands for active coronae. R magnitudes are extracted from the optical catalogue identifications, apart from sources 9, 10, 13 , 15 and 24 for which R is estimated from the flux calibrated spectra.}
\begin{tabular}{rcccccccccrrrcrll}
\hline
XGPS                &XGPS-I name         &       \multicolumn{2}{c}{XGPS}                   &2XMM name           &      \multicolumn{2}{c}{2XMM}                    &    \multicolumn{2}{c}{Optical}                            &$r_{90}$                 &$d$ x-o               &MOS cnts            &HR                  &Class               &R                   &Type                &Optical Id          \\   
id                  &                    &RA                  &DEC                 &                    &RA                  &DEC                 &RA                  &DEC                 &arcsec              &arcsec              &cts/ksec            &(MOS)                    &                    &                    &                    &                    \\   
 \hline
   1&XGPS-I J182833-103659&18 28 33.96&-10 36 59.5&2XMM J182833.7-103700&18 28 33.72&-10 37 00.5&           &           &   2.16&       & 427.00& 0.91&LMXB &     &          &                    \\   
   2&XGPS-I J182845-111711&18 28 45.94&-11 17 11.0&2XMM J182845.5-111710&18 28 45.51&-11 17 10.3&18 28 45.48&-11 17 10.7&   1.32&   0.50&  78.50&-0.86&AC   &11.71&M1e       &GSC2 S3003020461    \\   
   3&XGPS-I J183208-093906&18 32 08.93&-09 39 06.2&2XMM J183208.7-093905&18 32 08.78&-09 39 05.4&18 32 08.94&-09 39 06.0&   2.33&   2.45&  68.80& 0.79&HMXB?&16.80&Be?       &GSC2 S300302241761  \\   
   5&XGPS-I J182856-095413&18 28 56.04&-09 54 13.5&2XMM J182855.9-095414&18 28 55.96&-09 54 14.1&           &           &   1.73&       &  44.60& 0.86&UNID &     &          &                    \\   
   6&XGPS-I J182847-101334&18 28 47.76&-10 13 34.6&2XMM J182847.6-101337&18 28 47.69&-10 13 37.8&18 28 47.71&-10 13 35.5&   1.76&   2.32&  39.70&-0.97&AC   & 8.60&G0V       &HD 170248           \\   
   7&XGPS-I J182524-114525&18 25 24.48&-11 45 25.1&2XMM J182524.5-114525&18 25 24.58&-11 45 25.2&           &           &   2.26&       &  31.50& 0.67&UNID &     &          &                    \\   
   8&XGPS-I J183038-100248&18 30 38.11&-10 02 48.6&2XMM J183038.2-100246&18 30 38.24&-10 02 46.5&           &           &   2.29&       &  26.70& 0.68&UNID &     &          &                    \\   
   9&XGPS-I J183251-100106&18 32 51.65&-10 01 06.7&2XMM J183251.4-100106&18 32 51.48&-10 01 06.0&18 32 51.58&-10 01 05.4&   2.34&   1.55&  25.20& 0.27&CV   &21.57&          &A                   \\   
  10&XGPS-I J182745-120606&18 27 45.55&-12 06 06.8&                     &           &           &18 27 45.40&-12 06 06.6&   4.00&   2.20&  24.10& 0.56&HMXB?&21.60&Be        &A                   \\   
  11&XGPS-I J182506-120433&18 25 06.85&-12 04 33.6&2XMM J182506.7-120430&18 25 06.73&-12 04 30.3&18 25 06.87&-12 04 29.6&   2.63&   2.14&  17.60& 0.15&AC   &13.43&K0        &GSC2 S300111014094  \\   
  13&XGPS-I J182757-120941&18 27 57.24&-12 09 41.6&                     &           &           &18 27 58.03&-12 09 40.6&   4.00&  11.60&  16.10&-0.72&AC   & 9.40&G5        &TYC 5698-611-1      \\   
  14&XGPS-I J183116-100921&18 31 16.51&-10 09 21.2&2XMM J183116.5-100926&18 31 16.55&-10 09 26.3&18 31 16.51&-10 09 24.3&   2.04&   2.03&  14.90& 0.69&WR   &13.34&WN8       &USNO-A2.0 0750-13397443 \\   
  15&XGPS-I J182814-103728&18 28 14.23&-10 37 28.9&2XMM J182814.2-103726&18 28 14.22&-10 37 26.8&18 28 14.30&-10 37 28.8&   3.07&   2.31&  14.40& 0.86&HMXB?&20.30&          &2MASS J18281430-1037288  \\   
  16&XGPS-I J183035-102105&18 30 35.91&-10 21 05.9&2XMM J183035.7-102102&18 30 35.71&-10 21 02.8&18 30 35.85&-10 21 04.4&   2.22&   2.67&  14.20&-0.92&AC   &13.31&M4e/M5    &GSC2 S300302320885  \\   
  17&XGPS-I J182625-112854&18 26 25.73&-11 28 54.0&2XMM J182626.7-112854&18 26 26.74&-11 28 54.3&           &           &   2.31&       &  13.80& 0.81&UNID &     &          &                    \\   
  18&XGPS-I J182756-110450&18 27 56.84&-11 04 50.9&2XMM J182756.7-110448&18 27 56.80&-11 04 48.1&           &           &   2.32&       &  11.30& 0.91&UNID &     &          &                    \\   
  19&XGPS-I J182951-100441&18 29 51.57&-10 04 41.0&                     &           &           &18 29 51.77&-10 04 45.9&   4.00&   5.70&  10.90&-1.00&B    &15.40&$\sim$B   &USNO-A2.0 0750-13320307\\   
  20&XGPS-I J182639-114216&18 26 39.76&-11 42 16.4&2XMM J182639.6-114216&18 26 39.69&-11 42 16.1&18 26 39.56&-11 42 14.9&   2.44&   2.28&  10.60&-0.78&AC   &13.74&M2e       &GSC2 S300111036891  \\   
  21&XGPS-I J183103-095813&18 31 03.69&-09 58 13.5&2XMM J183103.7-095813&18 31 03.73&-09 58 13.6&18 31 03.73&-09 58 17.7&   3.02&   4.07&  10.20&-0.47&AC   &16.92&M6e       &GSC2 S30030222700   \\   
  22&XGPS-I J183031-094030&18 30 31.99&-09 40 30.5&2XMM J183031.9-094030&18 30 31.92&-09 40 30.8&           &           &   1.79&       &  10.00& 1.00&UNID &     &          &                    \\   
  23&XGPS-I J183017-095326&18 30 17.12&-09 53 26.3&2XMM J183017.0-095326&18 30 17.03&-09 53 26.3&18 30 17.01&-09 53 27.0&   1.64&   0.77&   9.40&-0.44&AC   &14.84&M2e       &GSC2 S30030226319   \\   
  24&XGPS-I J183113-094924&18 31 13.12&-09 49 24.5&2XMM J183113.0-094922&18 31 13.03&-09 49 22.6&18 31 13.02&-09 49 25.4&   2.80&   2.80&   9.20& 0.20&CV   &21.60&          &B                   \\   
  25&XGPS-I J182854-112656&18 28 54.86&-11 26 56.1&2XMM J182854.6-112656&18 28 54.68&-11 26 56.2&18 28 54.50&-11 26 56.2&   2.75&   2.69&   8.80& 0.18&LMXB?&15.87&          &USNO-B1.0 0785-0417035\\   
  27&XGPS-I J182911-105544&18 29 11.61&-10 55 44.9&2XMM J182911.6-105545&18 29 11.62&-10 55 45.4&18 29 11.46&-10 55 47.1&   2.62&   2.81&   8.50&-0.76&AC   &17.84&M5e       &GSC2 S30030233935   \\   
  28&XGPS-I J182827-111751&18 28 27.77&-11 17 51.8&2XMM J182827.5-111749&18 28 27.51&-11 17 50.0&           &           &   1.81&       &   8.40& 0.67&UNID &     &          &                    \\   
  29&XGPS-I J182740-113955&18 27 40.40&-11 39 55.3&2XMM J182740.3-113953&18 27 40.37&-11 39 53.7&18 27 40.40&-11 39 52.6&   1.47&   1.21&   7.60&-0.83&AC   &17.05&M5e       &GSC2 S300111038707  \\   
  30&XGPS-I J182740-102812&18 27 40.08&-10 28 12.4&2XMM J182740.0-102811&18 27 40.00&-10 28 11.1&18 27 40.08&-10 28 11.8&   2.73&   1.27&   7.40&-0.86&AC   &13.65&M1e       &GSC2 S300302037679  \\   
  31&XGPS-I J182930-105926&18 29 30.87&-10 59 26.3&2XMM J182930.8-105927&18 29 30.87&-10 59 27.9&           &           &   2.97&       &   7.30& 0.65&UNID &     &          &                    \\   
  33&XGPS-I J182628-120922&18 26 28.66&-12 09 22.5&2XMM J182628.3-120926&18 26 28.32&-12 09 26.2&18 26 28.48&-12 09 25.5&   3.07&   2.47&   6.90&-0.86&AC   &13.70&K0        &GSC2 S300111332915  \\   
  34&XGPS-I J183236-101144&18 32 36.64&-10 11 44.8&2XMM J183236.3-101146&18 32 36.39&-10 11 46.9&18 32 36.16&-10 11 44.6&   2.96&   4.15&   6.90& 0.75&AC?? &14.01&K1        &GSC2 S300302326629  \\   
  36&XGPS-I J183015-104538&18 30 15.94&-10 45 38.5&2XMM J183015.8-104538&18 30 15.81&-10 45 38.5&18 30 15.94&-10 45 38.4&   2.67&   1.94&   6.60& 0.89&HMXB?&12.74&Be        &GSC2 S300302371     \\   
  42&XGPS-I J182833-102652&18 28 33.60&-10 26 52.0&2XMM J182833.3-102650&18 28 33.38&-10 26 50.9&           &           &   2.72&       &   5.90& 0.47&UNID &     &          &                    \\   
  45&XGPS-I J182746-110255&18 27 46.33&-11 02 55.1&2XMM J182746.1-110250&18 27 46.20&-11 02 50.7&           &           &   1.89&       &   5.60& 0.21&UNID &     &          &                    \\   
  46&XGPS-I J183124-104751&18 31 24.77&-10 47 51.5&2XMM J183124.7-104748&18 31 24.70&-10 47 48.2&18 31 24.70&-10 47 46.0&   3.05&   2.16&   5.60&-1.00&AC   &     &B9V       &BD-10 4713B         \\   
  47&XGPS-I J182553-112713&18 25 53.86&-11 27 13.0&2XMM J182554.7-112711&18 25 54.76&-11 27 11.6&           &           &   2.52&       &   5.50& 0.92&UNID &     &          &                    \\   
  49&XGPS-I J182703-113714&18 27 03.66&-11 37 14.1&2XMM J182703.7-113713&18 27 03.75&-11 37 13.5&18 27 03.86&-11 37 13.9&   1.98&   1.65&   5.36&-0.21&AC   &15.49&G9        &GSC2 S300111041110  \\   
  55&XGPS-I J182813-103808&18 28 13.84&-10 38 08.7&2XMM J182813.8-103808&18 28 13.84&-10 38 08.1&18 28 13.82&-10 38 10.6&   3.53&   2.56&   5.00&-0.83&AC   &15.91&M4e       &GSC2 S300302033754  \\   
  56&XGPS-I J182828-102357&18 28 28.14&-10 23 57.3&2XMM J182827.9-102359&18 28 27.95&-10 23 59.1&18 28 28.13&-10 24 02.2&   2.75&   4.11&   4.90& 0.84&B    &14.78&B3        &GSC2 S300302038718  \\   
  57&XGPS-I J182938-094900&18 29 38.87&-09 49 00.8&2XMM J182938.8-094857&18 29 38.85&-09 48 57.6&           &           &   5.71&       &   4.90&-0.15&UNID &     &          &                    \\   
  60&XGPS-I J182511-115726&18 25 11.68&-11 57 26.3&2XMM J182511.5-115726&18 25 11.56&-11 57 27.0&18 25 11.56&-11 57 26.2&   3.49&   0.78&   4.80&-0.74&AC   &     &A9        &TYC 5698-4833 -1    \\   
  63&XGPS-I J182607-120627&18 26 07.17&-12 06 27.0&2XMM J182606.9-120626&18 26 06.95&-12 06 26.7&18 26 07.18&-12 06 25.2&   3.08&   3.66&   4.70&-0.31&AC?? &16.02&A8        &GSC2 S300111013026  \\   
  65&XGPS-I J182614-110549&18 26 14.69&-11 05 49.6&2XMM J182614.4-110548&18 26 14.46&-11 05 48.1&18 26 14.42&-11 05 50.2&   2.78&   2.14&   4.50& 0.11&CV   &19.41&          &USNO-B1.0 0789-0404378\\   
  69&XGPS-I J182706-112437&18 27 06.40&-11 24 37.4&2XMM J182706.3-112436&18 27 06.30&-11 24 36.4&           &           &   2.03&       &   4.20& 0.93&UNID &     &          &                    \\   

\hline
\end{tabular}
\label{xgps_tab}
\end{table*}
\end{landscape} 


\onecolumn

\newpage

\onecolumn

\begin{center}


\end{center}

\begin{table*}[h]
\begin{center}
\caption{Additional cross-identifications of optically identified XGPS sources.}
\begin{tabular}{rcl}
\hline
XGPS id & XGPS-I name & Identifications \\
\hline
1 & XGPS-I J182833-103659 & SAX J1828.5-1037 \\
2 & XGPS-I J182845-111711 & AX J182846-1116 \\
3 & XGPS-I J183208-093906 & AX J1832.1-0938, DENISJ183208.4-093905, 2MASS 18320893-0939058, \\
  &                       & IRAS 18293-094, NVSS 183209-093907, GPSR 22.154-0.154\\
14 & XGPS-I J183116-100921 & AX~J183116-1008, DENIS J183116.5-100924,2MASS 18311653-1009250, \\
   &                       & SSTGLMA G021.6064-00.1970 \\
15 & XGPS-I J182814-103728 & SSTGLMA G020.8457+00.2476 \\
25 & XGPS-I J182854-112656 & GSC2 S300111083138, DENIS J182854.6-112655, 2MASS 18285460-11265612\\
36 & XGPS-I J183015-104538 & USNO-A2.0 0750-13340834, 2MASS 18301593-1045384, SSTGLMA G020.9564-00.2565 \\
\hline
\end{tabular}
\label{crossids}
\end{center}
\end{table*}

\begin{table*}[h]
\begin{center}
\caption{Summary of X-ray spectral fits. \nh\ are in units of 10$^{22}\,$cm$^{-2}$. All errors are 90\% confidence level.}
\begin{tabular}{rlclll}
\hline
XGPS  & EPIC  & Nbr of source & Model & Fit  &Fits results  \\
id    & Cameras& photons &            & Quality & \\ \hline
3   & MOS1\&2      & 569  & phabs*powerl & C=37.8/40\,dof & \nh =2.69$^{+0.76}_{-0.67}$; $\Gamma$=1.45$^{+0.42}_{-0.40}$\\
3   & MOS1\&2      & 569  & phabs*mekal  & C=53.7/41\,dof &\nh =3.8 (fixed); kT=4.2$\pm$ 0.6\,keV\\
9   & pn           & 177  & tbabs*mekal (kT=20\,keV fixed)& $\chi^{2}$=13.3/11\,dof & \nh\ $\sim 0.8$ \\
14  & pn + MOS1\&2 & 362  & phabs*mekal & C=67.8/57\,dof &\nh =5.21$_{-1.25}^{+1.38}$; kT=2.14$_{-0.52}^{+1.09}$\,keV \\
14  & pn + MOS1\&2 & 362  & phabs*powerl & C=86.5/57\,dof & \nh =5.34$_{-1.68}^{+1.99}$; $\Gamma$=2.91$_{-0.86}^{+0.96}$ \\
15  & pn + MOS1\&2 & 279  & phabs*powerl & C=31.8/28\,dof & \nh =7.8$^{+5.9}_{-3.4}$; $\Gamma$=2.4$^{+1.5}_{-1.0}$ \\
24  & pn           & 54   & tbabs*mekal (kT=5\,keV fixed) & N/A & \nh = 0.2-2.4  \\
25  & pn + MOS1\&2 & 110  & phabs*mekal & C=21.6/31\,dof& \nh =0.48$_{-0.17}^{+0.31}$; kT$>$10\,keV \\
25  & pn + MOS1\&2 & 110  & phabs*powerl & C=12.5/12\,dof& \nh =0.39$_{-0.29}^{+0.44}$; $\Gamma$=1.07$^{+0.32}_{-0.32}$ \\
65  & pn           & 54   & tbabs*mekal (kT=5\,keV fixed) & N/A & \nh =0.8-2.4  \\
\hline
\end{tabular}
\label{xrayspectralsummary}
\end{center}
\end{table*}
\end{document}